\documentclass[aps,preprint,epsfig,rotate]{revtex4}

\begin{document}
%
\title{Metric gravity in the Hamiltonian form. Canonical transformations. 
       Dirac's modifications of the Hamilton method and integral invariants
       of the metric gravity}
 \author{Alexei M. Frolov}
 \email[E--mail address: ]{alex1975frol@gmail.com}

\affiliation{Department of Applied Mathematics, \\
 University of Western Ontario, London, Ontario N6H 5B7, Canada}

\date{\today}

\begin{abstract}
Two different Hamiltonian formulations of the metric gravity are discussed and applied to describe a free 
gravitational field in the $d$ dimensional Riemann space-time. Theory of canonical transformations, which 
relate equivalent Hamiltonian formulations of the metric gravity, is investigated in details. In particular, 
we have formulated the conditions of canonicity for transformation between the two sets of dynamical 
variables used in our Hamiltonian formulations of the metric gravity. Such conditions include the ordinary 
condition of canonicity known in classical Hamilton mechanics, i.e., the exact coincidence of the Poisson (or 
Laplace) brackets which are determined for the both new and old dynamical Hamiltonian variables. However, in 
addition to this any true canonical transformations defined in the metric gravity, which is a constrained 
dynamical system, must also guarantee the exact conservation of the total Hamiltonians $H_t$ (in the both 
formulations) and preservation of the algebra of first-class constraints. We show that Dirac's modifications 
of the classical Hamilton method contain a number of crucial advantages, which provide an obvious superiority 
of this method in order to develop various non-contradictory Hamiltonian theories of many physical fields, 
when a number of gauge conditions are also important. Theory of integral invariants and its applications to 
the Hamiltonian metric gravity are also discussed. For Hamiltonian dynamical systems with first-class 
constraints this theory leads to a number of peculiarities some of which have been investigated. 
\end{abstract}

\maketitle

%

\section{Introduction}

The main goal of this study is a deep analysis of canonical transformations of the Hamiltonian dynamical variables 
which are applied in the metric gravity. We investigate general properties of such transformations and formulate 
some criteria of canonicity. Another our aim is to discuss modifications made by Dirac \cite{Dir58}, \cite{Dir50} 
and \cite{Dir64} in the classical Hamilton method. We want to show that this Dirac's approach has many crucial 
advantages for the development and following improvement of various Hamiltonian formulations for the free 
gravitational field in metric gravity, where $2 d$ additional gauge conditions exist. We also introduce the 
integral invariants of the metric gravity and discuss applications of these invariants to the current and new 
Hamiltonian formulations of the metric gravity. 

In metric gravity the classical gravitational field is described as a symmetric tensor field which is defined in the 
$d$ dimensional Riemann space-time. As is well known the general theory of metric gravitational field(s) have been 
created more than 100 years ago by A. Einstein and it is based on his fundamental idea that the actual gravitational 
field has a tensor nature and it is defined in the four-dimensional (Riemann) space-time. Since then the gravitational 
field is designated as covariant components of the fundamental metric tensor $g_{\alpha\beta}$. In general, the metric 
gravity can be developed in the $d$ dimensional space-time (or $d-$space), where $d \ge 3$, while the time is always 
one-dimensional. In other words, there is no need to restrict ourselves to the four-dimensional case only, where $d$ 
= 4. In this study we also deal with the $d$-dimensional Riemann space-time. Everywhere below the notation 
$\overline{x}$, designates the $d-$dimensional vector which has the contravariant components $x^{\alpha}$, while its 
covariant components are designated as $x_{\alpha}$, where $\alpha = 0, 1, \ldots, d - 1$. The temporal component (or 
time component) of the coordinate vector $\overline{x}$ is $x^{0}$, while its spatial components are $x^{k}$ ($k = 1, 
2, \ldots, d - 1$). The same rule is applied everywhere in this study: components of $d-$vectors are labelled by Greek 
indices, while spatial components of these vectors are denoted by Latin indices. In respect to this agreement all 
components of the covariant fundamental tensor (or metric tensor) are designated as $g_{\alpha\beta}$ (see above), 
while the notation $g^{\alpha\beta}$ stands for the components of contravariant fundamental (or metric) tensor 
\cite{Kochin}, \cite{Dash}. The determinant of the $g_{\alpha\beta}$ is called the fundamental determinant and it is 
denoted by letter $g$. In the metric gravity the numerical value of $g$ is always real and negative, but the $- g$ 
value is positive, which allows one to operate with the expression such as $\sqrt{- g}$ and consider functions of 
$\sqrt{- g}$. Any suffix with a comma before it denotes differentiation according to the general scheme $F_{,\mu} = 
\frac{d F}{d x^{\mu}}$. In particular, the temporal derivative are always designated as $F_{,0} \Bigl( = 
\frac{d F}{d x^{0}} \Bigr)$. For an arbitrary metric-dependent functional $F(g_{\alpha\beta})$ the notation 
$F_{,\gamma}$ means $F_{,\gamma} = \Bigl(\frac{\partial F}{\partial g_{\mu\nu}}\Bigr) g_{\mu\nu,\gamma}$, etc. 

This paper has the following structure. The next two Sections play the role of `introductory part' for our current 
analysis. In particular, in the next Section we introduce the regular $\Gamma - \Gamma$ Lagrangian of the metric 
gravity. Then, by using this $\Gamma - \Gamma$ Lagrangian we define the momenta $\pi^{\mu\nu}$ which are also the 
dynamical (Hamiltonian) variables. These momenta are considered as variables which are dynamically conjugate to 
the corresponding generalized coordinates $g_{\alpha\beta}$. In general, the momenta $\pi^{\mu\nu}$ are defined as 
the partial derivatives of the $\Gamma - \Gamma$ Lagrangian in respect to the `velocities' $\frac{\partial 
g_{\mu\nu}}{\partial t} = g_{\mu\nu,0}$. The arising set of $2 d$ dynamical variables $\{ g_{\alpha\beta}, 
\pi^{\mu\nu} \}$ includes all variables which are needed to develop the Hamiltonian formulation of the metric 
gravity. In particular, the both canonical $H_C$ and total $H_{t}$ Hamiltonians of the metric gravity are written 
as the quadratic functions of all space-like momenta $\pi^{pq}$ and as a linear combination of temporal momenta 
$\pi^{0 \mu} (= \pi^{\mu 0})$. This essentially means that the free gravitational field is a constrained dynamical 
(Hamiltonian) system \cite{K&K}, \cite{Tyut} and all $d$ temporal momenta $\pi^{0\mu} (= \pi^{\mu 0})$ of this field 
can rigorously be defined only as the primary constraints. The commutators (or Poisson brackets) of these primary 
constraints with the canonical and total Hamiltonians produce $d$ secondary constraints. In general, the properly 
defined Poisson brackets always play a central role in Hamiltonian formulations of many physical theories. Briefly, 
we can say that the Poisson brackets is the most important working tool of any consistent Hamiltonian theory. 

In Section V by using our Poisson brackets we derive the Hamiltonian equations of motion for all dynamical variables 
of the metric gravity. Analogous Hamilton equations for the primary and secondary first-class constraints are also 
derived and discussed. The Dirac closure of this Hamiltonian formulation is explicitly demonstrated. Then we consider 
another Hamiltonian formulation of the metric gravity developed by Dirac \cite{Dir58}. For this Hamiltonian formulation 
we also derive the corresponding Hamilton equations of motion and determine the Poisson brackets between all essential 
first-class constraints (see, also our Appendix A). Then we define the canonical transformations of the dynamical 
Hamiltonian variables which relate the both these Hamiltonian formulations, i.e., \cite{Dir58} and \cite{K&K}. The 
criteria of canonicity for arbitrary transformations of the dynamical variables are formulated in the Section VI. Then 
we discuss modifications made by Dirac in the classical Hamiltonian method \cite{Dir58} - \cite{Dir64}. We have shown 
explicitly that these Dirac's modifications allowed him to create a logically closed and transparent Hamiltonian 
approach which has many advantages to study actual motions of various physical fields when numerous gauge conditions 
must also be taken into account. Here we also formulate the new principle of the ``complete reverse recovery'' which 
must be applied to any Hamiltonian formulation of the metric gravity to check its validity and correct relations with 
the original Einstein's field equations. This simple and physically clear principle can be used to ``separate the wheat 
from the chaff'' (Matthew, 3:13) in Hamiltonian gravity. It allows us to operate only with the true Hamiltonian 
formulations of metric gravity and discard the fake ones.

In Section VIII we introduce integral invariants of the metric gravity. In some sense this is a central part of this 
study, since the method of integral invariants allows one to create a real foundation for the Hamiltonian formulations 
of the metric gravity. In particular, by applying the integral invariants of metric gravity one can easy perform all 
steps of the rigorous Hamiltonian procedure and formulate various criteria of canonicity which can be applied actual 
transformations of dynamical variables. Concluding remarks can be found in the last Section. This paper also includes 
three Appendixes. Appendix A is a `pure technical' part, which contains derivations of some formulas. These formulas 
and expressions are of paramount importance for this study, but they could not be included in the main text, since 
this would damage the logic and harmony of our presentation. In Appendix B we discuss a number of tricky moments which 
traditionally complicate the correct definition of canonicity in classical and quantum mechanics. In Appendix C we 
explicitly derive some important formulas for the integral invariants. Along with the discussions of the latest 
achievements in Hamiltonian metric gravity, we also wanted to write a simple and transparent article which can be 
understood by any theoretical physicist who is familiar with the modern Hamiltonian methods developed for constrained 
dynamical systems. 

\section{Regular $\Gamma - \Gamma$ Lagrangian density of the metric gravity}

In this Section we introduce the regularized (or regular) Lagrangian density of metric gravity. As is well known 
(see, e.g., \cite{Carm} and \cite{LLTF}) the original Lagrangian density of the metric gravity coincides with the 
integrand in the Einstein-Hilbert integral-action $L_{EH}$ which equals to the product of scalar (or Gauss) 
curvature of the $d-$dimensional space $R = g^{\alpha\beta} R_{\alpha\beta}$ and factor $\sqrt{- g}$, which is 
Jacobian of the transformation from the flat space to the curved Riemann space  (see, e.g., \cite{Carm} and 
\cite{LLTF}). The invariant integral $\int R \sqrt{- g} d\Omega$ is called the gravitational action. The explicit 
form of Lagrangian density is $L_{EH} = \sqrt{- g} R = \sqrt{- g} g^{\alpha\beta} R_{\alpha\beta} = \sqrt{- g} 
g^{\alpha\beta} g^{\gamma\sigma} R_{\gamma\alpha\sigma\beta}$, where $R = g^{\alpha\beta} R_{\alpha\beta}$ is the 
scalar (or Gauss) curvature of $d-$dimensional space-time, while $R_{\alpha\beta}$ is the Ricci tensor 
\begin{eqnarray}
 R_{\alpha\beta} = \frac{\partial \Gamma^{\gamma}_{\alpha\beta}}{\partial x^{\gamma}} - \frac{\partial 
 \Gamma^{\gamma}_{\alpha\gamma}}{\partial x^{\beta}} + \Gamma^{\gamma}_{\alpha\beta} 
 \Gamma^{\lambda}_{\gamma\lambda} - \Gamma^{\lambda}_{\alpha\gamma} \Gamma^{\gamma}_{\beta\lambda} 
 \; , \; {\rm or} \; \; R_{\alpha\beta} = g^{\mu\nu} R_{\mu\alpha\nu\beta} = g^{\nu\mu} 
  R_{\nu\beta\mu\alpha} = R_{\beta\alpha} \; \label{equH1}  
\end{eqnarray}
In this equation and everywhere below in this study the notation $\Gamma^{\gamma}_{\alpha\beta} = \frac12 
g^{\gamma\nu} \Bigl( \frac{\partial g_{\nu\alpha}}{\partial x^{\beta}} + \frac{\partial g_{\nu\beta}}{\partial 
x^{\alpha}} - \frac{\partial g_{\alpha\beta}}{\partial x^{\nu}} \Bigr)$ are the Cristoffel symbols of the second 
kind (see, e.g., \cite{Kochin}). The Ricci tensor $R_{\alpha\beta} = g^{\gamma\sigma} R_{\gamma\alpha\sigma\beta}$ 
is simply related to the Einstein tensor $G_{\alpha\beta} = g^{\gamma\sigma} R_{\alpha\gamma\sigma\beta}$, since 
$R_{\alpha\beta} = - G_{\alpha\beta}$ \cite{Kochin}. In this notation the governing equations of the free 
gravitational field (famous Einstein's equations) are written in one of the following forms: $R_{\alpha\beta} = 0 
= G_{\alpha\beta}$. Any Hamiltonian formulation of the metric gravity must reproduce these original field equations 
exactly and unambiguously. This is the new fundamental principle of the ``complete reverse recovery'' and its 
applications to various new and old Hamiltonian formulations of the metric gravity allows one quickly ``to separate 
the wheat from the chaff'' (Mathew, 13, 24 - 30) (see below).   

An alternative (but equivalent!) form of the same Lagrangian density is written as: $L_{EH} = \sqrt{- g} 
g^{\alpha\beta} g^{\gamma\sigma} R_{\gamma\alpha \sigma\beta}$, where the notation $R_{\alpha\beta 
\gamma\sigma}$ designates the Riemann curvature tensor (or Riemann-Cristoffel tensor) which is 
\begin{eqnarray}
 R_{\alpha\beta \gamma\sigma} = \frac12 \Bigl[ \frac{\partial^{2} g_{\alpha\sigma}}{\partial x^{\beta} 
 \partial x^{\gamma}} + \frac{\partial^{2} g_{\beta\gamma}}{\partial x^{\alpha} \partial x^{\sigma}} 
  - \frac{\partial^{2} g_{\alpha\gamma}}{\partial x^{\beta} \partial x^{\sigma}} - \frac{\partial^{2} 
  g_{\beta\sigma}}{\partial x^{\alpha} \partial x^{\gamma}} \Bigr] + \Gamma_{\rho, \alpha\sigma} 
  \Gamma^{\rho}_{\beta\gamma} - \Gamma_{\rho, \beta\sigma} \Gamma^{\rho}_{\alpha\gamma} \; \; , \; 
 \label{two} 
\end{eqnarray} 
where $\Gamma_{\lambda, \mu\nu} = \frac12 \Bigl( \frac{\partial g_{\nu\alpha}}{\partial x^{\beta}} + 
\frac{\partial g_{\nu\beta}}{\partial x^{\alpha}} - \frac{\partial g_{\alpha\beta}}{\partial x^{\nu}} \Bigr)$ 
are the Cristoffel symbols of the first kind. The Riemann-Cristoffel tensor defined in Eq.(\ref{two}) is a 
covariant tensor of the fourth rank. As follows from the last equation the Einstein-Hilbert Lagrangian density 
$L_{EH} = \sqrt{- g} g^{\alpha\gamma} g^{\beta\sigma} R_{\alpha\beta \gamma\sigma}$ contains a number of 
second-order derivatives $\frac{\partial^{2} g_{\alpha\beta}}{\partial x^{\gamma} \partial x^{\lambda}}$ and 
cannot be used directly in the principle of least action. However, as it follows from Eqs.(\ref{equH1}) and 
(\ref{two}) all these derivatives of the second order are included in the Lagrangian density $L_{EH}$ only as 
a linear combination with the constant coefficients which do not contain any derivative of the metric tensor. 
Such a linearity of the invariant integral $\int R \sqrt{- g} d\Omega$ upon the second-order derivatives of the 
metric tensor can be used to transform this integral (by means of Gauss theorem) to the integral which does not 
include any second-order derivative. After a few simple transformations the invariant integral $S_g$ is reduced 
to the form 
\begin{eqnarray}
 \int R \sqrt{- g} d\Omega = \int g^{\alpha\beta} \Bigl( \Gamma^{\lambda}_{\alpha\gamma} 
 \Gamma^{\gamma}_{\beta\lambda} - \Gamma^{\gamma}_{\alpha\beta} \Gamma^{\lambda}_{\gamma\lambda}\Bigr) 
 \sqrt{- g} d\Omega + \int \frac{\partial \Bigl[\sqrt{- g} \Bigl( g^{\alpha\beta} 
 \Gamma^{\gamma}_{\alpha\beta} - g^{\alpha\gamma} \Gamma^{\beta}_{\alpha\beta} 
 \Bigr)\Bigr]}{\partial x^{\gamma}} d\Omega \; , \; \label{equH2}  
\end{eqnarray}
where the integrand of the first integral in the right-hand side of this equation contains only products of different 
powers of components of the metric tensor and their first-order derivatives, while the second integral has the form 
of a divergence of the vector-like quantity $\sqrt{- g} \Bigl( g^{\alpha\beta} \Gamma^{\gamma}_{\alpha\beta} - 
g^{\alpha\gamma} \Gamma^{\beta}_{\alpha\beta} \Bigr)$. It is clear that the second integral can be transformed (with 
the help of Gauss theorem) into an integral over a hyper-surface surrounding the $d-$dimensional volume over which 
the integration is carried out in other two integrals. When we vary the gravitational action $S_g$, the variation of 
this (second) term in the right-hand side of Eq.(\ref{equH2}) vanishes, since in respect to the principle of least 
action, the variation of the gravitational field at the limits of the region of integration must be equal zero. 

Now, from Eq.(\ref{equH2}) one finds  
\begin{eqnarray}
 \delta S_g = \delta \int L_{EH} d\Omega = \delta \int R \sqrt{- g} d\Omega = \delta \int L_{\Gamma-\Gamma} 
 d\Omega \; , \; {\rm or} \; \; \frac{\delta L_{EH}}{\delta g_{\mu\nu}} = \frac{\delta \Bigl(R \sqrt{- g} 
 \Bigr)}{\delta g_{\mu\nu}} = \frac{\delta L_{\Gamma - \Gamma}}{\delta g_{\mu\nu}} \; \; \label{LGG0}
\end{eqnarray}
where the notation $\delta$ means variation, while the notation $\frac{\delta F}{\delta g_{\mu\nu}}$ means the 
variational derivative (or Lagrange derivative) of the functional $F$. Also in this equation the symbol 
$L_{\Gamma-\Gamma} = \sqrt{- g} g^{\alpha\beta} \Bigl( \Gamma^{\lambda}_{\alpha\gamma} 
\Gamma^{\gamma}_{\beta\lambda} - \Gamma^{\gamma}_{\alpha\beta} \Gamma^{\lambda}_{\gamma\lambda} \Bigr)$ stands 
for the regularized (or regular) $\Gamma - \Gamma$ Lagrangian density of the metric gravity which plays a 
central role in numerous Hamiltonian approaches developed for the metric gravity. As follows from 
Eq.(\ref{LGG0}) the variational derivative of the $L_{\Gamma-\Gamma}$ Lagrangian density is a true tensor, 
while the original $L_{\Gamma-\Gamma}$ Lagrangian density is not a true scalar. The equality, Eq.(\ref{LGG0}), 
expresses the fact that we can replace the `singular' Einstein-Hilbert Lagrangian density $L_{EH} = \sqrt{- g} 
R$ by the regular $\Gamma - \Gamma$ Lagrangian density $L_{\Gamma-\Gamma} = \sqrt{- g} g^{\alpha\beta} \Bigl( 
\Gamma^{\lambda}_{\alpha\gamma} \Gamma^{\gamma}_{\beta\lambda} - \Gamma^{\gamma}_{\alpha\beta} 
\Gamma^{\lambda}_{\gamma\lambda} \Bigr)$ which is variationally equivalent to the original Einstein-Hilbert 
Lagrangian density and contains no second-order derivative. This $\Gamma - \Gamma$ Lagrangian density is 
also written in the following form
\begin{eqnarray}
 L_{\Gamma - \Gamma} &=& \frac14 \sqrt{-g} B^{\alpha\beta\gamma\mu\nu\rho} \Bigl(\frac{\partial 
 g_{\alpha\beta}}{\partial x^{\gamma}}\Bigr) \Bigl(\frac{\partial g_{\mu\nu}}{\partial x^{\rho}}\Bigr) = \frac14 
 \sqrt{-g} B^{\alpha\beta\gamma\mu\nu\rho} g_{\alpha\beta,\gamma} g_{\mu\nu,\rho} \; \; \nonumber \\
 &=& \frac14 \sqrt{-g} \Bigl( g^{\alpha\beta} g^{\gamma\rho} g^{\mu\nu} - g^{\alpha\mu} g^{\beta\nu} 
 g^{\gamma\rho} + 2 g^{\alpha\rho} g^{\beta\nu} g^{\gamma\mu} - 2 g^{\alpha\beta} g^{\gamma\mu} g^{\nu\rho} \Bigr) 
 g_{\alpha\beta,\gamma} g_{\mu\nu,\rho} \; \; , \; \label{LGG} 
\end{eqnarray}
where $B^{\alpha\beta\gamma\mu\nu\rho}$ is a homogeneous cubic polynomial of the contravariant components of the 
metric tensor $g^{\alpha\beta}$. The explicit definition of the $B^{\alpha\beta\gamma\mu\nu\rho}$ quantities 
follows directly from Eq.(\ref{LGG}). Below, we shall deal with the $\Gamma - \Gamma$ Lagrangian density only. In 
order to simplify the following formulas we shall designate this Lagrangian density by the letter $L$, i.e., $L = 
L_{\Gamma-\Gamma}$. Now, by using the $\Gamma - \Gamma$ Lagrangian density we can derive the explicit expressions 
for all contravariant components of momenta $\pi^{\mu\nu}$ and obtain the closed expression for the Hamiltonian(s) 
of the metric gravity. These important steps are made in the next Section.  

\section{Momenta. Canonical and total Hamiltonians of metric gravity}

In the previous Section we have introduced the $\Gamma - \Gamma$ Lagrangian density, Eq.(\ref{LGG}), of the metric 
gravity. At the second step of any standard Hamiltonian procedure, by using the known Lagrangian density we have to 
define the corresponding momenta. Our current derivation of momenta in this study is based on the approaches 
developed in the two earlier papers \cite{Dir58} and \cite{K&K} which still play a central role in all modern 
Hamiltonian formulations of the metric gravity. First, we need to re-write the formula, Eq.(\ref{LGG}), for the 
$\Gamma - \Gamma$ Lagrangian density to a slightly different form where all temporal derivatives of the covariant 
components of metric tensor, i.e., $g_{\alpha\beta,0}$, are explicitly separated from other similar derivatives 
(see, e.g., \cite{K&K}, \cite{Fro2021})  
\begin{eqnarray}
 L = \frac14 \sqrt{-g} B^{\alpha\beta 0\mu\nu 0} g_{\alpha\beta,0} g_{\mu\nu,0} + \frac12 \sqrt{-g} 
 B^{(\alpha\beta 0|\mu\nu k)} g_{\alpha\beta,0} g_{\mu\nu,k} + \frac14 \sqrt{-g} B^{\alpha\beta k \mu\nu l} 
 g_{\alpha\beta,k} g_{\mu\nu,l} \; , \; \label{LGGvel} 
\end{eqnarray}
where the notation $B^{(\alpha\beta\gamma|\mu\nu\rho)}$ means a `symmetrical' $B^{\alpha\beta\gamma\mu\nu\rho}$ 
quantity which is symmetrized in respect to the permutation of two groups of indexes, i.e.,
\begin{eqnarray}
 B^{(\alpha\beta\gamma|\mu\nu\rho)} &=& \frac12 \Bigl( B^{\alpha\beta\gamma\mu\nu\rho} + 
 B^{\mu\nu\rho\alpha\beta\gamma} \Bigr) = g^{\alpha\beta} g^{\gamma\rho} g^{\mu\nu} - g^{\alpha\mu} g^{\beta\nu} 
 g^{\gamma\rho} \nonumber \\ 
 &+& 2 g^{\alpha\rho} g^{\beta\nu} g^{\gamma\mu} - g^{\alpha\beta} g^{\nu\rho} g^{\gamma\mu} - g^{\alpha\rho} 
 g^{\beta\gamma} g^{\mu\nu} \; . \; \label{eq52}
\end{eqnarray}

The contravariant components of momentum $\pi^{\gamma\sigma}$ are defined as partial derivatives of the Lagrangian 
density, Eq.(\ref{LGGvel}), in respect to the corresponding velocities $g_{\gamma\sigma,0}$ (see, e.g., \cite{Dir64}). 
The expressions for the contravariant components of gravitational momenta (or momenta, for short) are  
\begin{eqnarray}
  \pi^{\gamma\sigma} = \frac{\partial L}{\partial g_{\gamma\sigma,0}} = \frac{1}{2} \sqrt{-g} B^{((\gamma\sigma) 
  0|\mu\nu 0)} g_{\mu\nu, 0} + \frac{1}{2} \sqrt{-g} B^{((\gamma\sigma) 0|\mu\nu k)} g_{\mu\nu, k} \; \; , \; 
 \label{mom}
\end{eqnarray}
where $B^{((\gamma\sigma) 0|\mu\nu 0)} = \frac12 \Bigl(B^{(\gamma\sigma 0|\mu\nu 0)} + B^{(\sigma\gamma 0|\mu\nu 
0)} \Bigr)$ is the symmetrized linear combination of the two $B^{(\gamma\sigma 0|\mu\nu 0)}$ quantities. The first 
term in the right-hand side of this equation is written in the form 
\begin{eqnarray}
 \frac{1}{2} \sqrt{-g} B^{((\gamma\sigma)0|\mu\nu 0)} g_{\mu\nu, 0} = \frac{1}{2} \sqrt{-g} g^{00} 
 \Bigl( e^{\mu\nu} e^{\gamma\sigma} - e^{\mu\gamma} e^{\nu\sigma} \Bigr) g_{\mu\nu, 0} = \frac{1}{2} 
 \sqrt{-g} g^{00} E^{\mu\nu\gamma\sigma} g_{\mu\nu, 0} \; , \; \; \label{B1}
\end{eqnarray}
where the notations $e^{\mu \nu}$ and $E^{\mu\nu\gamma\sigma}$ stands for the Dirac contravariant tensors of the 
second and fourth ranks, respectively. The explicit expressions for these tensors are
\begin{eqnarray}
 e^{\mu \nu} = g^{\mu \nu} - \frac{g^{0 \mu} g^{0 \nu}}{g^{00}} \; \; , \; \; {\rm and} \; \; \;
 E^{\mu \nu \gamma \rho} = e^{\mu \nu} e^{\gamma \rho} - e^{\mu \gamma} e^{\nu \rho} \; \; , \; \label{E}  
\end{eqnarray}
i.e., each component of these two tensors is a function of the contravariant components of the metric tensor only. 
For these tensors one finds the following symmetries in respect to permutations of their indexes: $e^{\mu \nu} = 
e^{\nu \mu}$ and $E^{\mu\nu\gamma\sigma} = E^{\gamma\sigma\mu\nu}$. Also, as follows directly from the formulas, 
Eq.(\ref{E}), the tensor $e^{\mu \nu}$ equals zero, if either index $\mu$, or index $\nu$ (or both) equals zero. 
Analogously, for the Dirac $E^{\mu\nu\gamma\sigma}$ tensor one finds that $E^{0\nu\gamma\sigma} = 0, E^{\mu 
0\gamma\sigma} = 0, E^{\mu\nu 0\sigma} = 0$ and $E^{\mu\nu\gamma 0} = 0$. Therefore, it is more productive to 
discuss the space-like quantities $e^{mn}$ and $E^{mnpq}$ only.   

The space-like $E^{p q k l}$ quantity is, in fact, the space-like Dirac tensor of the fourth rank. This space-like 
tensor $E^{p q k l}$ do not have components which are equal zero identically. Furthermore, this space-like tensor 
$E^{p q k l}$ is a positively defined, invertible tensor. Its inverse space-like tensor $I_{m n p q}$ is also 
positively defined and invertible space-like tensor of the fourth rank which is written in the form \cite{K&K} 
\begin{equation}
 I_{m n q p} = \frac{1}{d - 2} g_{m n} g_{p q} -  g_{m p} g_{n q} \; \; . \; \label{I}
\end{equation}
The relation between space-like tensors $I_{m n p q}$ and $E^{p q k l}$ is written in the form $I_{m n p q} E^{p 
q k l} = g^{k}_{m} g^{l}_{n} = \delta^{k}_{m} \delta^{l}_{n}$ (see, Appendix A), where $g^{\alpha}_{\beta} = 
\delta^{\alpha}_{\beta}$ is the substitution tensor \cite{Kochin} and the symbol $\delta^{\alpha}_{\beta}$ denotes 
the Kroneker delta ($\delta^{\alpha}_{\alpha}$ = 1 and $\delta^{\alpha}_{\beta} = 0$, if $\alpha \ne \beta$). From 
this definition one easily finds that $\delta^{\alpha}_{\beta} = \delta_{\alpha}^{\beta}$. 

In general, for the $B^{((\gamma\sigma) 0|\mu\nu 0)}$ coefficients in the formula, Eq.(\ref{mom}) one finds from 
Eqs.(\ref{B1}), (\ref{E}) and (\ref{I}) one finds for the two possible situations. First, for $\gamma = p$ and 
$\sigma = q$ these coefficients are always different from zero. In this `regular' case we obtain the following 
formula for space-like contravariant components of the momentum tensor 
\begin{eqnarray}
  \pi^{pq} &=& \frac{\partial L}{\partial g_{p q,0}} = \frac{1}{2} \sqrt{-g} B^{((p q) 0|\mu\nu 0)} g_{\mu\nu,0}  
  + \frac{1}{2} \sqrt{-g} B^{((p q) 0|\mu\nu k)} g_{\mu\nu, k} \nonumber \\
  &=& \frac{1}{2} \sqrt{-g} B^{((p q) 0|m n 0)} g_{m n,0} + \frac{1}{2} \sqrt{-g} B^{((p q) 0|m n k)} g_{m n, k} 
  \; \; \label{momenta}
\end{eqnarray}
for each pair of the spatial $(pq)-$indexes. In this case the $(pq;mn)-$matrix of the $\sqrt{-g} B^{((p q) 0|m n 0)} 
= g^{00} E^{pqmn}$ coefficients, which are located in front of the space-like $g_{m n, 0}$ velocities in the 
right-hand side of this equation, is invertible (see above). Therefore, in this case the field-velocity $g_{m n, 0}$ 
can be expressed as the linear combination of the space-like components $\pi^{pq}$ of momentum tensor, Eq.(\ref{I}): 
\begin{eqnarray}
 g_{mn, 0} &=& \frac{1}{g^{00}} \Bigl( \frac{2}{\sqrt{-g}} I_{m n p q} \pi^{pq} - I_{m n p q} B^{((pq) 0|\mu\nu 
 k)} g_{\mu\nu, k} \Bigr) = \frac{1}{g^{00}} I_{m n p q} \Bigl( \frac{2}{\sqrt{-g}} \pi^{pq} \nonumber \\
 &-& B^{((pq) 0|\mu\nu k)} g_{\mu\nu, k} \Bigr) \; \; , \; \label{veloc}
\end{eqnarray}
where the Dirac tensor $I_{m n p q}$ is defined by Eq.(\ref{I}). As follows from Eqs.(\ref{momenta}) and 
(\ref{veloc}) for all space-like components of the metric tensor $g_{pq}$ and corresponding momenta $\pi^{mn}$ 
one essentially finds no principal difference with those systems in classical mechanics which have Lagrangians 
written as quadratic functions of the velocities. Indeed, in metric gravity all space-like components of 
momenta and velocities are always related to each other by a few simple, linear equations, which however, take 
a multi-dimensional, or matrix form. The method described above is the direct and transparent generalization of 
Legendre's dual transformation for the tensor fields. 

In the second `singular' case, when $\gamma = 0$ (or $\sigma = 0$) in Eq.(\ref{mom}), the first term in the 
right-hand side of each of these equations equals zero. Therefore, these equations take the from of pure 
algebraic equations  
\begin{eqnarray}
 \pi^{0 \sigma} = \frac{\partial L}{\partial g_{0\sigma,0}} = \frac{1}{2} \sqrt{-g} B^{((0\sigma) 0|\mu\nu k)} 
 g_{\mu\nu, k} \; \; , \; \; {\rm and} \; \; \; \pi^{\sigma 0} = \frac{\partial L}{\partial g_{\sigma 0,0}} = 
 \frac{1}{2} \sqrt{-g} B^{((\sigma 0) 0|\mu\nu k)} g_{\mu\nu, k} \; \label{constr}
\end{eqnarray}
for $\sigma = 0, 1, \ldots, d - 1$. From this equations one finds finds $\pi^{0 \sigma} = \pi^{\sigma 0}$. Note 
also that these equations contain no velocities at all, i.e., we cannot express the $g_{0\sigma,0}$ velocities 
in terms of the momenta $\pi^{0\sigma}$ and vice versa. Each of the equations, Eq.(\ref{constr}), directly 
determines the momentum $\pi^{0\sigma}$ as a cubic polynomial of the contravariant components of the metric 
tensor $g^{\alpha\beta}$ which is multiplied by an additional factor $\sqrt{- g} g_{\mu\nu, k}$. In other words, 
the following $d$ functions
\begin{eqnarray}
 \phi^{0 \sigma} = \pi^{0 \sigma} - \frac{1}{2} \sqrt{-g} B^{((0 \sigma) 0|\mu\nu k)} g_{\mu\nu, k} \; = 
 \phi^{\sigma 0} \; \; , \; \label{primary}
\end{eqnarray}
where $\sigma = 0, 1, \ldots, d - 1$, must be equal zero during actual physical motions (or time-evolution) of 
the free gravitational field. In other words, during any actual motion (or time-evolution) of the free 
gravitational field the $d$ additional conditions (or constraints) $\phi^{0\sigma} = 0$ must be obeyed for the 
Hamiltonian dynamical variables, since otherwise such a motion is not possible. We have to emphasize that the 
equations $\phi^{0\sigma} = 0$ are correct only on the true Hamiltonian trajectories (or curves) $(x^{0}, 
g_{\alpha\beta}(x^{0}), \pi^{\mu\nu}(x^{0}))$ of the free gravitational field. Outside these trajectories, only 
some, or even none of these equations are satisfied. In \cite{Dir58} Dirac proposed to write these additional 
conditions in the symbolic form $\phi^{0\sigma} \approx 0$ (for $\sigma = 0, 1, \ldots, d - 1$) with a different 
sign $\approx$ from the usual (=). These `weak' equations are the primary constraints of the given Hamiltonian 
formulation (see, e.g., \cite{Dir58} and \cite{Dir64}). In other words, during time-evolution of the free 
gravitational field we always have $d$ primary constraints for the $d (d + 1)$ Hamiltonian variables $\{ 
g_{\alpha\beta}, \pi^{\mu\nu} \}$ and this number $d$ never changes, if one applies canonical transformations 
of the Hamiltonian dynamical variables (see below). This is a partial case of the law of inertia for the 
first-class constraints which is discussed below.     

Now, by applying the Legendre transformation to the known $\Gamma - \Gamma$ Lagrangian density $L$, of the metric 
gravity, Eq.(\ref{LGGvel}), and excluding all space-like velocities $g_{mn,0}$, we can derive the explicit 
formulas for the total $H_t$ and canonical $H_C$ Hamiltonians of the metric GR. Formally, these quantities are 
the Hamiltonian densities, but in this study we try to avoid any mention of Hamiltonian densities, since constant 
play with words `Hamiltonians' and `Hamiltonian densities' substantially complicates explanations and often leads 
to various confusions. In particular, the total Hamiltonian $H_t$ of the gravitational field in metric gravity 
derived from the $\Gamma - \Gamma$ Lagrangian density $L$, Eq.(\ref{LGG}), is written in the form 
\begin{eqnarray}
  H_t &=& g_{\alpha\beta,0} \pi^{\alpha\beta} - L =  g_{pq,0} \pi^{pq} + g_{0\sigma,0} \phi^{0\sigma} - L =  
 g_{pq,0} \pi^{pq} - L + g_{0\sigma,0} \phi^{0\sigma} \nonumber \\
   &=& H_C + g_{0 0,0} \phi^{0 0} + g_{0 k,0} \phi^{0 k} + g_{k 0,0} \phi^{k 0} = H_C + g_{0 0,0} \phi^{0 0} 
   + 2 g_{0 k,0} \phi^{0 k} , \; \label{eq1}
\end{eqnarray}
where $\phi^{0\sigma} = \pi^{0\sigma} - \frac{1}{2}\sqrt{-g} B^{\left( \left(0\sigma\right) 0\mid\mu\nu k\right)} 
g_{\mu\nu,k}$ are the primary constraints, while $g_{0\sigma,0}$ are the $\sigma$ velocities (or temporal velocities) 
and $H_C$ is the canonical Hamiltonian of the metric gravity
\begin{eqnarray}
 & &H_C = \frac{1}{\sqrt{-g} g^{00}} I_{mnpq} \pi^{mn} \pi^{pq} - \frac{1}{g^{00}} I_{mnpq} \pi^{mn} 
 B^{(p q 0|\mu \nu k)} g_{\mu\nu,k} \label{eq5} \\
 &+& \frac14 \sqrt{-g} \Bigl[ \frac{1}{g^{00}} I_{mnpq} B^{((mn)0|\mu\nu k)} B^{(pq0|\alpha\beta l)} - 
 B^{\mu\nu k \alpha\beta l}\Bigr] g_{\mu\nu,k} g_{\alpha\beta,l} \; \; , \; \nonumber
\end{eqnarray}
which does not contain any primary constraint $\phi^{0\sigma}$. The total Hamiltonian $H_t = H_C + g_{0\sigma,0} 
\phi^{0\sigma}$ is a scalar function, which is defined in the $d (d + 1)$ even-dimensional phase space $\Bigl\{ 
g_{\alpha\beta}, \pi^{\mu\nu} \Bigr\}$, where all components of the metric $g_{\alpha\beta}$ and momentum 
$\pi^{\mu\nu}$ tensors have been chosen as the basic Hamiltonian variables. The corresponding $d (d + 1)$ 
dimensional space of Hamiltonian variables must be endowed with a symplectic (or anti-symmetric) bilinear form 
(Poisson brackets), which turns this space into a symplectic, even-dimensional phase space. The definition of 
Poisson brackets between all basic dynamical variables, i.e., between coordinates $g_{\alpha\beta}$ and momenta 
$\pi^{\mu\nu}$, is discussed in the next Section. At the same time the spatial (covariant) components of the 
metric tensor $g_{mn}$ and spatial (contravariant) components of momenta $\pi^{pq}$ form another $d (d - 1)$ 
dimensional space, which is also transformed (by the same Poisson brackets) into a symplectic, even-dimensional 
phase space of the $d (d - 1)$ space-like Hamiltonian variables $\Bigl\{ g_{mn}, \pi^{pq} \Bigr\}$. 

\section{Poisson brackets}

In general, the Poisson brackets (or PB, for short) are the fundamental and crucially important tools of any correct 
Hamiltonian theory. The correct definition of these Poisson brackets is the central part of numerous Hamiltonian 
formulations developed for different physical systems of particles, fields and their combinations. As is well known 
(see, e.g., \cite{Gant} - \cite{Arnold}) the Poisson bracket is an antisymmetric, bi-linear form defined in the $2 
M$-dimensional phase space which is, in fact, a cotangent space to an $M-$dimensional manifold located in the 
position space. More accurate definition of the Poisson brackets can be found, e.g., in \cite{Arnold}. For arbitrary 
vectors $X, Y, Z$ from this phase space we can define the bi-linear form which is designated below as $[ X, Y ]$ and 
it is obeyed the four following rules (or axioms) 
\begin{eqnarray}
 &&[ X, Y ] = - [ Y , X ] \; \; \; \; \; ({\rm antisymmetry}) \; , \; \nonumber \\
 &&[ a_1 X_1 + a_2 X_2 , Y ] = a_1 [ X_1, Y ] + a_2 [ X_2, Y ] \; \; \; \; \; ({\rm linearity}\; {\rm in} \; 
 {\rm either} \; {\rm member}) \; , \; \nonumber \\  
 &&[ X Y, Z ] = [ X, Z ] Y + X [ Y, Z ] \; \; \; \; \; ({\rm the}\; {\rm product} \; {\rm law}) \; , \; \nonumber \\  
 &&[ X, [ Y, Z ]] + [ Y , [ Z, X ]] + [ Z , [ X, Y ]] = 0 \; \; \; \; \; ({\rm Jacobi} \; \; {\rm identity}) \; , \; 
 \nonumber 
\end{eqnarray}
where each of the $X, Y$ and $Z$ vectors belongs to the $2 M$ dimensional phase space. In metric gravity the $d ( d 
+ 1)-$dimensional phase space includes the $\frac{d(d + 1)}{2}$ generalized coordinates $g_{\alpha\beta}$ and 
$\frac{d(d + 1)}{2}$ momenta $\pi^{\mu\nu}$, or contravariant components of the momentum tensor $\pi$. An additional 
(or fifth) fundamental rule for the Poisson brackets, which is often called the `time-evolution' of the Poisson 
bracket, is written in the form 
\begin{eqnarray}
 \frac{\partial}{\partial t} [ X, Y ] = [ \frac{\partial X}{\partial t}, Y ] + [ X, \frac{\partial Y}{\partial t} ]
 \; \; , \; \; {\rm and} \; \; \;\frac{\partial}{\partial b} [ X, Y ] = [ \frac{\partial X}{\partial b}, Y ] 
 + [ X, \frac{\partial Y}{\partial b} ] \; \; \nonumber 
\end{eqnarray}
where $t$ is the temporal variable (or time, for short), while $b$ is an arbitrary numerical parameter.   

In the Hamiltonian version of metric gravity the basic dynamical variables are the generalized `coordinates' 
$g_{\alpha\beta}$ and momenta $\pi^{\mu\nu}$ defined above. In respect to this the Poisson brackets between the two 
functions of these dynamical variables are defined as follows: 
\begin{eqnarray}
  [ f_1, f_2 ] = \frac{\partial f_1}{\partial g_{\alpha\beta}} \frac{\partial f_2}{\partial \pi^{\alpha\beta}} - 
  \frac{\partial f_2}{\partial g_{\alpha\beta}} \frac{\partial f_1}{\partial \pi^{\alpha\beta}} = 
  \frac{\partial f_1}{\partial g_{\alpha\beta}} \frac{\partial f_2}{\partial \pi^{\alpha\beta}} -
  \frac{\partial f_1}{\partial \pi^{\alpha\beta}} \frac{\partial f_2}{\partial g_{\alpha\beta}} \; \; \; . 
  \; \; \label{PoisBrack}
\end{eqnarray}
The Poisson brackets between the generalized coordinates and momenta have the fundamental value for the purposes 
of this study. They are:
\begin{eqnarray}
 [ g_{\alpha\beta}, \pi^{\mu\nu}] = \frac{\partial g_{\alpha\beta}}{\partial g_{\gamma\sigma}} \frac{\partial 
 \pi^{\mu\nu}}{\partial \pi^{\gamma\sigma}} - \frac{\partial g_{\alpha\beta}}{\partial \pi^{\gamma\sigma}} 
 \frac{\partial \pi^{\mu\nu}}{\partial g_{\gamma\sigma}} = \Delta^{\mu\nu}_{\alpha\beta} = \frac12 
 \Bigl(g^{\mu}_{\alpha} g^{\nu}_{\beta} + g^{\nu}_{\alpha} g^{\mu}_{\beta}\Bigr) = \frac12 
 \Bigl(\delta^{\mu}_{\alpha} \delta^{\nu}_{\beta} +  \delta^{\nu}_{\alpha} \delta^{\mu}_{\beta}\Bigr) \; \; , 
 \; \label{eq15} 
\end{eqnarray}
where $g^{\mu}_{\alpha} = \delta^{\mu}_{\alpha} (= \delta^{\alpha}_{\mu}$) is the substitution tensor \cite{Kochin} 
and symbol $\delta^{\mu}_{\beta}$ is the Kronecker delta, while the notation $\Delta^{\mu\nu}_{\alpha\beta}$ stands 
for the gravitational (or tensor) delta-symbol. The three following properties of this delta-symbol are obvious and 
very useful in calculations of many Poisson brackets: (1) `horizontal' index symmetry $\Delta^{\mu\nu}_{\alpha\beta} 
= \Delta^{\nu\mu}_{\alpha\beta} = \Delta^{\mu\nu}_{\beta\alpha} = \Delta^{\nu\mu}_{\beta\alpha}$, (2) `vertical' 
index symmetry $\Delta^{\mu\nu}_{\alpha\beta} = \Delta^{\alpha\nu}_{\mu\beta} = \Delta^{\mu\alpha}_{\nu\beta} = 
\ldots = \Delta_{\mu\nu}^{\alpha\beta}$, and (3) the product property: $\Delta^{\nu\mu}_{\rho\sigma} 
\Delta^{\rho\sigma}_{\alpha\beta} = \Delta^{\mu\nu}_{\alpha\beta}$. By using these properties we can write that $[ 
g_{\alpha\beta}, \pi^{\mu\nu}] = \Delta^{\mu\nu}_{\alpha\beta} = \Delta_{\mu\nu}^{\alpha\beta} = [ g_{\mu\nu}, 
\pi^{\alpha\beta}]$. Note again that the total number of dynamical Hamiltonian variables in metric gravity is always 
even and equals $d (d + 1)$. The Poisson bracket, Eq.(\ref{eq15}), explains why in some papers the gravitational 
momenta $\pi^{\mu\nu}$ are called and considered as conjugate dynamical variables for the corresponding covariant 
components of metric tensor $g_{\mu\nu}$ (our coordinates). 

Other fundamental Poisson brackets between basic dynamical variables of the metric gravity equal zero identically, 
i.e., $[ g_{\alpha\beta}, g_{\mu\nu}] = 0$ and $[ \pi^{\alpha\beta}, \pi^{\mu\nu}] = 0$. In general, our dynamical 
variables depend upon one temporal and $(d - 1)$ spatial coordinates $x^{0}, x^{1}, \ldots, x^{d-1} = (x_0, 
\overline{x})$. In this case we have to apply the following definition of the Poisson brackets, e.g., 
\begin{eqnarray}
 [ g_{\alpha\beta}(\bar{x}, t), \pi^{\mu\nu}(\bar{x}^{\prime}, t)] = \Delta^{\mu\nu}_{\alpha\beta} 
 \delta^{d-1}(\bar{x} - \bar{x}^{\prime}) \; \; , \; \label{non-local} 
\end{eqnarray}
where $ \delta^{d-1}(\bar{y})$ is the usual delta-function in the position $(d - 1)-$space. Such a generalization 
of the Poisson brackets is straightforward and simple, but in this study we do not want to complicate our system 
of notations. In respect to this, below we shall always deal with the Poisson brackets of two quantities taken 
at the same spatial point.    

The explicit form of the fundamental Poisson brackets, Eq.(\ref{eq15}), allows one to derive the following 
formulas for slightly different Poisson brackets 
\begin{eqnarray}
 [ g^{\alpha\beta}, \pi^{\mu\nu}] = - \frac12 \Bigl( g^{\alpha\mu} g^{\beta\nu} + g^{\alpha\nu} g^{\beta\mu} 
 \Bigr) = - g^{\alpha\gamma} \Delta^{\mu\nu}_{\gamma\sigma} g^{\beta\sigma} = - [ \pi^{\mu\nu}, g^{\alpha\beta}] 
 \; \; {\rm and} \; \; [ g^{\alpha\beta}, g_{\mu\nu}] = 0 . \; \label{eq151} 
\end{eqnarray}
which contain the contravariant components of the metric tensor $g^{\alpha\beta}$. Now, by using the Poisson 
brackets, Eqs.(\ref{eq15}) and (\ref{eq151}), defined above we can determine the Poisson brackets of more 
complicated quantities and functions. As the first example we calculate the following Poisson bracket
\begin{eqnarray}
 [ g_{\alpha\beta} g^{\lambda\sigma}, \pi^{\mu\nu}] = [ g_{\alpha\beta}, \pi^{\mu\nu}] g^{\lambda\sigma} + 
 g_{\alpha\beta} [ g^{\lambda\sigma}, \pi^{\mu\nu}] = \Delta^{\mu\nu}_{\alpha\beta} g^{\lambda\sigma} - \frac12 
 g_{\alpha\beta} \Bigl( g^{\lambda\mu} g^{\sigma\nu} + g^{\lambda\nu} g^{\sigma\mu} \Bigr) \; . \label{eq151a} 
\end{eqnarray}
Let us assume that in this formula $\lambda = \beta$. In this case $g_{\alpha\beta} g^{\beta\sigma} = 
g^{\sigma}_{\alpha} = \delta^{\sigma}_{\alpha}$ and it is clear that $[ g^{\sigma}_{\alpha}, \pi^{\mu\nu}] = 0$. 
On the other hand, if $\lambda = \beta$, then for the right-hand side of Eq.(\ref{eq151a}) one finds: 
\begin{eqnarray}
 \Delta^{\mu\nu}_{\alpha\beta} g^{\beta\sigma} - \frac12 g_{\alpha\beta} \Bigl( g^{\beta\mu} g^{\sigma\nu} + 
 g^{\beta\nu} g^{\sigma\mu} \Bigr) = \frac12 \Bigl( \delta^{\mu}_{\alpha} g^{\nu\sigma} + \delta^{\nu}_{\alpha} 
 g^{\mu\sigma} \Bigr) - \frac12 \Bigl( \delta^{\mu}_{\alpha} g^{\nu\sigma} + \delta^{\nu}_{\alpha} g^{\mu\sigma} 
 \Bigr) = 0 \; , \;  
\end{eqnarray}
which means that for $\lambda = \beta$ the equation, Eq.(\ref{eq151a}), is written in the form $0 = 0$ and we  
have no contradiction here. Now, consider the following Poisson bracket $[ g_{\alpha\beta} g^{\alpha\beta}, 
\pi^{\mu\nu} ]$. As is well known from tensor calculus (see, e.g., \cite{Kochin}) $g_{\alpha\beta} g^{\alpha\beta} 
= d$, where $d$ is the dimension of tensor space. Therefore, this Poisson bracket is reduced to the equation 
\begin{eqnarray}
 0 = [ g_{\alpha\beta}, \pi^{\mu\nu}] g^{\alpha\beta} +  g_{\alpha\beta} [ g^{\alpha\beta}, \pi^{\mu\nu}] \; 
 \; {\rm or} \; \; \Delta^{\mu\nu}_{\alpha\beta} g^{\alpha\beta} - \frac12 g_{\alpha\beta} \Bigl( g^{\alpha\mu} 
 g^{\beta\nu} + g^{\alpha\nu} g^{\beta\nu} \Bigr) = 0 \; \; , \; \nonumber 
\end{eqnarray}
which is easily transformed to an obvious identity $g^{\mu\nu} - g^{\nu\mu} = 0$. Analogously, it is easy to find 
a number of remarkable relations between the Poisson brackets $[ g_{\alpha\beta}, \pi^{\mu\nu} ]$ and $[ 
g^{\alpha\beta}, \pi^{\mu\nu} ]$, temporal derivatives of the covariant and contravariant components of the metric 
tensor and Poisson brackets of these components with the canonical Hamiltonian $H_C$ which directly follows from 
Eq.(\ref{eq151a}) (see, also our `technical' Appendix A): 
\begin{eqnarray}
 &&[ g^{\sigma\gamma}, \pi^{\mu\nu} ] = - g^{\alpha\sigma} [ g_{\alpha\beta}, \pi^{\mu\nu} ] g^{\beta\gamma} = 
 - g^{\alpha\sigma} \Delta^{\mu\nu}_{\alpha\beta} g^{\beta\gamma} = - \frac12 \Bigl( g^{\sigma\mu} g^{\gamma\nu} 
 + g^{\sigma\nu} g^{\gamma\mu} \Bigr) \; , \; \label{PBOa} \\
 &&\frac{d g^{\sigma\gamma}}{d t} = - g^{\alpha\sigma} \Bigl(\frac{d g_{\alpha\beta}}{d t}\Bigr) g^{\beta\gamma} 
  \; \; \; , \; \; \; [ g^{\sigma\gamma}, H_C ] = - g^{\alpha\sigma} [ g_{\alpha\beta},  H_C ] g^{\beta\gamma} 
  \; . \; \label{PBO}
\end{eqnarray}

Second example is of great interest for analytical calculations of the Poisson brackets between components of 
momenta and some functions of the coordinates only.  Let $F(g_{\mu\nu}, g^{\lambda\sigma})$ be an arbitrary 
analytical function (or functional) of the co- and contravariant components of the metric tensor. The Poisson 
bracket of this $F(g_{\mu\nu}, g^{\lambda\sigma})$ function and components of momentum tensor $\pi^{\alpha\beta}$ 
takes the form
\begin{eqnarray}
 [ \pi^{\alpha\beta}, F(g_{\mu\nu}, g^{\lambda\sigma})] &=& - \frac{\partial F}{\partial g_{\mu\nu}} [ 
 g_{\mu\nu}, \pi^{\alpha\beta}] - \frac{\partial F}{\partial g^{\sigma\lambda}} [ g^{\sigma\lambda}, 
 \pi^{\alpha\beta}] \nonumber \\ 
 &=& \frac12 \Bigl(\frac{\partial F}{\partial g^{\sigma\lambda}}\Bigr) \Bigl( g^{\sigma\alpha} g^{\lambda\beta} 
 + g^{\sigma\beta} g^{\lambda\alpha} \Bigr) - \Bigl(\frac{\partial F}{\partial g_{\mu\nu}}\Bigr) 
 \Delta^{\alpha\beta}_{\mu\nu} \; . \label{eq152} 
\end{eqnarray}
In particular, for the $F(g_{\mu\nu})$ function (or functional) of the covariant components of the metric tensor 
only this Poisson bracket can be written in the form
\begin{eqnarray}
 [ \pi^{\alpha\beta}, F(g_{\mu\nu})] = - \frac{\partial F}{\partial g_{\mu\nu}} [ g_{\mu\nu}, \pi^{\alpha\beta}] 
 = - \Bigl(\frac{\partial F}{\partial g_{\mu\nu}}\Bigr) \Delta^{\alpha\beta}_{\mu\nu} \; . \label{eq152a} 
\end{eqnarray}
In the case when $F = \Phi(g_{\mu\nu}) g_{\rho\sigma, k}$ we obtain 
\begin{eqnarray}
 [ \pi^{\alpha\beta}, \Phi(g_{\mu\nu}) g_{\rho\sigma, k} ] = - \Bigl(\frac{\partial \Phi}{\partial g_{\mu\nu}}\Bigr) 
 \Delta^{\alpha\beta}_{\mu\nu} g_{\rho\sigma, k} - \Delta^{\alpha\beta}_{\rho\sigma} \Bigl( \Phi \Bigr)_{, k} = 
- \Bigl(\frac{\partial \Phi}{\partial g_{\alpha\beta}}\Bigr) g_{\rho\sigma, k} - \Bigl( \Phi \Bigr)_{, k} 
 \Delta^{\alpha\beta}_{\rho\sigma} \; \; \; \label{eq153a}
\end{eqnarray}
Here we have applied the integration by parts (see, also discussion at the end of this Section). The third example 
includes Poisson brackets between space-like components of the momenta $\pi^{mn}$ and components of the Dirac 
space-like tensor $e_{pq}$ and/or $e^{pq}$. They are 
\begin{eqnarray}
 [ \pi^{mn}, e_{pq} ] = - \Delta^{mn}_{pq} \; \; , \; \; {\rm and} \; \; [ \pi^{mn}, e^{pq} ] = \frac12 \Bigl( 
 g^{pm} g^{qn} + g^{pn} g^{qm} \Bigr) \; \; . \label{eq151A} 
\end{eqnarray}
As follows from the first PB in Eq.(\ref{eq151A}) and two other groups of Poisson brackets: $[ \pi^{mn}, \pi^{pq} 
] = 0, [ e_{mn}, e_{pq} ] = 0$ these $d (d - 1)$ variables $e_{mn}$ and $\pi^{pq}$ are the canonical Hamiltonian 
variables in the $d (d - 1)$-dimensional subspace of space-like dynamical variables of metric gravity. These PB 
play an important role in this study (see below). 

From the formula, Eq.(\ref{eq152a}), one also finds 
\begin{eqnarray}
 [ \pi^{\alpha\beta}, F(g_{\mu\nu})] = - \Bigl(\frac{\partial F}{\partial  g_{\mu\nu}}\Bigr) 
 \Delta^{\alpha\beta}_{\mu\nu} = - \Bigl(\frac{\partial F}{\partial g_{\alpha\beta}}\Bigr) 
 \Delta_{\alpha\beta}^{\mu\nu} = [ \pi^{\mu\nu}, F(g_{\alpha\beta})] \; \; , \; \;
\end{eqnarray}
or simply $[ \pi^{\alpha\beta}, F(g_{\mu\nu})] = [ \pi^{\mu\nu}, F(g_{\alpha\beta})]$. The principal moment 
here is the presence of tensor $\Delta$-symbol in these Poisson brackets. This equality simplifies 
calculations of a large number of Poisson brackets which are need to show canonicity of different sets of 
Hamiltonian dynamical variables. Also, by using the formula, Eq.(\ref{eq152a}), we can determine another group 
of important Poisson brackets between momenta and analytical functions of the fundamental determinant $g$ and 
its square root $\sqrt{g}$ (or $\sqrt{- g}$ as it is designated in the metric gravity). The general expression 
for the Poisson bracket between such a $F(g)$ function and $\pi^{\alpha\beta}$ is derived as follows
\begin{eqnarray}  
  [ F(g), \pi^{\alpha\beta}] = \Bigl( \frac{\partial F}{\partial g} \Bigr) \Bigl( \frac{\partial g}{\partial 
  g_{\mu\nu}} \Bigr) [ g_{\mu\nu}, \pi^{\alpha\beta}] = \Bigl( \frac{\partial F}{\partial g} \Bigr) g 
  g^{\mu\nu} \Delta^{\alpha\beta}_{\mu\nu} = \Bigl( \frac{\partial F}{\partial g} \Bigr) g g^{\alpha\beta} 
  \; \; \; . \; \label{eq154} 
\end{eqnarray}
or $[ \pi^{\alpha\beta}, F(g)] = - \Bigl( \frac{\partial F}{\partial g} \Bigr) g g^{\alpha\beta}$. Now, if $F(g) 
= g^{x}$, then one finds $[ \pi^{\alpha\beta}, F(g)] = -x F(g) g^{\alpha\beta}$. In particular, if $F(g) = 
\sqrt{- g}$ and $F(g) = \frac{1}{\sqrt{- g}}$ we obtain from the last equation 
\begin{eqnarray}
 [ \sqrt{- g}, \pi^{\alpha\beta}] = - \frac{1}{2 \sqrt{- g}} g g^{\alpha\beta} = \frac12 \sqrt{- g} 
 g^{\alpha\beta} \; \; {\rm and} \; \;
 [ \frac{1}{\sqrt{- g}}, \pi^{\alpha\beta}] = - \frac{1}{2 \sqrt{- g}} g^{\alpha\beta} \; , \; \label{eq155}
\end{eqnarray} 
respectively. Another Poisson bracket is often needed in operations with the both primary and secondary 
constraints: 
\begin{eqnarray}
 [ \pi^{0 \gamma}, \frac{g^{\sigma \lambda}}{g^{0 0}} ] = \frac{1}{2 g^{0 0}} \Bigl( g^{0 \sigma} 
 g^{\gamma \lambda} + g^{\gamma \sigma} g^{0 \lambda} - 2 g^{\sigma \lambda} g^{0 \gamma} \Bigr) \; \; . \; 
\end{eqnarray}
If $\lambda = 0$ here, then one finds 
\begin{eqnarray}
 [ \pi^{0 \gamma}, \frac{g^{\sigma 0}}{g^{0 0}} ] = \frac{1}{2 g^{0 0}} \Bigl( g^{\gamma \sigma} g^{0 0} 
 - g^{0 \sigma} g^{0 \gamma} \Bigr) \; \; . \; 
\end{eqnarray} 
Now, if we assume that $\sigma = 0$ here, then this Poisson bracket equals zero identically (as expected). 

To conclude our discussion of the Poisson brackets let us make the two following remarks. First, as it was shown 
in \cite{K&K} and \cite{PirSS} in metric gravity the Poisson bracket(s) between two primary constraints $\phi^{0 
\sigma}$ and $\phi^{0 \gamma}$, Eq.(\ref{primary}), are always equal zero, i.e., $[ \phi^{0 \sigma}, \phi^{0 
\gamma} ] = 0$. In fact, the explicit derivation of this formula is a very good and relatively simple exercise 
in calculations of Poisson brackets (see, also discussion below). Thus, in the metric gravity all primary 
constraints commute with each other. This drastically simplifies many important steps of our procedure which is 
described below. Second, we have to explain calculations of the Poisson brackets between momenta and expressions 
which include some spatial derivatives of the metric tensor such as $\Phi(g_{\mu\nu}) g_{\gamma\lambda,k}, 
\Phi(g_{\mu\nu}) g_{\gamma\lambda,k} g_{\rho\sigma,m}, g_{\gamma\lambda,k} g_{\rho\sigma,m}$, etc. These and 
other similar expressions arise very often in actual Hamiltonian formulations, and they can be found, e.g., in 
operations with the both canonical and total Hamiltonians, primary and secondary constraints and other 
expressions. Analytical calculations of such Poisson brackets by using `integration by parts' (see, the text 
around Eq.(\ref{eq153a})). To explain a few hidden details of such calculations let us consider the following 
Poisson brackets 
\begin{eqnarray}
 &&[\pi^{\alpha\beta}, \Phi(g_{\mu\nu}) g_{\gamma\lambda,p} g_{\rho\sigma,q}] = -\frac{\partial 
 \Phi(g_{\mu\nu})}{\partial g_{\mu\nu}} \Delta^{\alpha\beta}_{\mu\nu} g_{\gamma\lambda,p} g_{\rho\sigma,q} + 
 \Bigl[\Phi(g_{\mu\nu}) g_{\rho\sigma,q}\Bigr]_{,p} \Delta^{\alpha\beta}_{\gamma\lambda} + \Bigl[\Phi(g_{\mu\nu}) 
 g_{\gamma\lambda,p}\Bigr]_{,q} \Delta^{\alpha\beta}_{\rho\sigma} \; \; \nonumber \\ 
  &&= -\frac{\partial \Phi}{\partial g_{\mu\nu}} g_{\gamma\lambda,p} g_{\rho\sigma,q} \Delta^{\alpha\beta}_{\mu\nu}  
  + \Phi_{,p} g_{\rho\sigma,q} \Delta^{\alpha\beta}_{\gamma\lambda} + \Phi g_{\rho\sigma,qp} 
  \Delta^{\alpha\beta}_{\gamma\lambda} + \Phi_{,q} g_{\gamma\lambda,p} \Delta^{\alpha\beta}_{\rho\sigma} + \Phi 
   g_{\gamma\lambda,pq} \Delta^{\alpha\beta}_{\rho\sigma} \; \; , \; \label{AZet} 
\end{eqnarray} 
where $\Phi(x)$ is a scalar function of tensor argument(s). This formula can be simplified even further, but our 
goal here is to illustrate analytical computations of the Poisson brackets of momenta and some special functions 
and expressions which contain spatial derivatives of the metric tensor. In particular, the formula, Eq.(\ref{AZet}), 
explains the presence of second-order spatial derivatives of the metric tensor in some formulas below. 

All Poisson brackets mentioned above are crucially important for the goals of this study, since they define the 
unique symplectic structure which is closely related to our $d (d + 1)$-dimensional (tensor) phase space $\{ 
g_{\alpha\beta}, \pi^{\mu\nu} \}$, which is closely related to the original $d-$dimensional Riemann space in the 
metric gravity. In other words, such a simplectic structure is uniformly determined by the Poisson brackets 
between the covariant components of the fundamental metric tensor $g_{\alpha\beta}$ and contravariant components 
$\pi^{\mu\nu}$ of the momentum tensor. Finally, we have to note that there is an alternative approach to develop 
Hamiltonian formulations of the metric gravity which is based on the use of covariant components of momenta 
$\pi_{\mu\nu}$. In some sense this new approach is simpler than the method discussed above, but its applications 
lead to re-consideration of fundamental principles of the classical Hamiltonian procedure, operations in the dual 
phase space and analysis of combinations of the both straight and dual phase spaces for the tensor fields. This 
alternative approach and arising dual phase space are briefly considered below. 

\subsection{Covariant components of momenta. On the dual phase space} 

In actual physical theories of tensor fields an arbitrary tensor can be represented either by its covariant, or 
contravariant components. For an arbitrary Riemann space relations between co- and contravariant components of 
the same vector, or tensor-like quantity are determined by the co- and contravariant components of the fundamental 
metric tensor $g_{\alpha\beta}$ and $g^{\alpha\beta}$. Therefore, one can always represent the same tensor 
equations in the both covariant and contravariant forms. In general, many problems from tensor calculus can be 
simplified (even substantially), if they are re-written in the contravariant components of the same tensors and 
vice versa (some examples are considered in \cite{Kochin} and \cite{Dash}). The metric gravity can be one of such 
problems, since the both canonical and total Hamiltonians, Eqs.(\ref{eq5}) and (\ref{eq1}), contain multiple 
products of many contravariant components of the fundamental tensor $g^{\alpha\beta}$. Therefore, if we can 
properly define the covariant components of momenta $\pi_{\lambda\sigma}$, then our original problem can 
drastically be simplified. 

Let us define the covariant components of momenta by the relation $\pi_{\lambda\sigma} = g_{\lambda\mu} 
\pi^{\mu\nu} g_{\nu\sigma}$. Note here that in any Hamiltonian formulation of the metric gravity, the role of 
fundamental tensor $g_{\alpha\beta}$ is always twofold. First, it is traditionally used to raise and lower 
indices in vector and tensor expressions. On the other hand, in all Hamiltonian formulations of the metric 
gravity the components of the fundamental tensor $g_{\alpha\beta}$ are traditionally chosen as the generalized 
coordinates, i.e., dynamical variables which are dynamically conjugate to the corresponding momenta 
$\pi^{\mu\nu}$. Such a twofold role of the fundamental tensor $g_{\alpha\beta}$ (and $g^{\alpha\beta}$) in 
Hamiltonian metric gravity leads to an additional problem, since the momenta $\pi^{\mu\nu}$ do not commute with 
the coordinates $g_{\alpha\beta}$. In turn, this means that the following definitions of covariant components 
of momenta: $\pi_{\lambda\sigma} = g_{\lambda\mu} g_{\nu\sigma} \pi^{\mu\nu}, \pi_{\lambda\sigma} = 
g_{\lambda\mu} \pi^{\mu\nu} g_{\nu\sigma}$ and $\pi_{\lambda\sigma} = \pi^{\mu\nu} g_{\lambda\mu} g_{\nu\sigma}$ 
are not equivalent to each other. Indeed, it is easy to show that $\pi_{\lambda\sigma} = g_{\lambda\mu} 
g_{\nu\sigma} \pi^{\mu\nu} \ne g_{\lambda\mu} \pi^{\mu\nu} g_{\nu\sigma}$, since $[ g_{\nu\sigma}, \pi^{\mu\nu} 
] = \Delta^{\mu\nu}_{\nu\sigma} = \frac12 \Bigl( \delta^{\mu}_{\nu} \delta^{\nu}_{\sigma} + \delta^{\mu}_{\sigma} 
\Bigr) \ne 0$ in the general case. To avoid repetitive discussions of similar problems in this study we shall 
always define the covariant components of momenta by the relation: $\pi_{\lambda\sigma} = g_{\lambda\mu} 
\pi^{\mu\nu} g_{\nu\sigma}$ mentioned above. 

By using this definition of covariant momenta we can determine the following Poisson brackets 
\begin{eqnarray}
  [ g_{\alpha\beta}, \pi_{\mu\nu}] = \frac12 \Bigl( g_{\alpha\mu} g_{\beta\nu} + g_{\alpha\nu} g_{\beta\mu} 
  \Bigr) \; \; {\rm and} \; \; [ g^{\alpha\beta}, \pi_{\mu\nu}] = - \frac12 \Bigl( g^{\alpha}_{\mu} 
  g^{\beta}_{\nu} + g^{\alpha}_{\nu} g^{\beta}_{\mu} \Bigr) = - \Delta^{\alpha\beta}_{\mu\nu} \; \; 
  \label{eq153} 
\end{eqnarray}
and also $[ g^{\alpha\beta}, g^{\mu\nu}] = 0$ and $[ g_{\alpha\beta}, g^{\mu\nu}] = 0$. The formulas 
Eqs.(\ref{eq154}), (\ref{eq155}) and other can be re-derived for the covariant components of momentum 
$\pi_{\alpha\beta}$: 
\begin{eqnarray}
  [ F(g), \pi_{\alpha\beta}] = -\Bigl( \frac{\partial F}{\partial g} \Bigr) g g_{\alpha\beta} \; , \; 
  [ \sqrt{- g}, \pi_{\alpha\beta}] = - \frac12 \sqrt{- g} g_{\alpha\beta} \; , \; [ \frac{1}{\sqrt{- g}}, 
  \pi_{\alpha\beta}] = - \frac{1}{2 \sqrt{- g}}  g_{\alpha\beta} \; . \; \label{eq155a} 
\end{eqnarray}
These Poisson brackets are also important to perform analytical calculations in the Hamiltonian formulation 
of the metric gravity. As follows from these formulas the dynamical Hamiltonian variables $\{ g^{\alpha\beta}, 
\pi_{\mu\nu}\}$ form another set of Hamiltonian dynamical variables which is often called the dual set of 
Hamiltonian (dynamical) variables. In general, this dual set of Hamiltonian variables $\{ g^{\alpha\beta}, 
\pi_{\mu\nu}\}$ can also be used to develop another Hamiltonian formulation of the metric gravity which is 
simpler than the approach described above. Thus, for the tensor field in metric gravity we always have two 
sets of canonical Hamiltonian variables: (a) straight (or natural) set $\{ g_{\alpha\beta}, \pi^{\mu\nu}\}$, 
and (b) dual set of dynamical variables $\{ g^{\alpha\beta}, \pi_{\mu\nu}\}$ \cite{Fro2021}. Further analysis 
\cite{Fro2021} shows that the two similar sets of dynamical Hamiltonian variables (straight and dual sets) 
always arise and exist in any Hamiltonian formulation of the tensor field theory and they are related to each 
other by a special canonical transformation. Also, there is a beautiful formula \cite{Fro2021} for the Poisson 
brackets which unites the both straight and dual sets of dynamical variables defined for the same Hamiltonian 
system 
\begin{eqnarray}
  [ g_{\alpha\beta}, \pi^{\mu\nu}] = \Delta^{\mu\nu}_{\alpha\beta} = [ \pi_{\alpha\beta}, g^{\mu\nu}] \; \; 
  \; . \label{eq1551} 
\end{eqnarray}

Another interesting Poisson bracket in the metric gravity is 
\begin{equation}
  [ \pi_{\alpha\beta}, \pi^{\mu\nu}] = \frac12 \Bigl( \delta_{\alpha}^{\mu} \pi_{\beta}^{\nu} + 
  \delta_{\alpha}^{\nu} \pi_{\beta}^{\mu} + \delta_{\beta}^{\mu} \pi_{\alpha}^{\nu} + \delta_{\beta}^{\nu} 
  \pi_{\alpha}^{\mu} \Bigr) = - [ \pi^{\mu\nu}, \pi_{\alpha\beta} ] \; \; \; , \; \; \label{pipi}
\end{equation}
where $\pi_{\kappa}^{\rho} = g^{\rho \lambda} \pi_{\lambda \kappa} = g_{\kappa \lambda} \pi^{\lambda \rho}$. The 
last equality means that the co- and contravariant components of the momentum tensor do not commute with each 
other. On the other hand, if they commuted, then the direct and dual sets of simplectic dynamical variables in 
metric gravity would be equivalent to each other and there would be no real need to keep these two sets of 
dynamical variables (straight and dual). Indeed, in this case one can always express one set of dynamic variables 
in terms of another set and vice versa. Such cases include all Hamiltonian theories developed for the truly scalar 
fields and those fields which are represented by affine vectors and tensors. However, this is not true for the 
metric gravity and for other theories developed for actual tensor fields in multi-dimensional Riemann spaces of 
non-zero curvature. In general, to develop the truly correct and covariant Hamiltonian formulation for many 
dynamical system of tensor fields it is much better to deal with the mixed set of $2 d (d + 1)$ Hamiltonian 
dynamical variables. This big set is a unification of the two different $d (d + 1)-$dimensional sets of Hamiltonian 
dynamical variables: (a) the straight set $\{ g_{\alpha\beta}, \pi^{\mu\nu}\}$, and (b) the corresponding dual set 
$\{ g^{\alpha\beta}, \pi_{\mu\nu}\}$. Applications of the two sets of dynamical variables makes our Hamiltonian 
formulation complete, truly covariant and physically transparent. In addition to this, an instant use of the direct 
and dual sets of Hamiltonian dynamical variables allows one to write canonical transformations of the Hamiltonian 
dynamical variables in the most general and powerful form. 

\section{Hamilton equations of motion}

The main goal of any Hamiltonian formulation of some physical theory is to derive the correct Hamilton equations 
of motion by following the well established and physically transparent Hamilton procedure which has its internal 
logic based on Stoke's theorem in multi-dimensions. In general, the Hamilton method always provides a remarkable 
simplicity and universality in applications to actual dynamical systems and fields. Each of the Hamilton equations 
describes the complete time-evolution of one of the Hamiltonian dynamical variables. These correct Hamilton 
equations (or canonical equations) for the metric gravity are written in the following form 
\begin{eqnarray}
  \frac{d g_{\alpha\beta}}{d x^{0}} = [ g_{\alpha\beta}, H_t ] \; \; \; {\rm and} \; \; \; 
  \frac{d \pi^{\mu\nu}}{d x^{0}} = [ \pi^{\mu\nu}, H_t ] \; \; , \; \; \label{Hamtequat}   
\end{eqnarray}
where $x_0$ is the temporal variables and expressions in the right-hand sides of both equations are the Poisson 
brackets. In other words, the first-order time derivative of each of the Hamiltonian variables is proportional 
to the corresponding Poisson bracket of this variable with the total Hamiltonian $H_t$, Eq.(\ref{eq1}). The 
explicit form of these Hamiltonian equations and their solutions are discussed in \cite{Fro2021}. In particular, 
for space-like components of the metric tensor $g_{ij}$ one finds the following system of Hamilton equations 
\cite{Fro2021}:
\begin{eqnarray}
 \frac{d g_{ij}}{d x^{0}} &=& [ g_{ij}, H_{t} ] = [ g_{ij}, H_{C} ] = \frac{2}{\sqrt{-g} g^{00}} I_{(ij)pq} 
 \pi^{pq} - \frac{1}{g^{00}} I_{(ij)pq} B^{(p q 0|\mu \nu k)} g_{\mu\nu,k} \; \label{eq25} \\
 &=& \frac{2}{\sqrt{-g} g^{00}} I_{(ij)pq} \Bigl[ \pi^{pq} - \frac12 \sqrt{-g} B^{(p q 0|\mu \nu k)} 
  g_{\mu\nu,k} \Bigr] \; , \nonumber 
\end{eqnarray}
where the notation $I_{(ij)pq}$ designates the $(ij)-$symmetrized value of the $I_{ijpq}$ tensor defined in 
Eq.(\ref{I}), i.e., 
\begin{equation} 
 I_{(ij)pq} = \frac12 \Bigl( I_{ijpq} + I_{jipq} \Bigr) = \frac{1}{d - 2} g_{ij} g_{pq} - 
 \frac12 ( g_{ip} g_{jq} + g_{iq} g_{jp} ) \; \; \; . \label{eq26} 
\end{equation} 

Now, let us consider the Poisson brackets for the covariant components $g_{0\sigma}$ of the fundamental tensor. 
It is clear that the Poisson bracket of any $g_{0\sigma}$ component with the canonical Hamiltonian $H_C$, 
Eq.(\ref{eq5}), equals zero identically. Therefore, the Hamilton equations of motion for the covariant 
$g_{0\sigma} (= g_{\sigma 0}$) components of the metric tensor take the form 
\begin{eqnarray}
 \frac{d g_{0 \sigma}}{d x_0} = [ g_{0 \sigma}, H_{t}] = [ g_{0 \sigma}, H_{t} - H_C] = g_{0 \sigma,0} 
 \; \; \label{eq253}
\end{eqnarray}
and analogous equations for the $g_{\gamma 0}$ components. These formulas are, in fact, the definitions of 
the $\sigma-$velocities, where $\sigma = 0, 1, \ldots, d - 1$, which essentially coincide with the coefficients 
in front of the primary constraints in the total Hamiltonian, Eq.(\ref{eq1}). As follows from Eq.(\ref{PBO}) 
there is no need to derive and solve the equations of motion for the covariant components of the metric tensor 
$g^{\alpha\beta}$. Indeed, if we know the time evolution of all covariant components $g_{\mu\nu}$, then from 
Eq.(\ref{PBO}) one easily finds the exact description of time evolution for each $g^{\alpha\beta}$ component.  

In general, the Hamilton equations of motion for the contravariant components of momenta $\pi^{\alpha\beta}$ 
are significantly more complicated (see, e.g., \cite{Fro2021}) than analogous equations for the $g_{\alpha\beta}$ 
components (our coordinates). However, all these complications are pure technical and they are mainly related 
to a large number of Poisson brackets which must be determined in order to describe the complete time-evolution 
of all momentum variables $\pi^{\mu\nu}$. To understand the scale of this problem let us present here the 
Hamiltonian equations of motion for the contravariant space-like components of the momentum tensor $\pi^{ab}$:
\begin{eqnarray}
  & &\frac{d \pi^{ab}}{d x_0} = [ \pi^{ab}, H_{t} ] = [ \pi^{ab}, H_{C} ] = - \frac{1}{g^{00}} \Bigl[ 
 \frac{I_{mnpq}}{\sqrt{-g}}, \pi^{ab} \Bigr] \pi^{mn} \pi^{pq} \nonumber \\
 &+& \frac{1}{g^{00}} \Bigl[ I_{mnpq}, \pi^{ab} \Bigr] \pi^{mn} B^{(p q 0|\mu \nu k)} g_{\mu\nu,k} 
  + \frac{1}{g^{00}} I_{mnpq} \pi^{mn}\Bigl[ B^{(p q 0|\mu \nu k)}, \pi^{ab} \Bigr] g_{\mu\nu,k} + \ldots 
  \; . \label{eq255}
\end{eqnarray}
This formula indicates clearly that in the Hamilton equations in metric gravity which describe time-evolution 
of momenta are significantly more complicated than analogous equations for time-evolution of the generalized 
coordinates $g_{\alpha\beta}$. In general, the Poisson bracket $[ \pi^{ab}, H_{t} ]$ is determined term-by-term.

As follows from the Hamilton equations presented above the Hamilton method itself has a number of problems when 
it is applied to the metric gravity, or other dynamical systems with first-class constraints. First, we note that 
to write Hamilton equations of actual motion we need only the canonical Hamiltonian $H_C$ (not the total 
Hamiltonian $H_t$). Indeed, these equations are: 
\begin{eqnarray}
  \frac{d g_{m n}}{d x_0} = [ g_{m n}, H_{C} ] \; \; \; {\rm and} \; \; \; \frac{d \pi^{p q}}{d x_0} = [ 
  \pi^{p q}, H_{C} ] \; \; \label{eq255A}
\end{eqnarray}
and there are $d (d - 1)$ of these Hamilton equations of actual motion. Second, our Hamilton equations mentioned 
above, Eq.(\ref{eq255A}), do not contain temporal momenta $\pi^{0 \mu}$ and/or $\pi^{\nu 0}$ at all. This means 
that in these frames we cannot describe time-evolution of the temporal components of metric tensor $g_{0 \mu}$ 
and $g_{\nu 0}$ (our coordinates). Moreover, it is not entirely clear where we can get these equations from, 
since all these temporal momenta are included in our Hamiltonian formulation of the metric gravity only as primary 
constraints. Formally, to solve this problem introduce in the new Hamilton equations either the total Hamiltonian 
$H_t$, or the difference $H_t - H_C$, which is a linear combination of the primary constraints $\phi^{0 \gamma}$, 
Eq.(\ref{primary}). However, the coefficients in this linear combination are the $\sigma-$velocities, which are, 
in fact, arbitrary parameters of the method, rather than its dynamical variables. These arguments lead to an 
unambiguous conclusion that the Hamiltonian method itself must substantially be modified, if we want to apply it 
successfully to constrained dynamical systems, including the metric gravity. Such a modification was carried out 
by Dirac in his papers \cite{Dir58} - \cite{Dir64} and will be analyzed in detail in Section VII, but here we just 
want to mention its main steps. 

First of all, Dirac accepted all Hamilton equations from Eqs.(\ref{eq255A}) as the equations which correctly describe 
the actual motions in our dynamical system. Thus, these $d (d - 1)$ Hamilton equations have been incorporated in the 
new Dirac's modification of the classical Hamilton method. At the second step, Dirac rejected the $d$ equations, 
Eq.(\ref{eq253}), that are formally correct but practically useless. Instead, these equations have been replaced by 
an equal number of equations which describe time-evolution of the primary constraints $\phi^{0 \sigma}$ and define 
the new secondary constraints $\chi^{0 \sigma}$, i.e., $\frac{d \phi^{0 \sigma}}{d x_0} = [ \phi^{0 \sigma}, H_t] = 
[ \phi^{0 \sigma}, H_C] = \chi^{0 \sigma}$. Here we apply the fact that all primary constraints commute with each 
other, i.e., $[ \phi^{0\sigma}, \phi^{0\gamma}] = 0$ (see above). At the next (third) step Dirac explicitly derived 
the Hamilton equations which describe time-evolution of all $d$ secondary constraints $\chi^{0 \sigma}$: $\frac{d 
\chi^{0 \sigma}}{d x_0} = [ \chi^{0 \sigma}, H_t] = [ \chi^{0 \sigma}, H_C] = D^{\sigma}_{c} = a^{\alpha}_{\mu}(g) 
\chi^{0 \mu} + b^{\alpha}_{\mu} (g) \Bigl( f^{k}(g) \chi^{0 \mu} \Bigr)_{,k}$, where the function (or functional) 
$D^{\sigma}_{c}$ is the $\sigma-$component of Dirac closure which is a quasi-linear combination of the same secondary 
constraints and total spatial derivatives of expressions which contain the same secondary constraints. All other 
temporal derivatives of the Dirac closure will produce only similar quasi-linear combinations of secondary constraints 
and a few total spatial derivatives of them. The process of time-evolution is formally closed, since we will never see 
anything new in this chain. Briefly, this means that in Dirac's modification of the classical Hamiltonian method all 
$d$ primary and $d$ secondary constraints are considered as the new Hamiltonian dynamical variables. Note that the 
equations which describe the time-evolution of these new dynamical variables are written in a manifestly Hamilton form. 

Let us show how this procedure works in the case of metric gravity. The Poisson brackets between the primary constraints 
$\phi^{0\sigma}$ and canonical Hamiltonian are \cite{K&K}:
\begin{eqnarray}
 & & [ \phi^{0\sigma}, H_C ] = \frac{d \phi^{0\sigma}}{d x_0} = \chi^{0\sigma} = -\frac{g^{0\sigma}}{2 \sqrt{-g} 
 g^{00}} I_{mnpq} \pi^{mn} \pi^{pq} \nonumber \\
 &+& \frac{g^{0\sigma}}{2 g^{00}} I_{mnpq} \pi^{mn} U^{( pq0 \mid \mu\nu k )} g_{\mu\nu,k} + \Bigl[ 
 \pi_{,k}^{\sigma k} + \Bigl(\pi^{pk} e^{q \sigma} - \frac12 \pi^{pq} e^{k \sigma}) g_{pq,k} \Bigr] \nonumber \\
 &-& \frac{\sqrt{-g}}{8} \Bigl(\frac{g^{0\sigma}}{g^{00}} 
 I_{mnpq} B^{((mn) 0 \mid \mu \nu k)} B^{(pq0 \mid \alpha \beta t)} - g^{0\sigma} B^{\mu \nu k \alpha \beta t} 
 \Bigr) g_{\mu\nu,k} g_{\alpha\beta,t} \nonumber \\
 &+& \frac{\sqrt{-g}}{4 g^{00}} I_{mnpq} B^{((mn) 0 \mid\mu\nu k )} g_{\mu\nu,k} g_{\alpha\beta,t} 
 \Bigl[ g^{\sigma t} \Bigl( g^{00} g^{p \alpha} g^{q \beta} + g^{pq} g^{0 \alpha} g^{0 \beta} - 2 g^{\alpha q} 
 g^{0 p} g^{0 \beta} \Bigr) \nonumber \\
 &-& g^{\sigma p} \Bigl( 2 g^{00} g^{q \alpha} g^{t \beta} - g^{00} g^{\alpha \beta} g^{q t} + g^{\alpha\beta} 
 g^{0q} g^{0t} - 2 g^{q \alpha} g^{0 \beta} g^{0t} - 2 g^{t \alpha} g^{0 \beta} g^{0q} + 2 g^{qt} g^{0\alpha} 
 g^{0\beta} \Bigr) \nonumber \\ 
 &+& g^{0\sigma} ( 2 g^{\beta t} g^{\alpha p} g^{0q} - 2 g^{p\alpha} g^{q\beta} g^{0t} - 2 g^{pq} g^{t\beta} 
 g^{0\alpha} + 2 g^{pt} g^{q\beta} g^{0\alpha} + g^{pq} g^{\alpha\beta} g^{0t} - g^{tp} g^{\alpha\beta} g^{0q}) 
 \Bigr] \nonumber \\
 &-& \frac{\sqrt{-g}}{4} g_{\mu\nu,k} g_{\alpha\beta,t} \Bigl[ g^{\sigma t} ( g^{\alpha\mu} g^{\beta\nu} g^{0k} 
 + g^{\mu\nu} g^{\alpha t} g^{0\beta} - 2 g^{\mu\alpha} g^{k\nu} g^{0\beta} ) \nonumber \\
 &+& g^{0\sigma} ( 2 g^{\alpha t} g^{\beta\mu} g^{\nu k} - 3 g^{t\mu} g^{\nu k} g^{\alpha\beta} - 2 g^{\mu\alpha} 
 g^{\nu\beta} g^{kt} + g^{\mu\nu} g^{kt} g^{\alpha\beta} + 2 g^{\mu t} g^{\nu\beta} g^{k\alpha}) \nonumber \\
 &+& g^{\sigma\mu} \Bigl( (g^{\alpha\beta} g^{\nu t} - 2 g^{\nu\alpha} g^{t\beta}) g^{0k} + 2 ( g^{\beta\nu} g^{kt} 
 - g^{\beta k} g^{t\nu}) g^{0\alpha} + ( 2 g^{k\beta} g^{\alpha t} - g^{\alpha\beta} g^{kt}) g^{0\nu}\Bigr) 
 \Bigr] \nonumber \\
 &-& \frac{\sqrt{-g} g^{00}}{2} E^{pqt\sigma} \Bigl( \frac{1}{g^{00}} I_{mnpq} B^{((mn)0 \mid \mu\nu k)} g_{\mu\nu,k} 
  \Bigr)_{,t} - \frac{\sqrt{-g}}{2} B^{((\sigma 0) k \mid \alpha \beta t)} g_{\alpha\beta,kt} \; , \; \label{eqn8} 
\end{eqnarray}
where $U^{( pq 0 \mid \mu\nu k )}$ is the symmetrized form of the following expression  
\begin{eqnarray}
 U^{\alpha\beta 0 \mu\nu k} = B^{(\alpha\beta 0 \mid \mu \nu k)} - g^{0k} E^{\alpha \beta \mu \nu} + 2 g^{0\mu} 
 E^{\alpha \beta k \nu} \; \; \label{Sab}
\end{eqnarray}
and $\sigma = 0, 1, \ldots, d - 1$. Thus, the corresponding Poisson brackets of the primary constraints with the 
canonical Hamiltonians $H_C$ are the new functions of generalized coordinates $g_{\alpha\beta}$ (or $g^{\sigma\gamma}$) 
and momenta $\pi^{\mu\nu}$. In respect to the original terminology introduced by Dirac (see, e.g., \cite{Dir64}) these 
$\chi^{0\sigma}$ functions are the secondary constraints of this Hamiltonian formulation. Briefly, the definition of 
secondary constraints is written in the form: $\chi^{0\sigma} = [ \phi^{0\sigma}, H_C ]$, where $\sigma = 0, 1, \ldots, 
d - 1$. This means that in metric gravity we always have $d$ secondary constraints $\chi^{0\sigma}$ (= $\chi^{\sigma 
0}$). On actual Hamiltonian trajectories (and only on these trajectories) of the free gravitational field these 
secondary constraints must be equal zero, i.e., we can write the following weak equations $\chi^{0\sigma} \approx 0$ 
for $\sigma = 0, 1, \ldots, d - 1$. Note also that the Poisson brackets between the primary and secondary constraints 
are $[ \phi^{0\gamma}, \chi^{0\sigma} ] = \frac12 g^{\gamma\sigma} \chi^{0 0}$ \cite{Fro2021}. It can also be shown 
that all primary $\phi^{0\lambda}$ and secondary constraints $\chi^{0\sigma}$ which arise during this Hamiltonian 
formulation of the metric gravity are the first-class constraints \cite{Dir64}.

At the next step of the original Dirac procedure \cite{Dir58}, \cite{Dir50} and \cite{Dir64} we have to determine the 
temporal derivatives of all secondary constraints, i.e., $\frac{d \chi^{0\sigma}}{d x_{0}} = [ \chi^{0\sigma}, H_t] 
= [ \chi^{0\sigma}, H_C] + [ \chi^{0\sigma}, g_{0 0,0} \phi^{0 0} + 2 g_{0 k,0} \phi^{0 k}]$. The first Poisson 
brackets is 
\begin{eqnarray}
 \frac{d \chi^{0\sigma}}{d x_{0}} &=& [ \chi^{0\sigma}, H_{C} ] = -\frac{2}{\sqrt{-g}} I_{mnpq} \pi^{mn} 
 \Bigl(\frac{g^{\sigma q}}{g^{00}}\Bigr) \chi^{0p} + \frac12 g^{\sigma k} g_{00,k} \chi^{00} + \delta_{0}^{\sigma} 
 \chi_{,k}^{0k} \nonumber \\
 &+& \Bigl[ -2 \frac{1}{\sqrt{-g}} I_{mnpk} \pi^{mn} \frac{g^{\sigma p}}{g^{00}} + I_{mkpq} \Bigl(\frac{g^{\sigma 
 m}}{g^{00}}\Bigr) U^{(pq) 0 \mu\nu l} g_{\mu\nu,l} \Bigr] \chi^{0k} \nonumber \\
 &-& \Bigl[ g^{0\sigma} g_{00,k} + 2 g^{n\sigma} g_{0n,k} + \frac{g^{n\sigma} g^{0m}}{g^{00}} (g_{mn,k} 
 + g_{km,n} - g_{kn,m}) \Bigr] \chi^{0k} = D^{\sigma}_{c} \; , \; \label{close}
\end{eqnarray}
where $D^{\sigma}_{c}$ is the $\sigma$-component of the Dirac closure, $U^{(p q) 0 \mu\nu k}$ is the quantity $U^{p q 
0 \mu \nu k}$ from Eq.(\ref{Sab}) which is symmetrized upon all $p \leftrightarrow q$ permutations. The second Poisson 
bracket in the original expression for $\frac{d \chi^{0\sigma}}{d x_{0}}$ takes the form 
\begin{eqnarray}
 [ \chi^{0\sigma}, g_{00,0} \phi^{00} + 2 g_{0 k,0} \phi^{0 k}] = - \frac12 g^{0 \sigma} g_{00,0} \chi^{0 0} 
 - g^{0 k} g_{0 k,0} \chi^{0 0} = - \frac12 \Bigl( g^{0 \sigma} g_{00,0} + 2 g^{0 k} g_{0 k,0} \Bigr) 
  \chi^{0 0} \label{eq47}
\end{eqnarray}
and it is proportional to the secondary constraint $\chi^{0 0}$. Thus, the Poisson brackets $[ \chi^{0\sigma}, H_t]$ 
and $[ \chi^{0\sigma}, H_C]$ are written as a linear combinations with field-dependent coefficients (we call them 
quasi-linear combinations) of the secondary constraints $\chi^{0 \gamma}$ only. The $[ \chi^{0\sigma}, H_C]$ Poisson 
bracket is called the $\sigma$-component of Dirac closure $D^{\sigma}_{c}$, or the Dirac $\sigma-$closure for the 
Hamiltonian formulation of metric gravity. In some old papers the Dirac closure has been defined as the $[ 
\chi^{0\sigma}, H_t]$ Poisson bracket. The difference between these two definitions is proportional to the secondary 
constraint $\chi^{0 0}$ (see, Eq.(\ref{eq47})), and we do not have any principal contradiction between these two 
definitions of Dirac's closure. Also, note that this expression for Dirac closure, Eq.(\ref{close}), written in terms 
of secondary constraints only, is one of three possible results in the original Dirac procedure \cite{Dir50}, 
\cite{Dir64}. Briefly, this means that our Hamiltonian formulation of metric gravity does not lead either to any 
constraints of higher order, e.g., tertiary constraints, or to any inconsistency which can be fatal for the whole 
theory based on the $\Gamma - \Gamma$ Lagrangian, Eq.(\ref{LGG}) \cite{Dir64}. Finally, we need to say that in metric 
gravity the Dirac closure is a $d-$vector-like quantity in contrast with the Maxwell $d-$dimensional electrodynamics 
of the free EM-field, where the Dirac closure is a scalar which equals zero for the free EM-field \cite{Dir64}, 
\cite{FroUni}. 

Thus, in the metric gravity each primary constraint generates one secondary constraint and the Dirac's chain 
of first-class constraints ends at the secondary constraints. Finally, we have $d$ primary and $d$ secondary 
first-class constraints, i.e., the total number of the first-class constraints in metric gravity equals $2 
d$. In this sense there is an obvious similarity between the Hamiltonian approach for the Maxwell theory of 
multi-dimensional electromagnetic field (see, e.g., \cite{Dir64}, \cite{FroUni}) and Hamiltonian formulation 
of the metric gravity. Furthermore, all Hamiltonian formulations of different physical fields, which contain 
equal numbers of the primary and secondary first-class constraints, are quite similar to each other. The 
source of such a similarity can be traced back to the fact that the original Lagrangian density ($L_{\Gamma 
- \Gamma}$, Eq.(\ref{LGG}), in our case) is written as a quadratic-linear combination of velocities (or 
field-velocities). 

In conclusion we want to note that there is a direct relation which allows one to express the canonical 
Hamiltonian $H_C$ in terms of the secondary constraints $\chi^{0 \sigma}$ and total spatial derivatives 
\begin{eqnarray}
 H_C = - 2 g^{0 \sigma} \chi^{0 \sigma} + \Bigl( 2 g_{0 m} \pi^{m k} \Bigr)_{,k} + \Bigl[ \sqrt{- g} \Bigl( 
 g_{0\gamma} B^{((0 \gamma) k \mid \alpha \beta m)} - g_{0 n} B^{((n k) 0\mid \alpha \beta m)}  \Bigr) 
 g_{\alpha\beta, m} \Bigr]_{,k} \; \; . \; \; \label{HCconst}
\end{eqnarray}
This formula relates the canonical Hamiltonian $H_C$ which depends upon the space-like momenta $\pi^{m n}$ (they 
belong to the pure dynamical $d (d - 1)-$dimensional space) and secondary first-class constraints $\chi^{0 
\sigma}$ which belong to the pure constraint, or non-dynamical $2 d$ dimensional subspace (see below). From this 
point of view the equation, Eq.(\ref{HCconst}), is the `additionality' relation between the dynamical and 
constraint parts of the total Hamiltonian of the metric gravity.       

\section{Canonical transformations}

One of the main advantages of the Hamiltonian formulation(s) of any physical theory is a possibility to apply 
various canonical transformations of the Hamiltonian dynamical variables. In general, such canonical 
transformations can be used to simplify either the canonical Hamiltonian $H_C$, or to reduce this Hamiltonian 
to some special forms, e.g., to its natural form \cite{Fro2021}. In the Hamiltonian formulations of metric 
gravity the canonical transformations of Hamiltonian dynamical variables are often used to simplify the 
explicit form of secondary constraints. Indeed, the secondary constraints derived above in the form of 
Eq.(\ref{eqn8}) are very complex. Applications and even simple operations with secondary constraints written 
in such a form are very difficult. For instance, calculations of the Poisson brackets between primary and 
secondary constraints, of between any pair of secondary constraints produce formulas which are extremely 
cumbersome. For the first time, this has been noticed by Dirac in his fundamental paper \cite{Dir58}. To 
resolve these issues he used some canonical transformation of the original (Hamiltonian) dynamical variables 
which were originally introduced in \cite{PirSS}. At that time nobody performed similar transformations in 
metric gravity. This explains why Dirac in \cite{Dir58} started his transformations from the original $\Gamma 
- \Gamma$ Lagrangian density, Eq.(\ref{LGG}), which is also an ultimate source of the Hamiltonian theory. As 
is well known in classical mechanics we can always add the total temporal derivative to our original Lagrangian 
density without any change in the Lagrange equations of motion. The same rule is true in the general relativity 
and metric gravity, where we have also to take care about general covariance of all our formulas and expressions.   

Briefly, the relation between the Dirac Lagrangian density introduced in \cite{Dir58} and our $L_{\Gamma - 
\Gamma}$ Lagrangian density, Eq.(\ref{LGG}), is written in the form $L_D = L_{\Gamma - \Gamma} - L^{\star}$ 
\cite{FK&K}, where the additional Lagrangian density $L^{\star}$ takes the manifestly covariant form 
\begin{eqnarray}
  L^{\star} = \Bigl[ \Bigl(\sqrt{- g} g^{0 0}\Bigr)_{,\alpha} \frac{g^{0 \alpha}}{g^{0 0}} \Bigr]_{,0}  
  - \Bigl[ \Bigl(\sqrt{- g} g^{0 0}\Bigr)_{,0} \frac{g^{0 \alpha}}{g^{0 0}} \Bigr]_{,\alpha} =  
  \Bigl[ \Bigl(\sqrt{- g} g^{0 0}\Bigr)_{,k} \frac{g^{0 k}}{g^{0 0}} \Bigr]_{,0} - 
  \Bigl[ \Bigl(\sqrt{- g} g^{0 0}\Bigr)_{,0} \frac{g^{0 k}}{g^{0 0}} \Bigr]_{,k} \; . \label{LDLGG} 
\end{eqnarray}
This equation is reduced to the form 
\begin{eqnarray}
  L^{\star} = \frac12 \sqrt{- g} A^{\alpha \beta 0 \mu \nu k} g_{\alpha\beta,0} g_{\mu\nu,k} &=& 
  \frac12 \sqrt{- g} \Bigl( e^{\alpha\beta} e^{k \mu} g^{0 \nu} - e^{\mu\nu} e^{k \alpha} g^{0 \beta} + 
  e^{k \alpha} \frac{g^{0 \mu} g^{0 \nu} g^{0 \beta}}{g^{0 0}} \nonumber \\ 
  &-& e^{k \mu} \frac{g^{0 \alpha} g^{0 \nu} g^{0 \beta}}{g^{0 0}} \Bigr) g_{\alpha\beta,0} g_{\mu\nu,k} 
  \; \; . \; \; \label{AAB} 
\end{eqnarray}  
The $A^{\alpha \beta 0 \mu \nu k}$ coefficients defined in this equation has a few following symmetries. 
First, these coefficients are symmetric upon the $\alpha\beta \leftrightarrow \beta\alpha$ and $\mu\nu 
\leftrightarrow \nu\mu$ permutations. Second, the important property of the $A^{\alpha \beta 0 \mu \nu 
k}$ coefficients is their anti-symmetry with respect to interchange of the two pairs of Greek indices, 
i.e., $A^{\alpha \beta 0 \mu \nu k} = - A^{\mu \nu 0 \alpha \beta k}$. Third, these coefficients are 
linearly related with the coefficients $B^{(\alpha \beta 0 \mid \mu \nu k)}$ and Dirac tensor 
$E^{\alpha\beta\gamma\sigma}$ (both defined in Section III). The explicit form of this relation is
\begin{eqnarray}
  A^{\alpha \beta 0 \mu \nu k} = B^{(\alpha \beta 0 \mid \mu \nu k)} - g^{0 k} E^{\alpha \beta \mu \nu} 
  + 2 g^{0 \mu} E^{\alpha \beta k \nu} \; \; . \; \; \label{ABE}
\end{eqnarray} 

Finally, the relation between the Dirac's Lagrangian $L_D$ and our original $L_{\Gamma - \Gamma}$ 
Lagrangian of the metric gravity (see above) is written in the form \cite{FK&K}
\begin{eqnarray}
 L_D = L_{\Gamma - \Gamma} - L^{\star} = L_{\Gamma - \Gamma} - \frac12 \sqrt{- g} A^{\alpha \beta 0 \mu 
 \nu k} g_{\alpha\beta,0} g_{\mu\nu,k} \; , \; {\rm where} \; L^{\star} = \frac12 \sqrt{- g} 
 A^{\alpha \beta 0 \mu \nu k} g_{\alpha\beta,0} g_{\mu\nu,k} \; . \label{LD}
\end{eqnarray} 
From this equation one easily finds the following expression for the momenta $p^{\gamma\sigma}$ in the 
Dirac Hamiltonian formulation of the metric gravity 
\begin{eqnarray}
  \frac{\partial L_D}{\partial g_{\gamma\sigma,0}} = \frac{\partial L_{\Gamma - \Gamma} }{\partial 
  g_{\gamma\sigma,0}} - \frac{\partial L^{\star}}{\partial g_{\gamma\sigma,0}} \; \; , \; {\rm or} \; \; 
  p^{\gamma\sigma} = \pi^{\gamma\sigma} - \frac12 \sqrt{- g} A^{(\gamma \sigma) 0 \mu \nu k} g_{\mu\nu,k} 
  \; \; , \; \; \label{mDmGG} 
\end{eqnarray} 
where $p^{\gamma\sigma}$ are the new momenta (or Dirac's momenta), while $\pi^{\gamma\sigma}$ are the old 
momenta defined above in Section III. The last equation in Eq.(\ref{mDmGG}) is, in fact, the explicit 
definition of the Dirac's momenta which is conveniently to write in the two following forms: 
\begin{eqnarray}
 p^{p q} = \pi^{p q} - \frac12 \sqrt{- g} A^{(p q) 0 \mu \nu k} g_{\mu\nu,k} \; \; \; {\rm and} \; \; \;
 p^{0 \sigma} = \pi^{0 \sigma} - \frac12 \sqrt{- g} A^{(0 \sigma) 0 \mu \nu k} g_{\mu\nu,k} \; \; , \; \;
 \label{Dirmom} 
\end{eqnarray} 
where $A^{(0 \sigma) 0 \mu \nu k} g_{\mu\nu,k} = \frac12 \Bigl( B^{(\alpha \beta 0 \mid \mu \nu k)} + 
B^{(\beta \alpha 0 \mid \mu \nu k)} \Bigr)$ and $p^{0 \sigma} = p^{\sigma 0}$. Thus, we have the two sets 
of Hamiltonian dynamical variables for the two different Hamiltonian formulations of the metric gravity: $\{ 
g_{\alpha\beta}, \pi^{\mu\nu} \}$ (the old set) and $\{ g_{\alpha\beta}, p^{\mu\nu} \}$ (the new set). Since 
these two sets of dynamical variables are related to each other by a canonical transformation, then the three 
following conditions for the Poisson brackets must be obeyed: $[ g_{\alpha\beta}, g_{\mu\nu} ] = 0, [ 
g_{\alpha\beta}, p^{\mu\nu} ] = \Delta^{\mu\nu}_{\alpha\beta}$ and $[ p^{\alpha\beta}, p^{\mu\nu} ] = 0$, 
where all new variables are written in terms of the old variables. For old variables we already know that 
the following equations are true: $[ g_{\alpha\beta}, g_{\mu\nu} ] = 0, [ g_{\alpha\beta}, \pi^{\mu\nu} ] = 
\Delta^{\mu\nu}_{\alpha\beta}$ and $[ \pi^{\alpha\beta}, \pi^{\mu\nu} ] = 0$. In reality, applications of 
these canonicity conditions needs some additional explanations, since for all Hamiltonian systems such 
conditions are always derived and formulated in a different form which is based on the `alternative' Laplace 
(not Poisson!) brackets. Here we have to make one step aside and discuss the general canonicity conditions 
for an arbitrary transformation of the Hamiltonian dynamical variables. 

\subsection{General conditions of canonicity for transformations of the dynamical variables}

Let us assume that some Hamiltonian system is described by the $2 n$ independent dynamical variables $\{ q_k, 
p_k \}$, where $k = 1, \ldots, n$. In general, it is possible to replace these `old' dynamical variables by 
the new dynamical variables $\{ \tilde{q}_i, \tilde{p}_i \}$, where $i = 1, \ldots, n$:
\begin{eqnarray}
 \tilde{q}_{i} = \phi_{i}(t, q_{k}, p_{k}) \; \; \; \tilde{p}_{i} = \psi_{i}(t, q_{k}, p_{k}) \; \; , 
 \; \; \label{q-q}
\end{eqnarray}
but after such a transformation of variables we want to be sure that the new Hamiltonian system will be 
`dynamically equivalent' to our original Hamiltonian system. Transformations of the dynamical variables each of 
which transforms one Hamiltonian system into another Hamiltonian system, which is completely and unambiguously 
`dynamically equivalent' to the original system, are defined as the canonical transformations. In general, all 
canonical transformations of any Hamiltonian system form the closed algebraic structure, or group, for short 
(see, e.g., \cite{Fro2021}, \cite{Gant}). It was shown (by Jacobi) that for any time-dependent canonical 
transformation of the dynamical variables, Eq.(\ref{q-q}), the following canonicity condition (below, the main 
canonicity condition) must be obeyed 
\begin{eqnarray}
  \sum^{n}_{k=1} \tilde{p}_k d\tilde{q}_k - \tilde{H} \delta t = c \Bigl( \sum^{n}_{k=1} p_k dq_k 
  - H \Bigr) \delta t - \delta F(t, q_k, p_k) \; \; , \; \; \label{Jacob0}
\end{eqnarray} 
where $c (\ne 0)$ is some real number which does not depend upon the time $t$. The function $ F(t, q_k, p_k)$ 
is the Jacobi generating function, i.e., the function which generates this canonical transformation. Vice 
versa, one can easily show that, if Eq.(\ref{Jacob0}) holds for some transformation of the dynamical 
variables, then this transformation is canonical. For better understanding of equations from this subsection 
we use the explicit sing of summation. 

Moreover, since the valence $c (\ne 0)$ does not depend upon the time, then by establishing the criteria of 
canonicity, we can always restrict ourselves (for more details, see, e.g., \cite{Gant}) to the time-independent 
canonical transformations only, i.e., 
\begin{eqnarray}
 \tilde{q}_{i} = \phi_{i}(q_{k}, p_{k}) \; \; \; \tilde{p}_{i} = \psi_{i}(q_{k}, p_{k}) \; \; . 
 \; \; \label{q-q-t}
\end{eqnarray}
For a canonical time-independent transformation the main condition, Eq.(\ref{Jacob0}), is written in 
the form 
\begin{eqnarray}
  \sum^{n}_{k=1} \tilde{p}_k d\tilde{q}_k = c \sum^{n}_{k=1} p_k dq_k - \delta K(q_k, p_k) \; \; , 
  \; \; \label{Jacob1}
\end{eqnarray} 
where $\tilde{q}_{k}, \tilde{p}_{k}$ ($k = 1, \ldots, n$) are the new generalized coordinates and momenta, 
while $q_{i}, p_{i}$ ($i = 1, \ldots, n$) are the old coordinates and momenta (old dynamical variables). 
Also, in this equation $K(q_{k}, p_{k}) = F(\overline{t}, q_{k}, p_{k})$, i.e., it is a short Jacobi 
generating function of the coordinates and momenta only, which coincides with the Jacobi generating 
function $F(\overline{t}, q_{k}, p_{k})$ taken at some fixed time $t = \overline{t}$. The variation of 
$K(q_{k}, p_{k})$ is written in the form 
\begin{eqnarray}
  \delta K = - \sum^{n}_{i=1} \Bigl( \Phi_{i} \delta q_{i} + \Psi_{i} \delta p_{i} \Bigr) \; \; . \; 
  \; \label{Jacob2}
\end{eqnarray} 
On the other hand, by using the formula $\delta \tilde{q}_{k} = \frac{\partial \tilde{q}_{k}}{\partial q_{i}} 
\delta q_{i} + \frac{\partial \tilde{q}_{k}}{\partial p_{i}} \delta p_{i}$ in Eq.(\ref{Jacob1}) one finds the 
following expression for the $\delta K(q_{k}, p_{k})$ variation:  
\begin{eqnarray}
  \delta K = - \sum^{n}_{i=1} \Bigl[ \sum^{n}_{k=1} \Bigl( \tilde{p}_{k} \frac{\partial \tilde{q}_{k}}{\partial 
  q_{i}} \Bigl) - c p_{i} \Bigr] \delta q_{i} - \sum^{n}_{i=1} \Bigl[ \sum^{n}_{k=1} \Bigl( p_{k} \frac{\partial 
  \tilde{q}_{k}}{\partial p_{i}} \Bigr) \Bigr] \delta p_{i} \; \; . \; \; \label{Jacob3}
\end{eqnarray} 
By comparing Eqs.(\ref{Jacob2}) and (\ref{Jacob3}) one finds 
\begin{eqnarray}
 \Phi_{i} = \sum^{n}_{k=1} \tilde{p}_{k} \frac{\partial \tilde{q}_{k}}{\partial q_{i}} - c p_{i} \; \; \; \; 
 {\rm and} \; \; \; \; \Psi_{i} = \sum^{n}_{k=1} p_{k} \frac{\partial \tilde{q}_{k}}{\partial p_{i}} \; \; . 
 \; \; \label{Jacob4} 
\end{eqnarray} 

For canonical transformation(s) the expression in the left side of Eq.(\ref{Jacob2}) must be a total 
differential. From here one finds three following conditions: 
\begin{eqnarray}
 \frac{\partial \Phi_{i}}{\partial q_j} = \frac{\partial \Phi_{j}}{\partial q_i} \; \; \; , \; \; \; 
 \frac{\partial \Psi_{i}}{\partial p_j} = \frac{\partial \Psi_{j}}{\partial p_i} \; \; \; , \; \; \; 
 \frac{\partial \Phi_{i}}{\partial p_j} = \frac{\partial \Psi_{j}}{\partial q_i} \; \; \; , \; \; \; 
 \; \; . \; \; \label{Jacob50} 
\end{eqnarray} 
By substituting the functions $\Phi_{i}$ and $\Psi_{i}$ in these equations by their expressions from 
Eq.(\ref{Jacob4}) one finds after a few additional and simple transformations:   
\begin{eqnarray}
  && \sum^{n}_{k=1} \Bigl( \frac{\partial \tilde{q}_{k}}{\partial q_{i}} 
  \frac{\partial \tilde{p}_{k}}{\partial q_{j}} - \frac{\partial \tilde{q}_{k}}{\partial q_{i}} 
  \frac{\partial \tilde{q}_{k}}{\partial q_{j}} \Bigl) = 0 \; \; , \; {\rm or} \; \; \;  
  \{ q_{i}, q_{j} \} = 0 \; \; , \; \; \label{Jacob51} \\
  && \sum^{n}_{k=1} \Bigl( \frac{\partial \tilde{q}_{k}}{\partial p_{i}} 
  \frac{\partial \tilde{p}_{k}}{\partial p_{j}} - \frac{\partial \tilde{q}_{k}}{\partial p_{i}} 
  \frac{\partial \tilde{q}_{k}}{\partial p_{j}} \Bigl) = 0 \; \; , \; {\rm or} \; \; \;  
  \{ p_{i}, p_{j} \} = 0 \; \; , \; \; \label{Jacob53} \\
  && \sum^{n}_{k=1} \Bigl( \frac{\partial \tilde{q}_{k}}{\partial q_{i}} 
  \frac{\partial \tilde{p}_{k}}{\partial p_{j}} - \frac{\partial \tilde{q}_{k}}{\partial p_{i}} 
  \frac{\partial \tilde{q}_{k}}{\partial q_{j}} \Bigl) = c \delta_{ij} \; \; , \; {\rm or} 
  \; \; \; \{ q_{i}, p_{j} \} = c \delta_{ij} \; , \; \; \label{Jacob55}
\end{eqnarray} 
where $\delta_{ij}$ is the Kronecker symbol and $c$ is some numerical constant. The constructions (or sums) 
which appear in these three equations are the Laplace brackets which are well known in classical mechanics 
(see, e.g., \cite{Gant}, \cite{Gold}). The standard notation for the Laplace brackets (see, e.g., \cite{Gant}, 
\cite{Gold}) is $\{ , \}$. Each of these sums includes $2 n$ functions ($\tilde{q}_{k}$ and $\tilde{p}_{k}$) 
and two variables only, e.g., either $q_{i}, q_{j}$, or $p_{i}, p_{j}$, or $q_{i}, p_{j}$. As follows from 
Eqs.(\ref{Jacob51}) - (\ref{Jacob55}) some transformation of the dynamical variables will be canonical if 
(and only if) the three groups of following conditions are obeyed: $\{ q_{i}, q_{j} \} = 0 , \{ p_{i}, p_{j}
\} = 0$ and $\{ q_{i}, p_{j} \} = c \delta_{ij}$, where $c \ne 0$ and $(i,j) = 1, \ldots, n$, for all $2 n$ 
new dynamical variables $\tilde{q}_{k}, \tilde{p}_{k}$ ($k = 1, \ldots, n$). 

In reality, the original Laplace brackets are not convenient in applications. However, as follows from the 
Appendix B these brackets can be replaced by the Poisson brackets, each of which is the adjoint to the 
corresponding Laplace bracket. In terms of the Poisson brackets the same criteria of canonicity are written 
in a different form (for more details, see our Appendix B): $[\tilde{q}_{i}, \tilde{q}_{j} ] = 0 \; \; , \; 
\; [ \tilde{p}_{i}, \tilde{p}_{j} ] = 0 \; \; , \; \; [ \tilde{q}_{i}, \tilde{p}_{j} ] = c \delta_{ij}$, 
where $(i,j) = 1, \ldots, n$ and $c$ is the valence of this canonical transformation. These numerical values 
of the Poisson brackets taken for $c = 1$ are used below as the criteria of canonicity for the transformation 
of dynamical variables. To simplify the text below we shall call these brackets by the canonical, univalent 
set of the Poisson brackets, or CUSPB, for short. 

\subsection{Applications to the metric gravity}

Let us apply the formulas derived above to the metric gravity by considering a transformation from the 
old set of dynamical variables $\{ g_{\alpha\beta}, \pi^{\mu\nu} \}$ to the new set of such variables 
$\{ g_{\alpha\beta}, p^{\mu\nu} \}$. First, we note that the generalized coordinates $g_{\alpha\beta}$ 
are identical in the both sets. For the brackets defined in the previous subsection this means that 
$\{ g_{\alpha\beta}, g_{\gamma\rho} \} = 0$ and $[ g_{\alpha\beta}, g_{\gamma\rho} ] = 0$. Furthermore, 
for the univalent ($c = 1$) transformations of dynamical variables we have for the new momenta $p^{\mu\nu} 
= \pi^{\mu\nu} + f^{\mu\nu}(g_{\gamma\rho})$, where $f^{\mu\nu}(g_{\gamma\rho})$ is a tensor function of 
generalized coordinates only. From here one finds that $[ g_{\alpha\beta}, p^{\mu\nu} ] = [ g_{\alpha\beta}, 
\pi^{\mu\nu} ] = \Delta^{\mu\nu}_{\alpha\beta}$. In other words, the first and last Poisson brackets from 
CUSPB are obeyed automatically for our transformation of the Hamiltonian dynamical variables. The only 
non-trivial bracket in CUSPB (see, Eq.(\ref{adjnt32}) in the Appendix B) is the second Poisson bracket 
between two new momenta which takes the following form in our tensor notations:   
\begin{eqnarray} 
  [ p^{\alpha\beta}, p^{\mu\nu} ] = 0 \; \; , \; {\rm or} \; \; [ \pi^{\alpha\beta} - \frac12 
  \sqrt{- g} A^{(\alpha \beta) 0 \sigma \rho k} g_{\sigma\rho,k}, \pi^{\mu\nu} -  \frac12 
  \sqrt{- g} A^{(\mu \nu) 0 \lambda \kappa k} g_{\lambda\kappa,k} ] = 0  \; \; , \; \label{p-p}
\end{eqnarray} 
which is instantly reduced to the equation 
\begin{eqnarray} 
  [ \pi^{\alpha\beta}, \sqrt{-g} A^{(\mu \nu) 0 \sigma \rho m} g_{\sigma\rho,m} ] =  
  [ \pi^{\mu\nu}, \sqrt{- g} A^{(\alpha \beta) 0 \sigma \rho m} g_{\sigma\rho,m} ] 
  \; \; . \; \label{pi-pi}
\end{eqnarray} 
The transformation of dynamical variables will be canonical, if (and only if) this equation is obeyed.  
To proof the validity of this equation one has to perform direct calculations of the Poisson brackets 
in the both sides of Eq.(\ref{pi-pi}). In reality, the both sides of Eq.(\ref{pi-pi}) are compared 
with each other and identical terms (in the both sides) are cancelled. Finally, this equation is 
reduced to the form of identity such as $0 = 0$. 

In those cases when either $\alpha = 0$, or $\beta = 0$ (or both) one obtains from Eqs.(\ref{ABE}) and 
(\ref{LD}) the following equation: 
\begin{eqnarray} 
  p^{0 \gamma} = \pi^{0 \gamma} - \frac12 \sqrt{- g} B^{((0 \gamma) 0 \mid \mu \nu k)} g_{\mu\nu,k} 
  , \; \; \label{PrimetD} 
\end{eqnarray} 
which defines the momenta with one (or two) temporal component(s), or primary constraints $p^{0 \gamma} 
\approx 0$ in the Dirac's Hamiltonian formulation of the metric gravity. For these momenta the 
canonicity conditions, Eq.(\ref{p-p}), must also be obeyed. After a few simple transformations the 
essential canonicity conditions for the $p^{0 \gamma}$ and $p^{0 \sigma}$ momenta take one of the 
following forms  
\begin{eqnarray} 
 [ p^{0 \gamma}, p^{0 \sigma} ] = 0 \; \; , \; \; {\rm or} \; \: [ \pi^{0 \gamma}, \sqrt{- g} B^{((0 
 \sigma) 0 \mid \mu \nu k)} g_{\mu\nu,k} ] = [ \pi^{0 \sigma}, \sqrt{- g} B^{((0 \gamma) 0 \mid \mu 
 \nu k)} g_{\mu\nu,k} ] \; \; , \; \; \label{PrimeD}
\end{eqnarray} 
which simply means that all primary constraints in the Dirac's Hamiltonian formulations commute with 
each other. The same statement is true for the original Hamiltonian formulation of metric GR \cite{K&K} 
discussed above. This fact has been checked in \cite{PirSS}. On the other hand, we have to note that 
the fact that all primary constraints in the metric gravity commute with each other follows directly 
from the canonicity of the complete Dirac's set of Hamiltonian dynamical variables. 

At this point it is very convenient to introduce the universal notation $\phi^{\mu\nu}$ for the momenta, 
or for contravariant components of the momenta. In Dirac's Hamiltonian formulation these momenta are 
$\phi^{\mu\nu} = p^{\mu\nu}$, while in the Hamiltonian formulation from \cite{K&K} these momenta are 
$\phi^{\mu\nu} = \pi^{\mu \nu} - \frac12 \sqrt{- g} A^{(\mu \nu) 0 \alpha \beta k} g_{\alpha\beta,k}$. 
In this notation the canonical Hamiltonians $H_C$ in the both formulation of metric gravity are 
represented in the same `universal' form \cite{FK&K}
\begin{eqnarray} 
 & &H_C = \frac{1}{\sqrt{-g} g^{00}} I_{mnpq} \phi^{mn} \phi^{pq} - \frac{1}{g^{00}} \phi^{mn} 
 \Bigl( g^{0 l} g_{m n,l} - 2 g^{0\alpha} g_{\alpha n,m} \Bigr) \; \; \; \label{DirH-C} \\
 &+& \frac14 \sqrt{-g} \Bigl[ \frac{1}{g^{00}} \Bigl( g^{0 k} E^{(m n) \mu \nu} - 2 g^{0\mu} 
 E^{(m n) k \nu} \Bigr) \Bigl(g^{0 l} g^{\alpha}_{m} g^{\beta}_{n} - 2 g^{0 \alpha} g^{l}_{m} 
 g^{\beta}_{n} \Bigr) - B^{\mu\nu k \alpha\beta l}\Bigr] g_{\mu\nu,k} g_{\alpha\beta,l} \; \; , 
 \; \nonumber
\end{eqnarray} 
where $g^{\alpha}_{\beta} = \delta^{\alpha}_{\beta}$ is the substitution tensor defined above. In 
the both formulations the primary constraints commute with each other, i.e., $[ \phi^{0\gamma}, 
\phi^{0\sigma} ] = 0$. The knowledge of the canonical Hamiltonian $H_C$ and all primary constraints 
allows one to restore the total Hamiltonian $H_{t}$:  
\begin{eqnarray}   
  H_t = H_C + g_{0 0,0} \phi^{0 0} + 2 g_{0 k,0} \phi^{0 k} = H_C + g_{0 0,0} p^{0 0} + 
  2 g_{0 k,0} p^{0 k} \; \; . \; \; \label{DirH-t}   
\end{eqnarray}  
In particular, in the Dirac's Hamiltonian formulation we obtain $ H_t = H_C + g_{0 0,0} p^{0 0} + 2 g_{0 k,0} 
p^{0 k}$. It appears that the total Hamiltonian $H_t$ does not change during canonical transformations of the 
dynamical variables, i.e., $H^{K\&K}_t(g_{\alpha\beta}, \pi^{\mu\nu}) = H^{Dir}_t(g_{\alpha\beta}, p^{\mu\nu})$ 
\cite{FK&K}. In other words, the total Hamiltonian is an obvious and unique invariant of this theory. In respect 
to this, the Hamilton equations do not change its form during canonical transformations and we can write, e.g.,  
\begin{eqnarray} 
  g_{\alpha\beta,0} = [ g_{\alpha\beta}, H^{K\&K}_t] \; , \; \pi^{\mu\nu}_{,0} = [ \pi^{\mu\nu}, 
  H^{K\&K}_t] \; \Leftrightarrow \; g_{\alpha\beta,0} = [ g_{\alpha\beta}, H^{Dir}_t] \; , 
  \; p^{\mu\nu}_{,0} = [ p^{\mu\nu}, H^{Dir}_t] \label{HK&KtoHD}
\end{eqnarray} 
i.e., these two sets of Hamilton equations are equivalent to each other. In other words, the Hamilton equations 
conserve their form during canonical transformation of the dynamical variables. In fact, this was the first 
definition (or criterion) of canonicity for the transformations of dynamical variables which has been formulated 
by Sir William R. Hamilton himself in 1834 and 1835. We have shown that his criterion works for the Hamiltonian 
approach to the metric gravity. However, the metric gravity is a dynamical system with constraints. It is clear 
that the Hamilton criterion, as well as other criteria of canonicity known for the transformations of dynamical 
variables in classical mechanics, must be supplemented by some statement(s) about the algebra of constraints (see 
below). Therefore, we need to derive the explicit expressions for the secondary constraints, their Poisson 
brackets with the canonical and/or total Hamiltonians, primary constraints, etc. 

All secondary constraints in the Dirac's Hamiltonian formulation are derived from the equations $\chi^{0\sigma} = 
[ \phi^{0\sigma}, H_t] = [ \phi^{0\sigma}, H_C]$. The explicit expressions are   
\begin{eqnarray}  
  \chi^{0\sigma} = - \frac{g^{0\sigma}}{2 \sqrt{-g} g^{00}} I_{mnpq} \phi^{mn} \phi^{pq} + 
  g^{\sigma}_{m} \Bigl[ (\phi^{mk})_{,k} + \Bigl( \phi^{pk} e^{qm} - \frac12 \phi^{pq} 
  e^{km}\Bigr) g_{pq,k} \Bigr] \nonumber \\ 
  + \frac12 \sqrt{- g} g^{0\sigma} \Bigl[ - g_{mn,kl} E^{mnkl} + \frac14 g_{mn,k} g_{pq,l} 
  \Bigl( - E^{mnpq} e^{kl} + 2 E^{klpn} e^{mq} + E^{pqnl} e^{mk} \Bigl)\Bigr] \; \; . \; 
  \label{Dirchi} 
\end{eqnarray}
This formula is very compact and contains only two lines (compare with the formula, Eq.(\ref{eqn8})). It indicates 
clearly that Dirac's idea to apply canonical transformations of the Hamiltonian dynamical variables in order to 
simplify secondary first-class constraints works perfectly. Now, by using the explicit form of the primary 
$\phi^{0 \gamma} = p^{0\gamma}$ and secondary $\chi^{0 \sigma}$ constraints in Dirac's formulation one finds 
\begin{eqnarray}  
 [ \phi^{0 \gamma}, \chi^{0 \sigma}]_{Dirac} = [ p^{0 \gamma}, \chi^{0 \sigma}]_{Dirac} = 
 \frac12 g^{\gamma\sigma} \chi^{0 0}_{Dirac} \; \; \; , \; \; \label{phichi}
\end{eqnarray} 
i.e., the formula which exactly coincides (by its form) with the formula $[ \phi^{0 \gamma}, \chi^{0 
\sigma}]_{K\&K} = \frac12 g^{\gamma\sigma} \Bigl(\chi^{0 0}\Bigr)_{K\&K}$ mentioned above. It is very 
interesting, since the explicit forms of all primary and secondary constraints are substantially
different in these two formulations. This and other similar facts directly follow from the canonicity 
of our transformation of the Hamiltonian dynamical variables. The time-evolution of the secondary 
constraints leads to the following formula
\begin{eqnarray}  
  & & \frac{d \chi^{0\sigma}}{d x_0} = \chi^{0\sigma}_{,0} = [ \chi^{0\sigma}, H_C] = - \Bigl[ 
  \frac{2}{\sqrt{-g} g^{00}} I_{pqmk} g^{\sigma m} \phi^{pq} + g^{0 \sigma} g_{0 0,k} + 
  2 g^{\sigma p} g_{0 p,k} \nonumber \\
  &+& \Bigl(\frac{g^{\sigma p} g^{0 q}}{g^{0 0}}\Bigr) \; \; \Bigl( g_{p q,k} + g_{q k,p} - 
  g_{p k,q} \Bigr)\Bigr] \chi^{0 k} - g^{\sigma}_{0} (\chi_{0 k})_{,k} + \frac12 g^{\sigma k} 
  g_{0 0,k} \chi^{0 0} = D^{\sigma}_{c} \; , \; \label{DirclD} 
\end{eqnarray} 
where $D^{\sigma}_{c}$ is the $\sigma-$component of the Dirac closure derived in the Dirac's Hamiltonian 
formulation of the metric gravity. All components of the Dirac closure are quasi-linear combination of 
secondary constraints and some total spatial derivatives of these secondary constraints. Again, this 
formula is very compact and transparent. Finally, we want to present the formula, which allows one to 
express the canonical Hamiltonian $H_C$ in terms of secondary constraints and some (total) spatial 
derivatives, Eq.(\ref{HCconst}), can also be derived in the Dirac Hamiltonian formulation. The formula 
takes the form, which is slightly different from Eq.(\ref{HCconst}) above:   
\begin{eqnarray} 
 H_C = -2 g_{0\lambda} \chi^{0\lambda} &+& \Bigl(2 g_{0 m} \phi^{mk} \Bigr)_{,k} - \Bigl[\sqrt{- g} 
 E^{mnpq} g_{mn,q} \nonumber \\
 &-& \sqrt{- g} g_{\mu\nu,k} \; \Bigl(\frac{g^{0\mu}}{g^{00}}\Bigr) \; \Bigl( g^{\nu p} g^{0 k} - 
 g^{\nu k} g^{0 p} \Bigr)\Bigr]_{,p} \; \label{HDirconst} 
\end{eqnarray} 
This formula also represents the canonical Hamiltonian $H_C$ (in the Dirac formulation) written as a 
quasi-linear combination of the secondary constraints $\chi^{0 \sigma}$ and a few total spatial 
derivatives of some expressions which include the same secondary constraints.   

Now, we can complete our discussion of canonical transformations in the metric gravity. There are three 
general rules which regulate changes in the primary and secondary constraints during such transformations. 
The first rule is simple and it is called the law of inertia for the first-class constraints. Indeed, by 
performing a number of canonical transformations between different sets of dynamical variables we have found 
that the total number of the primary $\phi^{0 \sigma}$ constraints $N_p$ never changes during such 
transformations. The same statement is true for the total number of secondary $\chi^{0 \sigma}$ constraints 
$N_s$ and for the sum $N_p + N_s$. We have to emphasize here that all primary and secondary constraints which 
arise in the Hamiltonian formulations of metric gravity are first-class. The second rule of `form-invariance' 
is even simpler: the internal structure of all first-class constraints must be conserved during canonical 
transformations of the Hamiltonian dynamical variables. The preservation of form-invariance for all first-class 
constraints is crucial to prove that any two Hamiltonian formulations developed for the same constrained 
dynamical system are equivalent to each other. The third rule essentially follows from the second rule: all 
Poisson brackets between the first-class constraints and canonical/total Hamiltonians, other constraints, etc, 
must also be form-invariant during canonical transformations. For simple Poisson brackets, e.g., for the 
$[\phi^{0\sigma}, \chi^{0\gamma}]$ brackets, this rule leads to the exact coincidence of corresponding 
expressions. The three rules mentioned here essentially mean preservation of the algebra of first-class 
constraints. Thus, the canonical transformations in the metric gravity must guarantee a complete preservation 
of the form-invariance for the total Hamiltonian $H_t$ and for the algebra of first-class constraints. 

The formulas derived in this Section allow one to apply the Dirac's Hamiltonian formulation of metric gravity 
to analyze and solve various gravitational problems. In some cases, however, one needs to know analytical 
expressions for other Poisson brackets, e.g., the Poisson bracket between two secondary constraints $[ 
\chi^{0\gamma}, \chi^{0\sigma} ]$ is of great interest, but it has never been obtain in previous papers. This 
Poisson bracket is determined in our `technical' Appendix A. In general, calculations of this and other 
Poisson brackets can be performed with the use of our formulas and method described in Section IV. 

\section{Dirac's modifications of the classical Hamilton method}

In this Section we want to reconsider modifications which were made by Dirac in the classical Hamilton method 
\cite{Dir58}, \cite{Dir50}, \cite{Dir64}. This will eventually lead us to the new universal criterion of canonicity 
for Hamiltonian formulations of metric gravity. First, we note again that in any Hamiltonian formulation of metric 
gravity we always have $\frac{d (d + 1)}{2}$ generalized coordinates $g_{\alpha\beta}$ and $\frac{d (d + 1)}{2}$ 
momenta $p^{\mu\nu}$. These coordinates and momenta are the Hamiltonian dynamical variables of our problem (metric 
gravity). The total number of these variables equals $d (d + 1)$ which is an even number for any $d-$dimensional 
Riemann space-time. Note also that our original Lagrangian $L_{\Gamma - \Gamma}$, Eq.(\ref{LGGvel}), is a quadratic 
function upon velocities of the space-like components of the metric tensor $g_{m n}$, i.e., upon $g_{m n,0}$. On the 
other hand, the same $L_{\Gamma-\Gamma}$ Lagrangian is a linear function of the $d$ remaining $g_{0 \gamma,0}$ (= 
$g_{\gamma 0,0}$) velocities which are also called the temporal velocities. 

By using the standard Legendre transformation (see Section III) one can pass from the Lagrangian $L_{\Gamma 
- \Gamma}$ to the Hamiltonian $H_C$ which is quadratic function of space-like momenta $\pi^{m n}$. This 
Hamiltonian is called the canonical Hamiltonian, and it is an explicit function of the space-like dynamical 
$\{ g_{m n}, p^{p q} \}$ variables (there are $d(d-1)$ of such variables) and $d$ `temporal' coordinates $g_{0 
\sigma}$ only. The Hamiltonian $H_C$ is of great interest for the whole metric gravity, since it describes 
the actual motions of a free gravitational field. However, we have to note that this Hamiltonian $H_C$ does 
not depend upon any of the temporal momenta, i.e., it does not include any of the $p^{0 \sigma}$ (or 
$p^{\sigma 0}$) momenta. This means that all Poisson brackets such as $[ g_{0 \sigma}, H_C] = 0$ and $[ 
g_{\sigma 0}, H_C] = 0$ equal zero identically, and canonical Hamiltonian $H_C$ does not describe time-evolution 
of the temporal coordinates $g_{0 \sigma}$ and/or $g_{\sigma 0}$ coordinates, in principle. Briefly, we can say 
that in the canonical Hamiltonian $H_C$ these temporal coordinates $g_{0 \mu}$ (and $g_{\mu 0}$) are rather 
parameters than actual dynamical variables. 

For normal applications of the Hamilton method we must have a Hamiltonian that contains all momenta, 
including temporal ones. Such a complete, or total Hamiltonian $H_t$ will describe time-evolution of 
all $d (d + 1)$ dynamical variables of the problem $\{ g_{\alpha\beta}, p^{\mu\nu} \}$ and all 
functions of these variables, including the canonical Hamiltonian $H_C$, new coordinates and momenta, 
which can be introduced by some canonical transformations, etc. Formally, this total Hamiltonian can 
be derived from our quadratic-linear Lagrangian $L_{\Gamma - \Gamma}$ by using the Legendre transform 
which is described in detail in Section III. However, for our quadratic-linear Lagrangian $L_{\Gamma 
- \Gamma}$, Eq.(\ref{LGGvel}), the Legendre transform works with some singularities. The two main 
singularities must be mentioned here, since they play crucial roles in Dirac's modification of the 
classical Hamilton method. First, as follows from the definition of momenta and from the general 
technique of Legendre transformations, we cannot obtain, in principle, the explicit expressions of 
the velocities $v_{\gamma} (= g_{0\gamma,0})$ written in terms of momenta $p^{0\gamma}$ and vice 
versa. Instead, we obtain the following algebraic equations: $p^{0 \gamma} \approx f(g_{\alpha\beta}, 
g^{0 \gamma})$, or $\phi^{0 \gamma} = p^{0 \gamma} - f(g_{\alpha\beta}, g^{0 \gamma}) \approx 0$ 
which are called the primary constraints (see, e.g., \cite{Dir64} and references therein). Second, 
in respect to the procedure of Legendre transformation, this moment must be multiplied by the 
corresponding velocity $g_{0\gamma,0} (= v_{\gamma})$, which is not a dynamical variable of our 
Hamiltonian method. This velocity is rather a parameter (arbitrary parameter) of the updated Legendre 
procedure. 

Thus, we have derived the total Hamiltonian $H_t$, Eq.(\ref{eq1}), which is written as the sum of the 
canonical Hamiltonian $H_C$ and primary constraints $\phi^{0 \alpha}$. The coefficients in front of the 
primary constraints equal to the corresponding velocities $v_{\alpha}$, i.e., $H_t = H_C + g_{0 0,0} 
\phi^{0 0} + 2 g_{0 k,0} \phi^{0 k} = H_C + v_{\alpha} \phi^{0 \alpha}$. Now, the time-evolution of any 
dynamical variable, or any function/functional, or quantity, which depends upon the complete set of 
dynamical variables $\{ g_{\alpha\beta}, p^{\mu\nu} \}$, are determined by the Poisson bracket of this 
variable (or function) with the total Hamiltonian $H_t$, e.g., $g_{\alpha\beta,0} = [ g_{\alpha\beta}, 
H_t ]$. Note that this new (total) Hamiltonian of the metric gravity acts in the $d (d + 1)$ dimensional 
space of the dynamical variables, in contrast with the canonical Hamiltonian $H_C$ which formally operates 
in the $d (d - 1)$ dimensional space of the space-like dynamical variables. In general, introduction of 
the new Hamiltonian $H_t$ always brings some new motions that did not exist in the original Hamiltonian 
system with the canonical Hamiltonian $H_C$.

Immediately, the two following questions arise: (1) what is the sense of these `additional' motions, and 
(2) how can they affect the actual motions of our field, which are determined by the canonical Hamiltonian 
$H_C$. To understand this and answer the questions raised, let us consider the time-evolution of the 
canonical Hamiltonian $H_C$. First of all, we can write the following general formula which describes 
time-evolution of the canonical Hamiltonian  
\begin{eqnarray} 
  H_{C}(t + \Delta) = H_{C}(t) + \frac{\Delta}{1!} \Bigl(\frac{d H_{C}}{d t}\Bigr) + 
   \frac{\Delta^{2}}{2!} \Bigl(\frac{d^{2} H_{C}}{d t^{2}}\Bigr) + \frac{\Delta^{3}}{3!} \Bigl(\frac{d^{3} 
   H_{C}}{d t^{3}}\Bigr) + \ldots \; \; , \; \label{H-CDelta}
\end{eqnarray} 
where $\Delta$ is a small time interval. In this equation the first-order time derivative of $H_C$ is written 
in the form: 
\begin{eqnarray} 
   \frac{d H_{C}}{d t} = [ H_C, H_t ] =  [ H_C, H_C + v_{\alpha} \phi^{0\alpha} ] = v_{\alpha} [ H_C, 
   \phi^{0\alpha} ] = - v_{\alpha} \; \chi^{0\alpha} \; \; , \; \label{1stDer}
\end{eqnarray} 
where $\phi^{0\alpha}$ and $\chi^{0\alpha}$ are the primary and secondary first-class constraints, respectively. 
The explicit formulas for the $\phi^{0\alpha}$ and $\chi^{0\alpha}$ constraints are presented above (see, 
Eqs.(\ref{PrimetD}) and (\ref{Dirchi})). Also in this equation and everywhere below the notation $v_{\alpha} 
(= g_{0\alpha,0})$ is an arbitrary, in principle, velocity of the temporal $(0\alpha)$-component of the metric 
tensor. In Dirac's theory this and other similar velocities, e.g., $v_{\beta} (= g_{0\beta,0}), v_{\gamma} (= 
g_{0\gamma,0})$, etc, are considered as arbitrary parameters of the method. 

The second time-derivative of the canonical Hamiltonian $H_C$ is
\begin{eqnarray} 
   \frac{d^{2} H_{C}}{d t^{2}} = \Bigl[ \frac{d H_{C}}{d t}, H_t ] =  [ - v_{\alpha} \chi^{0\alpha}, H_C 
   + v_{\beta} \phi^{0\beta} ] = - v_{\alpha} D^{\alpha}_{c} - \frac12 \; v_{\alpha} v_{\beta} \; 
   g^{\alpha\beta} \chi^{0 0} \; \; \; \label{2ndDer}
\end{eqnarray} 
where $D^{\alpha}_{c}$ is the $\alpha$ component of the Dirac closure (see, Eq.(\ref{DirclD})), while 
$\chi^{0 0}$ is the secondary constraint defined above (see, Eq.(\ref{Dirchi})). The third time-derivative 
of the canonical Hamiltonian takes the form 
\begin{eqnarray} 
   \frac{d^{3} H_{C}}{d t^{3}} = [ \frac{d^{2} H_{C}}{d t^{2}}, H_t ] =  
   - v_{\alpha} [ D^{\alpha}_{c}, H_C ] &+& \frac12 v_{\alpha} v_{\beta} [ g^{\alpha\beta} \chi^{0 0}, H_C ] 
   + \frac12 v_{\alpha} v_{\beta} v_{\gamma} [ g^{\alpha\beta} \chi^{0 0}, \phi^{0 \gamma} ] \nonumber \\ 
   &-& v_{\alpha} v_{\gamma} [ D^{\alpha}_{c}, \phi^{0\gamma} ] \; \; . \; \label{3rdDer}
\end{eqnarray} 
In principle, such a chain of time derivatives $\frac{d^{n} H_{C}}{d t^{n}}$ is infinite (in contrast with the
$n-$dimensional Maxwell electrodynamics \cite{FroUni}), but we have to note that all values in the right-hand 
sides of these equations are always represented as finite, linear (or quasi-linear) combinations of the 
secondary, first-class constraints only. Furthermore, the coefficients in front of each term in these 
expressions depends upon the $v_{\alpha}, v_{\beta}, v_{\gamma}$ and other similar velocities, which ``are 
completely arbitrary and at our disposal'' \cite{Dir64}. In other words, these velocities are arbitrary 
parameters in the Dirac's modification of the classical Hamilton method. It is clear that similar 
transformations which depend upon arbitrary parameters cannot affect the actual (Hamiltonian) motion of the 
original dynamical system, e.g., a free gravitational field, in our case. Instead, they produce some changes 
in the Hamiltonian dynamical variables, which do not correspond to a change of physical state. Generators of 
such `fictional' transformations are the secondary first-class constraints (as Dirac predicted in \cite{Dir64}). 
This follows directly from Eqs.(\ref{1stDer}) - (\ref{3rdDer}) and other similar equations for higher-order 
derivatives in that chain. In field theory similar transformations of the dynamical variables are well known, 
and in earlier papers they were called gauge transformations, or simply gauges. Dirac could obtain and write 
(see, e.g., \cite{Dir64}) all essential equations for the actual motion and for the corresponding gauge 
generators in the united form of Hamilton equations. This explains the overall significance of Dirac's 
modification of the classical Hamilton method.  

In Dirac method the complete system of Hamiltonian equations for the metric gravity is written in the form 
\begin{eqnarray} 
  \frac{d g_{p q}}{d t} = g_{p q, 0} = [ g_{p q}, H_C ] \; \; \; {\rm and} \; \; \;
  \frac{d p^{m n}}{d t} = \Bigl(p^{m m}\Bigr)_{,0} = [ p^{p q}, H_C ] \; \; \; . \label{metrgr1}
\end{eqnarray} 
These $d (d - 1)$ Hamilton equations describe the actual motion of a free gravitational field. Solutions of 
these equations cannot become $v-$dependent at any moment of time-evolution (see discussion above). In addition 
to these equations we also have $d$ Hamilton equations which describe time-evolution of the primary constraints: 
\begin{eqnarray} 
  \frac{d \phi^{0 \alpha}}{d t} = [ \phi^{0 \alpha}, H_C ] = \chi^{0 \alpha} \; \; , \; \; \label{metrgr2} 
\end{eqnarray} 
where $\chi^{0 \alpha}$ are the secondary constraints. The following group of $d$ Hamilton equations describe 
time-evolution of the secondary constraints: 
\begin{eqnarray} 
  \frac{d \chi^{0 \alpha}}{d t} = [ \chi^{0 \alpha}, H_C ] = D^{\alpha}_{c} \; \; , \; \; \; \label{metrgr3} 
\end{eqnarray} 
where $D^{\alpha}_{c} = a^{\alpha}_{\mu} (g) \chi^{0 \mu} + b^{\alpha}_{\mu} (g) \Bigl( f^{k}(g) \chi^{0 \mu} 
\Bigr)_{,k}$ is the $\alpha-$component of the Dirac closure, which is a quasi-linear combination of the 
secondary constraints $\chi^{0 \sigma}$ and some total spatial derivatives of expressions which also include the 
same secondary constraints. All classical theory of the free gravitational field (in metric gravity) is summed 
up in these Dirac's equations, Eqs.(\ref{metrgr1}) - (\ref{metrgr3}), written here in a manifestly Hamiltonian 
form. Note also that there is a simple procedure which allows one to simplify (drastically) the explicit form of 
Dirac closure in the metric gravity. Indeed, in metric gravity on the shell of primary constraints we can always 
determine $d$ field-dependent coefficients ${\cal C}^{\alpha}_{\beta}(g)$ for which the following equations are 
satisfied: 
\begin{eqnarray} 
  [ \chi^{0 \alpha} + {\cal C}^{\alpha}_{\beta} \phi^{0 \beta} , H_C ] = \Lambda_{\alpha} \chi^{0 \alpha} 
  \; \; , \; \; \; \label{Frolov1} 
\end{eqnarray} 
where $\alpha = 0, 1, \ldots, d - 1, \beta = 0, 1, \ldots, d - 1$ and $\Lambda_{\alpha}(g)$ are some algebraic, 
field-dependent expressions, which are often called `eigenvalues' (or factor-eigenvalues) of the Dirac closure. 
Derivation of equations for the unknown ${\cal C}^{\alpha}_{\beta}(g)$ coefficients in Eq.(\ref{Frolov1}) is 
straightforward. In this procedure the Dirac closure becomes `diagonal' and each component of Dirac closure, 
e.g., $D^{\alpha}_{c}$ always contains only one secondary constraint, e.g., $\chi^{0 \alpha}$ in Eq.(\ref{Frolov1}). 
All secondary constraints in this procedure are uniquely determined as factor-eigenvectors which are defined on the 
shell of primary constraints. In this version of Dirac's approach we do not need to say many words to describe the 
internal structure of Dirac closure. 

Also, it is important to remember that in metric gravity we always have $[ \phi^{0 \alpha}, \phi^{0 \beta} ] = 0$, 
which means the pair-wise commutativity of the primary constraints. These generalized Hamilton equations, 
Eqs.(\ref{metrgr1}) - (\ref{metrgr3}), form a complete and unambiguous set of equations, which govern the behaviour 
of a free gravitational field in the $d (d + 1)-$dimensional space of dynamical variables, or in the original 
$d-$dimensional Riemann space. The Hamiltonian equations from the first group, Eq.(\ref{metrgr1}), are the canonical 
Hamilton equations of actual motion for true dynamical variables. Analogous equations from the second group, 
Eqs.(\ref{metrgr2}) - (\ref{metrgr3}), are the Hamilton equations for gauge generators. These equations determine 
the actual gauge generators for the given dynamical system, i.e., for the free gravitational field in our case. All 
equations from the second group describe certain changes in the dynamical variables, i.e., coordinates and momenta, 
which do not affect the real physical state. 

Thus, our original $d (d + 1)$-dimensional space of dynamical variables in the metric gravity splits into the 
$d (d - 1)$-dimensional space of dynamical variables, which describe actual motions, and $2 \; d$-dimensional 
space of variables, which are transformed in some way with time, but this does not make any changes in the real 
physical state. Formally, we can write this in the form: $S[d (d + 1)] = S[d (d - 1)] \oplus S[2 \; d]$, where 
all spaces are even-dimensional. If an additional temporal variable $t$ is introduced in our analysis, then all 
these three spaces become odd-dimensional and Hamilton method works perfectly in each of these spaces. Note that 
the Hamilton equations in the form of Eqs.(\ref{metrgr1}) - (\ref{metrgr3}) are more useful and informative for 
the field people, than the equivalent original system of the $d (d + 1)$ Hamilton equations:
\begin{eqnarray} 
  \frac{d g_{\alpha\beta}}{d t} = g_{\alpha\beta, 0} = [ g_{\alpha\beta}, H_t ] \; \; \; {\rm and} \; \; \;
  \frac{d p^{\mu\nu}}{d t} = \Bigl(p^{\mu\nu}\Bigr)_{,0} = [ p^{\mu\nu}, H_t ] \; \; \; . \label{metrgrD}
\end{eqnarray} 
The replacement of this system of Hamilton equations by much more useful system of slightly different Hamilton 
equations, Eqs.(\ref{metrgr1}) - (\ref{metrgr3}), is the main advantage of the Dirac's modifications made in 
the classical Hamilton method. Another advantage follows from the fact that the governing equations for all 
gauge generators are also written in the form of Hamilton equations. The third advantage is obvious: from now 
on all calculations in the metric gravity are reduced to analytical calculations of the Poisson brackets only.  

Finally, we can formulate the complete and pure formal criterion of canonicity for some transformation between any 
two equivalent Hamiltonian formulations of the metric gravity. Based on arguments and equations presented in this 
and previous Sections, the universal criterion of canonicity for the metric gravity can be formulated in the 
following form. Some transformation of the dynamical Hamilton variables in metric gravity is canonical if (and only 
if) it transforms our original system of Hamilton equations, Eqs.(\ref{metrgr1}) - (\ref{metrgr3}), into a new 
system of similar Hamilton equations: 
\begin{eqnarray} 
  & & \frac{d \tilde{g}_{p q}}{d t} = \tilde{g}_{p q, 0} = [ \tilde{g}_{p q}, \tilde{H}_C ] \; \; \; {\rm and} 
   \; \; \; \frac{d \tilde{p}^{m n}}{d t} = \Bigl(\tilde{p}^{m m}\Bigr)_{,0} = [ \tilde{p}^{p q}, \tilde{H}_C ]  
   \; \; , \; \; \label{ametrgr} \\
 & & \frac{d \tilde{\phi}^{0 \alpha}}{d t} = [ \tilde{\phi}^{0 \alpha}, \tilde{H}_C ] = \tilde{\chi}^{0 \alpha} 
  \; \; \; {\rm and} \; \; \; \frac{d \tilde{\chi}^{0 \alpha}}{d t} = [ \tilde{\chi}^{0 \alpha}, \tilde{H}_C ] 
  = \tilde{D}^{\alpha}_{c} \; \; , \; \; \label{bmetrgr}  
\end{eqnarray} 
where the sign $\; \tilde{} \;$ means the new variable and/or function, while all new functions $\tilde{H}_C, 
\tilde{\chi}^{0 \alpha}$ and $\tilde{D}^{\alpha}_{c} = \tilde{a}^{\alpha}_{\mu} (\tilde{g}) \tilde{\chi}^{0 \mu} 
+ \tilde{b}^{\alpha}_{\mu}(\tilde{g}) \Bigl( \tilde{f}^{k}(g) \tilde{\chi}^{0 \mu} \Bigr)_{,k} $, which appear in 
these equations, must have the same structure as the old functions $H_C, \chi^{0 \alpha}$ and $D^{\alpha}_{c}$ in 
Eqs.(\ref{metrgr1}) - (\ref{metrgr3}). This new system of equations represents the form-invariance of the Hamilton 
equations derived by Dirac for the metric gravity, which is a constrained dynamical system with the first-class 
constraints only. Also, for the true canonical transformation in the metric gravity the following equations for 
the Poisson brackets must be obeyed: 
\begin{eqnarray} 
  & & [ \tilde{g}_{\alpha \beta}, \tilde{g}_{\mu \nu} ] = 0 \; \; , \; \; [ \tilde{g}_{\alpha \beta}, 
  \tilde{p}^{m n} ] = \Delta^{m n}_{\alpha \beta} \; \; , \; \; [ \tilde{p}^{m n}, \tilde{p}^{p q} ] = 0 
  \; \; , \; \; [ \tilde{\phi}^{0 \gamma}, \tilde{\phi}^{0 \sigma} ] = 0  \nonumber \\  
  & & [ \tilde{g}_{\mu \nu}, \tilde{\phi}^{0 \sigma} ] = \Delta^{0 \sigma}_{\mu \nu} \; \; \; , \; \; \; 
  [ \tilde{p}^{m n}, \tilde{\phi}^{0 \sigma} ] = 0 \; \; . \; \; \label{Ametrgr} 
\end{eqnarray} 
This criterion of canonicity can be generalized to other Hamiltonian dynamical systems with first-class 
constraints.

Note also that in this Section we discuss only one version of the complete Dirac's approach \cite{Dir58}, 
which has been developed to deal with a free gravitational field in the metric gravity. Generalization of 
this procedure to other fields with non-trivial gauge invariance is also possible. For instance, the same 
approach works perfectly for a free electromagnetic field even in multi-dimensions (see, e.g., \cite{Dir64}, 
\cite{Tyut} and \cite{FroUni}). Our preliminary results indicate clearly that the quantum version of this 
approach is applicable (with some changes) to the modern united electroweak theory.  

\subsection{On complete reverse recovery of the original field equations} 

In the previous Sections, we have carefully derived the Hamiltonian equations for a free gravitational 
field and all primary and secondary first-class constraints. The main purpose of our analysis was to 
obtain the correct equations of motion (or time-evolution) of a free gravitational field and obtain 
all important  gauge conditions. Here the following question immediately arises: what are these correct 
Hamiltonian equations of motion? Where and how was the criterion of correctness established? The answer 
is clear, and we have to recognize as correct only such Hamiltonian equations and first-class constraints 
which uniformly lead us back to the original (or maternal) equations of motion already known for our field.
For a free gravitational field the maternal field equations are the Einstein's equations $G_{\alpha\beta} 
= 0$ (or $R_{\alpha\beta} = 0$) mentioned in Section II. For a free electromagnetic field (or EM-field, 
for short) the maternal equations are the Maxwell equations in vacuum. Therefore, any correct Hamiltonian 
approach for EM-field must be able to produce the governing Maxwell equations at any spatial point ${\bf 
x}$ and at any moment of time $t$. To explain how this works, let us consider the Hamiltonian formulation 
for the Maxwell electromagnetic field in the $(n + 1)-$dimensional (flat) space-time. 

We can start directly form the explicit form of the corresponding $EM-$Hamiltonian (all missing details, 
definitions and notations can be found in \cite{Dir64} and \cite{FroUni}). Also, form the definition of 
momenta $B^{\mu} = F^{\mu 0}$ to this point we already have one primary constraint $\phi = B^{0} \approx 
0$, since the both $F^{\mu \nu}$ and $F_{\mu \nu}$ tensors are always antisymmetric. The fundamental 
Poisson brackets are: $[ A_{\mu}({\bf x}), B^{\nu}({\bf x}^{\prime}) ] = g^{\nu}_{\mu} \delta^{n}({\bf x} 
- {\bf x}^{\prime}), [ A_{\mu}({\bf x}), A_{\nu}({\bf x}^{\prime}) ] = 0$ and $[ B^{\mu}({\bf x}), 
B^{\nu}({\bf x}^{\prime}) ] = 0$. The Hamiltonian of a free electromagnetic field in the $(n + 
1)$-dimensional space-time is written in the form \cite{Dir64}, \cite{FroUni}:    
\begin{eqnarray}
 H = \int \Bigl( \frac14 F^{p q} F_{p q} - \frac12 F^{p 0} F_{p 0} + F^{q 0} A_{0,q} \Bigr) d^{n}{\bf x} = 
 \int \Bigl( \frac14 F^{p q} F_{p q} + \frac12 B^{p} B^{p} - A_{0} B^{p}_{,p} \Bigr) d^{n}{\bf x} \; \; 
 , \; \label{HamltA3}
\end{eqnarray}
where all notations are exactly the same as in \cite{Dir64} and \cite{Fro2021}. The corresponding Hamiltonian 
density takes the form
\begin{eqnarray}
  {\cal H} = \frac14 F^{p q} F_{p q} + \frac12 B^{p} B^{p} - A_{0} B^{p}_{,p} \; \; . \; \label{Hamltden1}
\end{eqnarray}
Integration by parts of the first term in the Hamiltonian, Eq.(\ref{HamltA3}), leads to the following expression 
for the Hamiltonian density Eq.(\ref{Hamltden1}):    
\begin{eqnarray}
 {\cal H} = - \frac12 \Bigl(F^{p q}\Bigr)_{q} A_{p} + \frac12 B^{p} B^{p} - A_{0} B^{p}_{,p} = \frac12 \Bigl( 
 \frac{\partial^{2} A_{q}}{\partial x_{p} \partial x_{q}} - \frac{\partial^{2} A_{p}}{\partial x_{q} \partial 
 x_{q}} \Bigr) A_{p} + \frac12 B^{p} B^{p} - A_{0} B^{p}_{,p} \; , \; \label{Hamltden3}
\end{eqnarray}
where $p = 1, 2, \ldots, n$ and $q = 1, 2, \ldots, n$, i.e., all these indexes are space-like. First of 
all, by determining the Poisson bracket $[ B^{0}, {\cal H}] = B^{p}_{,p}$ one finds the secondary 
constraint $\chi = B^{p}_{,p} \approx 0$. In standard notation this means that $\frac{\partial}{\partial 
t}\Bigl(\frac{\partial A_{p}}{\partial x_{p}}\Bigr) \approx 0$, or $\frac{\partial A_{p}}{\partial 
x_{p}} = C$, where $C$ is a numerical constant which does not depend upon ${\bf x}$ and/or $t$. This 
secondary constraints $\chi$ commute with the Hamiltonian density ${\cal H}$ and Dirac closure equals 
zero identically.  

By using the Hamiltonian density ${\cal H}$, Eq.(\ref{Hamltden3}), we obtain the following system of canonical 
Hamilton equations 
\begin{eqnarray}
 \frac{d A_{p}}{d t} = [ A_{p}, {\cal H} ] = \frac{\partial {\cal H}}{\partial B^{p}} = \frac12 \; (2 B^{p}) 
 = B^{p} \; \; \; \label{caneq1}
\end{eqnarray}
and 
\begin{eqnarray}
 \frac{d B^{p}}{d t} = [ B^{p}, {\cal H} ] = - \frac{\partial {\cal H}}{\partial A_{p}} = - \frac12 \Bigl[ 2  
   \Bigl( \frac{\partial^{2} A_{q}}{\partial x_{q} \partial x_{p}} - \frac{\partial^{2} A_{p}}{\partial x_{q} 
   \partial x_{q}} \Bigr)\Bigr] = \frac{\partial^{2} A_{p}}{\partial x_{q} \partial x_{q}} - \frac{\partial^{2} 
   A_{q}}{\partial x_{q} \partial x_{p}} \; \; . \; \label{caneq2}
\end{eqnarray} 
Combination of these two equations one finds 
\begin{eqnarray}
 \frac{d^{2} A_{p}}{d t^{2}} = \frac{\partial^{2} A_{p}}{\partial x_{q} \partial x_{q}} - \frac{\partial^{2} 
 A_{q}}{\partial x_{q} \partial x_{p}} \; \; . \; \label{eqmot2}
\end{eqnarray}
Now, taking into account the condition which follows from secondary constraint: $\frac{\partial A_{q}}{\partial 
x_{q}} = C$, we can reduce this equation to the form of $n$-dimensional wave equation:
\begin{eqnarray}
 \frac{\partial^{2} A_{p}}{\partial t^{2}} - \frac{\partial^{2} A_{p}}{\partial x_{q} \partial x_{q}} = 0 \; 
 \; , \; {\rm or} \; \; \frac{\partial^{2} {\bf A}}{\partial t^{2}} - \Delta {\bf A} = 0 \; , \; \label{caneq2}
\end{eqnarray}
where ${\bf A} = (A_1, A_{2}, \ldots, A_n)$ is the $n-$dimensional vector potential of the EM-field. This is 
the Maxwell equations for a free electromagnetic field in the $(n + 1)-$dimensional space-time. The 
$n-$dimensional Laplace operator $\Delta$ in this equation is $\Delta = \frac{\partial^{2} }{\partial x_{q} 
\partial x_{q}}$. Thus, we have recovered all Maxwell equations of the free radiation field. Note here that 
if someone does not recognize any constraints at all, or these constraints were determined with mistakes, 
then such a Hamiltonian formulation of the electromagnetic theory does not allow one to recover the 
corresponding Maxwell equations. Formally, in this case any relation with the original Maxwell theory of 
radiation will be lost. In reality, this means that our Hamiltonian, Eq.(\ref{HamltA3}), does not describe 
the Maxwell EM-field in vacuum, or this Hamiltonian formulation is not valid for the Maxwell electromagnetic 
field and cannot be used in applications to this field. This is the principle of complete reverse recovery 
of the original field equations in application to the Maxwell theory of EM-field. 

Now, consider the Hamiltonian formulation of the metric gravity which has been developed by Dirac in \cite{Dir58}. 
In contrast with a free Maxwell EM-field, for a free gravitational field, everything becomes significantly more 
complicated, but our principle of the complete reverse recovery works in this case too. Recently, we have shown 
that the equations for the second-order temporal derivatives of the space-like co-variant components $g_{m n}$ of 
the metric tensor, which follow from the Hamilton equations obtained in the Dirac Hamiltonian formulation of the 
metric gravity, essentially coincide with the corresponding Einstein's equations for the same components. However, 
at that time this article was already completed and it was not possible to add a few new chapters into it. In 
addition to this, it takes a long time to transform a set of difficult formulas into a logically perfect text. 
Therefore, our results in this direction will be published some time later and elsewhere. Here we just want to 
present a few important details of the procedure used.  

Note that there are three peculiarities in the Einstein's equations for a free gravitational field ($R_{\alpha \beta} 
= 0$ plus $d$ additional conditions $R^{\gamma}_{\alpha ,\gamma} = \frac12 \frac{\partial R}{\partial x^{\alpha}}$), 
which are crucially important for our present purposes. First, these Einstein's equations are written as a system 
of differential equations which contains the first- the second-order temporal derivatives of the metric tensor 
$g_{\alpha\beta}$. Second, all second-order temporal derivatives from the $g_{0 \beta}$ and $g_{\beta 0}$ components 
of metric tensor cancel out from these Einstein's equations. Third, the temporal second-order derivatives of the 
spatial components of metric tensor are explicitly included in the Einstein's equations. In fact, each second-order 
derivative $\frac{d^{2} g_{m n}}{d t^{2}} = \frac{d^{2} g_{m n}}{d x^{2}_{0}}$, arises in Einstein's equations only 
from the $R_{0 m 0 n}$ components of the Riemann curvature tensor. From here it is easy to find that all these 
second-order temporal derivatives $\frac{d^{2} g_{m n}}{d t^{2}}$ are included in Einstein's equations only as 
separated terms, and each of these terms has the same numerical coefficient $-\frac12$ in front. As follows from here 
one can reduce the Einstein's equations for the $g_{m n}$ components to the form $\frac{d^{2} g_{m n}}{d t^{2}} 
= Q(g_{p q}, \frac{d g_{p q}}{d t}, g_{0 \alpha}, \frac{d g_{0 \alpha}}{d t})$. These equations must coincide (or be 
equivalent) with the analogous equations for the $\frac{d^{2} g_{m n}}{d t^{2}}$ derivatives which follow from the 
Hamilton equations in the Dirac formulation of the Hamiltonian metric gravity. In fact, we have to calculate the 
Poisson bracket $\frac{d^{2} g_{m n}}{d t^{2}} = [[ g_{m n}, H_t ], H_t ] = [[ g_{m n}, H_C ], H_t ]$ and exclude 
all momenta by using the Hamilton equations, primary and secondary first-class constraints derived in the Dirac 
Hamiltonian formulation (see above). Such calculations are quite complex, extremely time-consuming and very 
sensitive, since any mistake made either in the Hamiltonian, or in one of the first-class constraints substantially 
complicates further calculations. Also after such a mistake one can lose any relation with the original (or 
maternal) theory and cannot move forward, until this mistake was found and corrected. Nevertheless, after many
weeks of calculations, we are happy to report that Hamiltonian formulation of the metric gravity, which has been 
developed by Dirac in \cite{Dir58}, successfully passed this our test of recovery (at least partially). Now, by 
using this Hamiltonian formulation we are able to recover the original field equations for all covariant space-like 
components $g_{m n}$ of the metric tensor. An alternative Hamiltonian formulation of metric gravity which obviously 
fails this test is mentioned in Section IX.

To conclude this discussion we have to note that any correct Hamiltonian formulation of an arbitrary, in principle, 
field theory must reproduce (exactly and unambiguously) the original governing equations of this field. The correct 
Hamilton equations of motions and explicit form of all first-class constraints are crucially important to reach this 
goal. If this is not the case, then such a Hamiltonian theory is wrong and has nothing to do with the maternal field 
theory. 

\section{Invariant integrals of the metric gravity} 

This Section is a central part of our study, since here we define a number of integral invariants of the metric 
gravity, i.e., we reach an absolute top of the classical Hamilton mechanics. Obviously, in one Section we cannot 
even outline the main problems which exist in the theory of integral invariants and its applications to the 
Hamiltonian metric gravity. Therefore, the following presentation of this theory will be very brief. Moreover, 
here we restrict ourselves only to a description of the extension of integral invariants for dynamical 
(Hamiltonian) systems with constraints. More details of the theory of integral invariants and its applications 
to the Hamiltonian formulations of metric gravity will be presented in our next article \cite{Fro2023}.

First of all we need to define the one-dimensional, or line integrals in the metric gravity. In general, the 
one-dimensional integral in multi-dimensional Riemann spaces is defined as follows  
\begin{eqnarray}
  \int \pi^{\alpha\beta} \; d g_{\alpha\beta} &=& \int \pi^{\alpha\beta} \; \Bigl(\frac{\partial 
   g_{\alpha\beta}}{\partial x^{\beta}}\Bigr) dx^{\beta} = \int \pi^{\alpha\beta} \Bigl( 
   \Gamma_{\alpha,\beta \gamma} + \Gamma_{\beta,\alpha \gamma} \Bigr) dx^{\gamma} \nonumber \\ 
   &=& \int \pi^{\alpha\beta} \Bigl( g_{\alpha\lambda} \Gamma^{\lambda}_{\beta \gamma} + g_{\beta\lambda} 
  \Gamma^{\lambda}_{\alpha \gamma} \Bigr) dx^{\gamma} \; \; , \; \; \label{7eq1}  
\end{eqnarray}  
where $\Gamma_{\alpha,\beta \gamma}$ are the Cristoffel symbols of the first kind, while $\Gamma^{\alpha}_{\beta 
\gamma}$ are the Cristoffel symbols of the second kind (see, e.g., \cite{Kochin}). The integrand in this integral 
is not a tensor. This means that the line (or one-dimensional) integral substantially depends on the curve along 
which it is calculated and also on the initial and final points chosen on this curve. If the start and end points 
coincide with each other, then such an integral is called a closed loop integral, or an integral taken along a 
closed loop. Below, similar closed loop integrals are designated by the sign $\oint$. In general, the complete 
theory of line (or one-dimensional) integrals in multi-dimensional Riemann spaces is very complex. However, for 
our current analysis of the integral invariants in the Hamiltonian formulations of metric gravity we do need to 
use the formula, Eq.(\ref{7eq1}). In fact, all line integrals can be considered in the $\frac{d (d + 
1)}{2}$-dimensional pseudo-Euclidean (orthogonal) space, which is formally identical (or isomorphic) to our 
original $d-$dimensional Riemann space-time (for more details, see, e.g., \cite{Kochin}, \cite{Dash}).

This is a very good news, since all integrals and integral forms defined in multi-dimensional pseudo-Euclidean 
spaces can be handled in a familiar way (see, e.g., \cite{HCartan}, \cite{Fland}). In particular, by applying 
the usual definition of the closed loop integrals in pseudo-Euclidean spaces we can consider the two following 
integrals in the $\frac{d (d - 1)}{2}$-dimensional space: 
\begin{eqnarray}  
 I = \oint \Bigl[ \pi^{m n} \; d g_{m n} - H_C dt \Bigr] \; \; \; {\rm and} \; \; \; I_{P} = 
 \oint \pi^{m n} \; d g_{m n} \; \; , \; \; \label{7eq2} 
\end{eqnarray}  
where $\pi^{m n}$ are the space-like components of momenta, while $H_C$ is the canonical Hamiltonian which has 
been defined in Section III. For now we restrict ourselves to the consideration of the space-like components of 
momenta $\pi^{m n}$ and coordinates $g_{m n}$. In other words, below we shall deal with the $\frac{d (d - 
1)}{2}$ dimensional Euclidean position space (and $d (d - 1)$-dimensional phase space) instead of the original 
$(d - 1)$-dimensional sub-space in our original $d$-dimensional Riemann space-time. The coordinates in this 
position space coincide with the covariant components of the fundamental space-like tensor $g_{mn}$. The first 
integral in Eq.(\ref{7eq2}) is called the Poincare-Cartan integral invariant, while the second integral is the 
Poincare integral invariant \cite{Cartan}, \cite{Poin} which is often called the main integral invariant of 
mechanics (or Hamiltonian mechanics). This Poincare integral invariant has a fundamental value for the 
Hamiltonian formulation of metric gravity as well as for the general theory of canonical transformations in 
metric gravity and for analysis and solution of other gravitational problems. Indeed, it is relatively easy 
to prove the following statement. If for some system of first-order differential equations written for the 
space-like components of the metric tensor $g_{m n}$ and momenta $\pi_{m n}$:  
\begin{eqnarray}  
  \frac{d g_{m n}}{d t} = Q_{mn}(t, g_{a b}, \pi^{p q}) \; \; \; , \; \; \; \frac{d \pi^{m n}}{d t} = 
  P^{m n}(t, g_{a b}, \pi^{p q}) \; \; \;  \label{7eq3}
\end{eqnarray}  
the Poincare integral $I_P$, Eq.(\ref{7eq2}), is invariant, then this system of equations, Eq.(\ref{7eq3}), is 
Hamiltonian in the moment of time $t$ which is located between $t - \delta$ and $t + \delta$, where $\delta$ is 
a very small positive number. This term `Hamiltonian' means here that the functions $Q_{mn}(t, g_{a b}, \pi^{p 
q})$ and $P^{m n}(t, g_{a b}, \pi^{p q})$ from the right-hand side of Eqs.(\ref{7eq3}) are represented as the 
partial derivatives (or Poisson brackets) of some scalar function $H$, i.e., 
\begin{eqnarray}
  Q_{mn}(t, g_{a b}, \pi^{p q}) = \frac{\partial H}{\partial \pi^{mn}} = [ g_{mn}, H ] \; \; \; , \; \; \; 
  P^{m n}(t, g_{a b}, \pi^{p q}) = - \frac{\partial H}{\partial g_{mn}} = [ \pi^{mn}, H ] \; \; , \; \label{7eq4}
\end{eqnarray}  
where the notation $[ a, b]$ stands for the Poisson bracket defined by Eq.(\ref{PoisBrack}). An uniform 
reconstruction of the explicit form of this function $H$ (or Hamiltonian) is not an easy task, but if we know 
that the Poincare-Cartan integral is also an integral invariant, then the unknown Hamiltonian exactly coincides 
\cite{Cartan} with the canonical Hamiltonian $H_C$ mentioned in the first integral from Eq.(\ref{7eq2}).  

Let us discuss the following fundamental question. We shall assume that the integral $I$, Eq.(\ref{7eq2}), is 
an integral invariant for our dynamical system and $H_C$ is our Hamiltonian which describes the actual motion 
of this system, i.e., the equations of time-evolution take the form of Hamilton equations, Eq.(\ref{metrgr1}), 
for our $d (d-1)$ dynamical variables $\{ g_{m n}, \pi^{p q} \}$, where $[ g_{m n}, g_{p q} ] = 0, [ \pi^{m n}, 
\pi^{p q} ] = 0$ and $[ g_{m n}, \pi^{p q} ] = \Delta^{p q}_{m n}$. Now, we want to extend our phase space by 
adding a set of $2 d$ new dynamical variables  $\{ g_{0 \gamma}, \pi^{0 \gamma} \}$. Here we assume the usual 
permutation symmetry for all `additional' coordinates and momenta: $g_{0 \gamma} = g_{\gamma 0}$ and $\pi^{0 
\gamma} = \pi^{\gamma 0}$. The total dimension of this new phase space will be $d (d + 1)$, which corresponds 
to the $d-$dimensional Riemann space-time, and it is the main working space for the general relativity and 
metric gravity. We require that in this new extended phase space the integral defined by the last expression 
in Eq.(\ref{7eq2A}) must also be an integral invariant with the new Hamiltonian $H_t$. Furthermore, the new 
`extended' Hamiltonian $H_t$ must be closely related with the canonical Hamiltonian $H_C$ from Eq.(\ref{7eq2}).   

Based on the formulas derived in Section III we can transform the Poincare-Cartan integral invariant from 
Eq.(\ref{7eq2}) into the following form 
\begin{eqnarray}  
 I = \oint \Bigl[ \pi^{m n} \; d g_{m n} &-& H_C dt \Bigr] = \oint \Bigl\{\Bigl[ \pi^{m n} \; d g_{m n} + 
 \pi^{0 \gamma} \Bigl(\frac{d g_{0 \gamma}}{d t}\Bigr) d t \Bigr] - \Bigl[ H_C + \pi^{0 \gamma} 
 \Bigl(\frac{d g_{0 \gamma}}{d t}\Bigr) \Bigr] d t \Bigr\} \nonumber \\ 
 &=& \oint \Bigl[ \pi^{\alpha \beta} \; d g_{\alpha \beta} - \Bigl( H_C + v_{\gamma} \pi^{0 \gamma} \Bigr) dt 
 \Bigr] = \oint \Bigl[ \pi^{\alpha \beta} \; d g_{\alpha \beta} - H_t dt \Bigr] \; \; , \; \; \label{7eq2A} 
\end{eqnarray} 
where $H_t = H_C + v_{\gamma} \pi^{0 \gamma} = H_C + g_{0 \gamma,0} \pi^{0 \gamma}$ and $g_{0 \gamma,0} = 
v_{\gamma}$ ($\gamma = 0, 1, \ldots, d - 1$) are the corresponding velocities, while $\pi^{0 \gamma}$ are the 
temporal momenta which must be equal zero along each Hamilton trajectory of the actual motion. In other words, 
we have $d$ primary constraints $\pi^{0 \gamma} \approx 0$ in the metric gravity. Otherwise, i.e., if $\pi^{0 
\gamma} \ne 0$ for some $\gamma$, then Eq.(\ref{7eq2A}) does not hold. Another crucial fact which has been used 
to transform, Eq.(\ref{7eq2A}), follows from the formulas for canonical Hamiltonian $H_C$, Eqs.(\ref{eq5}) and 
(\ref{DirH-C}), which do not contain any of the temporal momenta $\pi^{0 \gamma}$, but it may include some of 
the $g_{0 \mu}$ and/or $g_{\mu 0}$ coordinates. If these conditions are obeyed, then from Eq.(\ref{7eq2A}) we 
can derive the following equality: 
\begin{eqnarray}  
  \oint \Bigl[ \pi^{m n} \; d g_{m n} - H_C dt \Bigr] = I = 
  \oint \Bigl[ \pi^{\alpha \beta} \; d g_{\alpha \beta} - H_t dt \Bigr] \; \; , \; \; \label{7eq3A} 
\end{eqnarray}
which means that the both these integrals, i.e., integrals on the left and right sides of this equation, 
are the true integral invariants and their numerical values equal to each other. In other words, these two 
integral invariants coincide with each other, i.e., they are not independent, and for constrained dynamical 
systems we always have to deal with this complication. 

The formula, Eq.(\ref{7eq3A}), has a number of consequences for Hamiltonian formulations of the metric gravity, 
but here we consider just one of them. First, as follows from the left-hand side of Eq.(\ref{7eq3A}) the set of 
dynamical variables $\{ g_{mn}, \pi^{pq} \}$ will be canonical, if the Poisson brackets between these dynamical 
variables coincide with CUSPB, i.e., $[ g_{mn}, g_{pq}] = 0, [ g_{mn}, \pi^{pq}] = \Delta^{pq}_{mn}, [ \pi^{mn}, 
\pi^{pq}] = 0$ for all possible $(mn)-$ and $(pq)$-pairs. In other words, it is a necessary and sufficient 
condition of canonicity in this case. However, if we apply the same arguments to the integral in right-hand side 
of Eq.(\ref{7eq3A}) we can only say that the exact coincidence of the Poisson brackets between dynamical variables 
$[ g_{\alpha\beta}, g_{\mu\nu}] = 0, [ g_{\alpha\beta}, \pi^{\mu\nu}] = \Delta^{\mu\nu}_{\alpha\beta}, [ 
\pi^{\alpha\beta}, \pi^{\mu\nu}] = 0$ with the standard CUSPB values is only necessary (but not sufficient!) 
condition of canonicity. In order to obtain the sufficient conditions of canonicity we must also guarantee that 
all temporal momenta $\pi^{0 \gamma}$ and/or $\pi^{\gamma 0}$ do not change with time $t$ along the true Hamilton 
trajectories. This means that all time-derivatives of the temporal momenta must be equal zero at all times, i.e., 
we have a number of additional equations such as $\pi^{0 \gamma} \approx 0, [ \pi^{0 \gamma}, H_t] \approx 0, [ [ 
\pi^{0 \gamma}, H_t], H_t] \approx 0$, etc. To obey all these equations for the primary, secondary and other 
constraints we have to follow Dirac's modifications made in the classical Hamilton method for constrained dynamical 
systems (see above). Otherwise, if some of these conditions do not hold, then the numerical value of the integral 
in the right-hand side of Eq.(\ref{7eq3A}) will be different from $I$, i.e., this integral is not invariant in 
this case and we have an obvious contradiction here. This explains why the criteria of canonicity derived for 
constrained dynamical systems always include two parts: (a) coincidence of the Poisson brackets between dynamical 
variables with the standard CUSPB values, and (b) conservation of the algebra of first-class constraints. This 
presumes the form-invariance of all first-class constraints and Poisson brackets of these constraints with each 
other, and with the canonical/total Hamiltonians and other essential functions of dynamical variables.    

\section{Conclusions}

Thus, we have investigated the two different Hamiltonian formulations \cite{Dir58} and \cite{K&K} of the metric 
gravity in $d-$dimensional Riemann space, where $d \ge 3$. These two Hamiltonian formulations are related to 
each other by a canonical transformation of dynamical variables in the $d(d + 1)$-dimensional phase space and 
each of them allows one to restore the complete $d-$dimensional diffeomorphism as the correct (and well known) 
gauge invariance of the free gravitational field in the metric gravity. By using the known canonical 
transformation between these two Hamiltonian formulations of the metric gravity we have investigated the basic 
properties of other similar canonical transformations and derived some useful criteria of canonicity for an 
arbitrary transformation of the Hamiltonian dynamical variables in the $d(d + 1)-$dimensional phase space. The 
results of our study are important in numerous applications, since in metric gravity canonical transformations 
of Hamiltonian dynamical variables are often used to simplify either the canonical $H_C$ and/or total 
Hamiltonian(s) $H_t$, or secondary constraints, or to reduce the canonical Hamiltonian $H_C$ to some special 
form, e.g., to its normal form, which is well known in classical mechanics.  

In general, all criteria of canonicity for transformations of dynamical variables in the metric gravity require 
the exact coincidence of the Poisson (or Laplace) brackets for the new and old dynamical (Hamiltonian) variables. 
Briefly, if the Poisson brackets of the new dynamical variables (expressed in the old dynamical variables) do not 
coincide with their canonical values, then such a transformation is not canonical. This is the universal criterion 
of canonicity which is known from classical mechanics of Hamiltonian dynamical systems. However, in all Hamiltonian 
formulations of metric gravity we always deal with the constrained dynamical systems. Therefore, all criteria of 
canonicity, which are valid for such systems, must contain the second part which deals with the algebra of 
first-class constraints, form-invariance of the canonical/total Hamiltonian(s) and/or form-invariance of the 
Hamilton equations. For instance, the true canonical transformation in the metric gravity must keep 
form-invariance of the Hamilton equations derived in the Dirac's modification of the classical Hamilton method. 
It can be illustrated by the transformation of Eqs.(\ref{metrgr1}) - (\ref{metrgr3}) into Eqs.(\ref{ametrgr}) - 
(\ref{bmetrgr}) during our canonical transformation of the Hamiltonian dynamical variables. This is the first 
criterion of canonicity in the metric which is relatively simple and ready to be applied to actual problems. The 
second similar criterion \cite{FK&K} requires the exact coincidence of the total Hamiltonian $H_t$ and preservation 
of the algebra of constraints for both (old and new) Hamiltonian formulations of the metric gravity. 
 
We have also reconsidered modifications made by Dirac \cite{Dir58}, \cite{Dir50}, \cite{Dir64} in the classical 
Hamilton approach. It is shown that these modifications are crucial to improve overall efficiency of the new 
Hamiltonian method for dynamical systems with constraints, including various physical fields with additional 
gauge conditions, or gauges, for short. The main advantage of the new Dirac's approach is a possibility to write 
all governing equations in the united, manifestly Hamilton form (see, Eqs.(\ref{metrgr1}) - (\ref{metrgr3}) and 
Eqs.(\ref{ametrgr}) - (\ref{bmetrgr}) above). The original Dirac's idea that all motions in Hamiltonian dynamical 
systems with first-class constraints can always be separated into actual motions and special motions along 
constraints (or gauge-consistent motions) was extremely productive. Now, by using this Dirac's modification of 
the classical Hamilton method we can describe time-evolution of a large number of actual and model fields. 
Furthermore, we can make a conjecture that the free fields which represents all currently known fundamental 
interactions can unambiguously be described by the this version of Hamiltonian method, which was originally 
developed and later modified by Dirac. 

In this study we have also considered the method of integral invariants and applied it to investigate canonical 
transformations between different Hamiltonian formulations of the metric gravity. This method was originally 
proposed and developed by Poincare and Cartan \cite{Cartan}. Since then it was transformed into a very powerful 
approach, which currently is an absolute tool in the Hamilton classical mechanics. In reality, the invariance 
of the Poincare-Cartan integral can be chosen as a foundation of the whole Hamiltonian mechanics. Indeed, if 
this integral is invariant for some dynamical system, then such a system is Hamiltonian and its time-evolution 
is described by a system of Hamilton equations. For Hamiltonian dynamical systems with constraints the general 
theory of integral invariants must be modified, but its overall power still remains outstanding.   

Unfortunately, the limited space of this article did not allow us to discuss other important directions of the 
Hamiltonian formulations of metric gravity. In particular, we could not consider the explicit derivation of the 
gauge generators which are defined by chain of the first-class constraints \cite{K&K}, \cite{Cast} (see also 
\cite{Fro2021}). Also, in this study, we didn't even mention various non-canonical quasi-Hamiltonian formulations 
of the metric gravity. However, we can make a reference to an excellent review article \cite{K&K2011} which 
contains a detailed analysis of this problem and a large number of references to papers published up to the 
beginning of 2011. Here we want to note that any of these non-canonical Hamiltonian formulations uses a set of 
dynamical ADM-variables, which were introduced in \cite{ADM}. This fact has been noticed and criticized by 
Bergmann, Dirac and many others. Our calculations of the corresponding Poisson brackets can be found in 
\cite{FK&K} and \cite{Fro2021}. However, since early 1960's there were no explanation of this remarkable fact 
and its consequences neither from ADM people, nor from their followers (see, e.g., \cite{Regg} - \cite{Wald2}). 
Then, in 1985 it was suddenly detected  that ADM formulation of the Hamiltonian metric gravity cannot restore, 
in principle, the total four-dimensional diffeomorphism \cite{IshKuch} which is the correct and well known gauge 
symmetry of a free gravitational field in four-dimensional space-time. Recently, we have found another crucial 
problem for ADM gravity and similar non-canonical `Hamiltonian' formulations. Indeed, in Dirac's Hamiltonian 
formulation, we could restore the original Einstein's equations for a free gravitational field. Analogous ADM 
Hamiltonian formulation uses, in part, the same dynamical variables (12 of 20 variables), but there are some 
fundamental mistakes in all secondary constraints. Therefore, the extra terms which present in the restored 
field equations for ADM formulation do not cancel each other (as they do in Dirac's formulation), but remain 
and even multiply. Finally, in Dirac's Hamiltonian formulation we obtain the maternal Einstein equations with no 
additional terms, while for ADM Hamiltonian formulation we have similar equations with many extra terms in them. 
As follows from this fact the ADM Hamiltonian formulation either describes some different (i.e., non-Einstein's) 
field, or it is an absolutely wrong theoretical construction which does not represent and real and/or model 
field (if the arising system of extended Einstein-like equations is not closed). In the future, under better 
circumstances, we plan to discuss these (and other) issues which currently exist in the Hamiltonian formulations 
of the metric gravity.  
 
I am grateful to my friends N. Kiriushcheva, S.V. Kuzmin and D.G.C. (Gerry) McKeon (all from the University of 
Western Ontario, London, Ontario, Canada) for helpful discussions and inspiration. 

\appendix
\section{Some useful relations in the Hamiltonian version of metric gravity}
\label{A} 

In this `pure technical' Appendix we derive a few equations and relations which are crucially important for 
our Hamiltonian formulation of the metric gravity. First, let us define the space-like tensor $I_{mnpq}$ by 
the equation $I_{mnpq} = a g_{mn} g_{pq} - g_{mp} g_{nq}$, Here we want to show that $I_{mnpq} E^{pqkl} = 
\delta^{k}_{m} \delta^{l}_{n}$ in the case when $a = \frac{1}{d - 2}$. First of all, we note that 
$g_{\alpha\beta} g^{\alpha\beta} = d$, where $d$ is the total dimension of our Riemann space. From here one 
finds
\begin{eqnarray}
  g_{\alpha\beta} e^{\alpha\beta} = g_{\alpha\beta} g^{\alpha\beta} - g_{\alpha\beta} \frac{g^{0\alpha} 
  g^{0\beta}}{g^{00}} = d - g^{0}_{\alpha} \frac{g^{0\alpha}}{g^{00}} = d - 1 \; \; . \; \; \label{Ap1eq01}  
\end{eqnarray}
On the other hand, since $e^{0 \beta} = 0, e^{\alpha 0} = 0$ and $e^{0 0} = 0$, we can write the last 
equality in a different form 
\begin{eqnarray}
 d - 1 = g_{\alpha\beta} e^{\alpha\beta} = g_{pq} e^{pq} + g_{p 0} e^{p 0} +  g_{0 q} e^{0 q} + g_{0 0} 
 e^{0 0} = g_{pq} e^{pq} \; \; . \; \; \label{Ap1eq02}
\end{eqnarray}
In other words, we obtain $g_{pq} e^{pq} = d - 1 =  g_{\alpha\beta} e^{\alpha\beta}$. Furthermore, one can  
derive a similar rule to lower the index in the $e^{\alpha\beta}$ tensor: $g_{\alpha\beta} e^{\beta\gamma} = 
e^{\gamma}_{\alpha} = g_{p q} e^{q m} = e^{m}_{p} = g^{m}_{p} = \delta^{m}_{p}$. 

Now, we can prove the statement formulated above. The formula for the $I_{mnpq} E^{pqkl}$ product takes the form
\begin{eqnarray}
  I_{mnpq} E^{pqkl} &=& ( a g_{mn} g_{pq} - g_{mp} g_{nq} ) \bigl( e^{pq} e^{kl} - e^{pk} e^{ql} \Bigr) = 
  a (d - 1) g_{mn} e^{kl} - g_{mp} \delta^{p}_{n} e^{kl} \nonumber \\ 
  &-& a g_{mn} e^{pk} \delta^{l}_{p} + \delta^{k}_{m} \delta^{l}_{n} = [ a (d - 1) - 1 - a ] g_{mn} e^{kl} + 
 \delta^{k}_{m} \delta^{l}_{n} \; \; . \; \; \label{Ap1eq03}
\end{eqnarray}
From here one finds that if $a (d - 2) = 1$ (or $a = \frac{1}{d - 2}$), then the first term in the last equation 
equals zero identically and $I_{mnpq} E^{pqkl} =  \delta^{k}_{m} \delta^{l}_{n}$. 

Another important relation which we want to prove here is the connection between the $[ g_{\alpha\beta}, 
\pi^{\mu\nu} ]$ and $[ g^{\alpha\beta}, \pi^{\mu\nu} ]$ Poisson brackets. Formally, the covariant components of 
the metric tensor $g_{\alpha\beta}$ are the only generalized coordinates in the metric gravity. However, there 
is an obvious relation between covariant and contravariant components of the metric tensor: $g_{\alpha\beta} 
g^{\beta\gamma} = g_{\alpha}^{\gamma} = \delta^{\gamma}_{\alpha}$. From here one finds the following relation 
between the two Poisson brackets
\begin{eqnarray}
  [ \pi^{\mu\nu}, g_{\alpha\beta} ] g^{\beta\gamma} + g_{\alpha\beta} [ \pi^{\mu\nu}, g^{\beta\gamma} ] = 0 \; 
  \; {\rm or} \; \; [ g_{\alpha\beta}, \pi^{\mu\nu} ] g^{\beta\gamma} = - g_{\alpha\beta} [ g^{\beta\gamma}, 
  \pi^{\mu\nu} ] \; \; . \; \; \label{Ap1eq1}
\end{eqnarray}
By multiplying the both sides of this equation by the tensor $g^{\alpha\sigma}$ we obtain the following relation 
\begin{eqnarray}
 g^{\alpha\sigma} [ g_{\alpha\beta}, \pi^{\mu\nu} ] g^{\beta\gamma} = - g_{\alpha\beta} g^{\alpha\sigma} 
 [ g^{\beta\gamma}, \pi^{\mu\nu} ] = - g_{\beta}^{\sigma} [ g^{\beta\gamma}, \pi^{\mu\nu} ] =  
 - [ g^{\gamma\sigma}, \pi^{\mu\nu} ] \; \; , \; \; \label{Ap1eq2}
\end{eqnarray} 
or 
\begin{eqnarray} 
  [ g^{\gamma\sigma}, \pi^{\mu\nu} ] = - g^{\alpha\sigma} \; \Delta_{\alpha\beta}^{\mu\nu} \; g^{\beta\gamma} = 
  - \frac12 \Bigl( g^{\mu\sigma} g^{\nu\gamma} + g^{\mu\gamma} g^{\nu\sigma} \Bigr) = - [ \pi^{\mu\nu}, 
  g^{\gamma\sigma} ] \; \; , \; \; \label{Ap1eq3}  
\end{eqnarray} 
i.e., the result which exactly coincides with the formulas, Eqs.(\ref{eq151}) and (\ref{PBOa}), from the main text. 
The derivation of two other formulas from Eqs.(\ref{PBO}) is absolutely analogous. 

The last formula, which we want to derive in this Appendix, is the Poisson brackets between two secondary, 
fist-class constraints of the metric gravity, i.e., $[ \chi^{0\sigma}, \chi^{0\gamma} ]$. This formula is needed 
to complete our Hamiltonian formulation of the metric gravity. Furthermore, it is of great interest in a number 
of gravitational problems. This formula has never been produced in earlier studies, since its direct derivation 
is not an easy task. Below, we apply a different approach \cite{Fro2021} which is based on the Jacobi identity:
\begin{eqnarray} 
  &[&\chi^{0\sigma}, \chi^{0\gamma} ] = [ \chi^{0\sigma}, [ \phi^{0\gamma}, H_C ] ] = 
  - [ \phi^{0\gamma}, [ H_C, \chi^{0\sigma} ] ] - [ H_C, [ \chi^{0\sigma}, \phi^{0\gamma} ] ] \nonumber \\
  &=& [ \phi^{0\gamma}, [ \chi^{0\sigma}, H_C ] ] + [ H_C, [ \phi^{0\gamma}, \chi^{0\sigma} ] ] =
  [ \phi^{0\gamma}, D^{\sigma}_C ] - \frac12 [ g^{\gamma\sigma}, H_C ] \; \; \; , \; \; \label{chi-chi}
\end{eqnarray} 
where $D^{\sigma}_C$ is the $\sigma-$component of the Dirac's closure, Eq.(\ref{DirclD}), while $H_C$ is the 
canonical Hamiltonian of the metric gravity, Eq.(\ref{DirH-C}). Here we apply the expressions derived in the 
Dirac's Hamiltonian formulation of metric gravity. 

Analytical calculations of the both terms in the last equation from Eq.(\ref{chi-chi}) are relatively easy, since 
only a few Poisson brackets really contribute. Here we just present the final result for the second term in 
Eq.(\ref{chi-chi}): 
\begin{eqnarray} 
 - \frac12 [ g^{\gamma\sigma}, H_C ] &=& \frac{1}{4 \sqrt{- g} g^{0 0}} I_{mnpq} \Bigl[\Bigl( g^{\gamma m} 
 g^{\sigma n} + g^{\gamma n} g^{\sigma m} \Bigr) \phi^{p q} + \phi^{m n} \Bigl( g^{\gamma p} g^{\sigma q} + 
 g^{\gamma q} g^{\sigma p} \Bigr)\Bigr] \nonumber \\
 &-& \frac{1}{4 g^{0 0}} \Bigl( g^{\gamma m} g^{\sigma n} + g^{\gamma n} g^{\sigma m} \Bigr) \Bigl( g^{0 k} 
 g_{m n,k} - 2 g^{0 \alpha} g_{\alpha n,m} \Bigr) \; \; . \; \label{1st-term}
\end{eqnarray} 
This expression can be simplified, e.g., by using the identities such as $g_{m n} g^{m \gamma} = g_{\alpha n} 
g^{\alpha \gamma} - g_{0 n} g^{0 \gamma} = g^{\gamma}_{n} - g_{0 n} g^{0 \gamma}$, but here we do not want to
make similar simplifications. 

In order to determine the first Poisson bracket in Eq.(\ref{chi-chi}) we note that the $\sigma$-component of 
the Dirac closure can be written in the compact form of quasi-linear combination of secondary first-class 
constraints $D^{\sigma}_C = V^{\sigma}_{\lambda} \chi^{0 \lambda} = V^{\sigma}_{k} \chi^{0 k} + V^{\sigma}_{0} 
\chi^{0 0}$, where $V^{\sigma}_{\lambda} = V^{\sigma}_{\lambda}( g_{p q}, \phi^{m n}, g^{0 \mu})$ is the 
structure functions (or functional) \cite{FK&K}. Therefore, for the Poisson bracket in Eq.(\ref{1st-term}) 
we can write $[ \phi^{0 \gamma}, D^{\sigma}_C] = [ \phi^{0 \gamma}, V^{\sigma}_{\lambda}] \; \; \chi^{0 
\lambda} + V^{\sigma}_{\lambda} \; \; [ \phi^{0 \gamma}, \chi^{0 \lambda} ]$, where $\lambda = (0, k)$. In 
these notations the explicit formula for the second term is written in the form: 
\begin{eqnarray} 
  V^{\sigma}_{\lambda} [ \phi^{0 \gamma}, \chi^{0 \lambda} ] &=& - \Bigl[ \frac{g^{\sigma m}}{\sqrt{- g} 
  g^{0 0}} I_{pqmk} \phi^{p q} + \frac12 g^{\sigma 0} g_{0 0,k} + g^{\sigma p} g_{0 p,k} + \frac{g^{\sigma p} 
  g^{0 q}}{2 g^{0 0}} \Bigl( g_{p q,k} + g_{q k,p} \nonumber \\
   &-& g_{p k,q} \Bigr)\Bigr] g^{\gamma k} \chi^{0 0} - \frac12 g^{\sigma}_{0} (g^{\gamma k} \chi^{0 0})_{,k} 
   + \frac12  g^{0 \gamma} g^{\sigma k} g_{0 0,k} \chi^{0 0} \; \; , \; \label{2nd-term} 
\end{eqnarray} 
where $g^{\alpha}_{\beta} = \delta^{\alpha}_{\beta}$ is the substitution tensor. Analogous formula for the 
first term is derived either by applying the following expression $[ \phi^{0 \gamma}, V^{\sigma}_{\lambda}] 
= - \frac{\partial V^{\sigma}_{\lambda}}{\partial g_{0 \gamma}}$, or directly. The final expression takes 
the form
\begin{eqnarray} 
 [ \phi^{0 \gamma}, V^{\sigma}_{\lambda} ] \chi^{0 \lambda} &=& \Bigl[ \frac{1}{\sqrt{- g} g^{0 0}} 
 \Bigl( g^{0 \sigma} g^{\gamma m} +  g^{\gamma \sigma} g^{0 m} - 3 g^{0 \gamma} g^{\sigma m} \Bigr) 
 I_{pqmk} \phi^{p q} \Bigr] \chi^{0 k} \nonumber \\
&-& \delta^{\gamma}_{0} 
 \Bigl(g^{0 \sigma} \chi^{0 k}\Bigr)_{,k} 
 - \delta^{\gamma}_{p} \Bigl(g^{\sigma p} \chi^{0 k}\Bigr)_{,k} 
 + \frac12 \delta^{\gamma}_{0} \Bigl(g^{\sigma k} \chi^{0 k}\Bigr)_{,k} \; \; \; \label{3rd-term}
\end{eqnarray} 
The complete formula for the $[ \chi^{0\sigma}, \chi^{0\gamma} ]$ Poisson bracket is the algebraic sum of three 
expressions from Eqs.(\ref{1st-term}) - (\ref{3rd-term}).  

\section{On canonicity of the Hamiltonian dynamical variables}
\label{B}

Let us briefly discuss the criteria of canonicity for the different sets of dynamical variables which describe 
the same Hamiltonian system. In almost all applications of similar criteria in classical mechanics it is 
important to know that the new set of dynamical variables will be canonical, if the old set of dynamical 
variables was canonical. For simplicity, below the new dynamical variables $\tilde{q}_{i}$ and $\tilde{p}_{j}$ 
are designated by the upper signs $\; \tilde{} \;$, while old dynamical variables are denoted as $q_{i}$ and 
$p_{j}$. Now, we can define the following $2 n \times 2 n$ Jacobi matrix $\hat{M}$ of some transformation 
of the dynamical variables: 
\begin{eqnarray} 
    \hat{M} = \left( \begin{array}{cccccc}
 \frac{\partial \tilde{q}_1}{\partial q_1} & \ldots & \frac{\partial \tilde{q}_1}{\partial q_n} & 
 \frac{\partial \tilde{q}_1}{\partial p_1} & \ldots & \frac{\partial \tilde{q}_1}{\partial p_n} \\
  \ldots & \ldots & \ldots & \ldots & \ldots & \ldots \\ 
 \frac{\partial \tilde{q}_n}{\partial q_n} & \ldots & \frac{\partial \tilde{q}_n}{\partial q_n} & 
 \frac{\partial \tilde{q}_n}{\partial p_1} & \ldots & \frac{\partial \tilde{q}_n}{\partial p_n} \\
 \frac{\partial \tilde{p}_1}{\partial q_1} & \ldots & \frac{\partial \tilde{p}_1}{\partial q_n} & 
 \frac{\partial \tilde{p}_1}{\partial p_1} & \ldots & \frac{\partial \tilde{p}_1}{\partial p_n} \\
  \ldots & \ldots & \ldots & \ldots & \ldots & \ldots \\ 
 \frac{\partial \tilde{p}_n}{\partial q_n} & \ldots & \frac{\partial \tilde{p}_n}{\partial q_n} & 
 \frac{\partial \tilde{p}_n}{\partial p_1} & \ldots & \frac{\partial \tilde{p}_n}{\partial p_n} \\
 \end{array} \right) 
 \; \; = \; \; \left( \begin{array}{cc} 
 \frac{\partial \tilde{{\bf q}}}{\partial {\bf q}} & \frac{\partial \tilde{{\bf q}}}{\partial {\bf p}} \\
 \frac{\partial \tilde{{\bf p}}}{\partial {\bf q}} & \frac{\partial \tilde{{\bf p}}}{\partial {\bf p}} \\
  \end{array} \right) \; \; . \; \, \label{Jacobi}
\end{eqnarray}
where $\frac{\partial \tilde{{\bf q}}}{\partial {\bf q}}, \frac{\partial \tilde{{\bf q}}}{\partial {\bf p}}, 
\frac{\partial \tilde{{\bf p}}}{\partial {\bf q}}$ and $\frac{\partial \tilde{{\bf p}}}{\partial {\bf p}}$ 
are the $n \times n$ matrices. Another matrix which we need here is the unit simplectic $2 n \times 2 n$ 
matrix $\hat{J}$:     
\begin{eqnarray}
  \hat{J} = \left( \begin{array}{cccccc}
  0 & \ldots & 0 & -1 & \ldots & 0 \\
  \ldots & \ldots & \ldots & \ldots & \ldots & \ldots \\ 
  0 & \ldots & 0 & 0 & \ldots & -1 \\
  1 & \ldots & 0 & 0 & \ldots & 0 \\
  \ldots & \ldots & \ldots & \ldots & \ldots & \ldots \\ 
  0 & \ldots & 1 & 0 & \ldots & 0 \\
\end{array} \right) 
 \; \; = \; \; \left( \begin{array}{cc} 
  0 & -E \\
  E &  0 \\
  \end{array} \right) \; \; . \; \, \label{B2}
\end{eqnarray}
where $E$ is the $n \times n$ unit matrix. It easy to show that the matrix $\hat{J}$ is invertable and it 
obeys the following equation: $\hat{J}^{2} = - \hat{J} \hat{J}^{-1}$, or $\hat{J}^{-1} = - \hat{J}$. 
This fundamental property of the unit simplectic matrix $\hat{J}$ substantially determines many known 
properties of the canonical transformations and predicts a number of necessary steps in the Hamilton method. 
As follows from Eq.(\ref{B2}) the matrix $\hat{J}$ is not self-adjoint, but its product with the imaginary 
unit $\imath$, i.e., the matrix $\imath \hat{J}$ is a truly self-adjoint matrix. This fact is crucial for 
correct definition of the Poisson brackets in quantum mechanics (see below). 

It can be shown that for the matrix $\hat{M}$, which represents some canonical transformation of the 
Hamilton dynamical variables, the following condition is always obeyed: $\hat{M}^{\prime} \hat{J} \hat{M} 
= c \hat{J}$, where $c (\ne 0)$ is a real and/or complex number which is called the valence of this 
transformation. Also in this equation $\hat{J}$ is the unit simplectic matrix and $\hat{M}^{\prime}$ is 
the matrix adjoint to the original matrix $\hat{M}$. Vice versa, if some matrix of differential 
transformation of variables $\hat{M}$ obeys the equation $\hat{M}^{\prime} \hat{J} \hat{M} = c \hat{J}$, 
then such a differential transformation of dynamical variables, which is represented by the matrix 
$\hat{M}$, is canonical with the valence $c$. Thus, we have formulated the criterion of canonicity in 
terms of the Jacobi matrix which is easily determined for an arbitrary transformation of the Hamiltonian 
dynamical variables. Let us discuss this matrix criterion with some additional details. First of all, by 
using the properties of the $\hat{J}$ matrix mentioned above it is possible to show (in two steps) that 
the two following equations (or conditions) follow from each other   
\begin{eqnarray}
  \hat{M}^{\prime} \hat{J} \hat{M} = c \hat{J} \; \; \Longleftrightarrow \; \; \hat{M} \hat{J} 
  \hat{M}^{\prime} = c \hat{J} \; \; . \; \label{simplec1}
\end{eqnarray}
which means that these two equations are equivalent to each other. Indeed, at the first step we multiply 
the both sides of the equation $\hat{M}^{\prime} \hat{J} \hat{M} = c \hat{J}$ by the $[\hat{M}^{\prime}]^{-1}$ 
matrix from the left and by the $[\hat{M}]^{-1}$ matrix from the right. These operations lead to the new 
equation $\frac{1}{c} \hat{J} = [\hat{M}^{\prime}]^{-1} \hat{J} [\hat{M}]^{-1}$. At the second step we just 
need to reverse the both sides of this equation. By taking into account that $\hat{J}^{-1} = - \hat{J}, 
[\hat{M}^{-1}]^{-1} = \hat{M}$ and $\{ [\hat{M}^{\prime}]^{-1} \}^{-1} = \hat{M}^{\prime}$, we find the 
equality: $c \hat{J} = \hat{M} \hat{J} \hat{M}^{\prime}$, which exactly coincides with the second equation in 
Eqs.(\ref{simplec1}). On the other hand it is easy to check that, if we start from the second equation in 
Eqs.(\ref{simplec1}), then the two analogous and simple steps allow one to derive the first equation in 
Eqs.(\ref{simplec1}). 

The matrix $\hat{M}$ (and/or $\hat{M}^{\prime}$) which obeys any of these equations, Eqs.(\ref{simplec1}), is 
called the true simplectic matrix with numerical valence $c$. In general, all even-dimensional (non-singular) 
simplectic matrices form the closed simplectic group $Sp(k,R)$ (or $Sp(k,C)$), where $k = 2 n$. Finally, we 
arrive to the following theorem: some non-singular transformation of Hamilton dynamical variables $\tilde{q}_{i} 
= \tilde{q}_{i}(t, q_{i}, p_{i}), \tilde{p}_{i} = \tilde{p}_{i}(t, q_{i}, p_{i})$, where $i = 1, \ldots, n$, will 
be canonical, if (and only if) its Jacobi matrix $\hat{M}$, Eq.(\ref{Jacobi}), is the true symplectic matrix with 
the valence $c$. In this case the condition,  Eq.(\ref{simplec1}), must be obeyed identically for all old 
dynamical variables and time $t$. 

Further investigation shows that the matrix elements of the $\hat{M}^{\prime} \hat{J} \hat{M}$ matrix 
coincide with the corresponding Lagrange brackets written in old coordinates and momenta, while the 
matrix elements of the $\hat{M} \hat{J} \hat{M}^{\prime}$ matrix coincide with the corresponding Poisson 
brackets which are also written in old coordinates and momenta. Thus, we have an obvious duality between 
the Lagrange and Poisson brackets, which can be illustrated by a simple rule which is applied to form 
the $(ij)-$matrix elements of the adjoint matrix $\hat{M}^{\prime}$ (from $\hat{M}$) and vice versa. 
This rule is simple: in each partial derivative, which is included in the Jacobi matrix of the canonical 
transformation (or its adjoint), the letters and indices at the top and bottom are swapped, while the 
symbol $\; \tilde{} \;$ always stays at the top. Let us consider the following example. As mentioned 
above the matrix equation $\hat{M}^{\prime} \hat{J} \hat{M} = c \hat{J}$ for the canonical transformation 
is equivalent to the following system of equations:
\begin{eqnarray}
 \{ q_{i}, q_{j} \} = 0 \; \; , \; \;  \{ p_{i}, p_{j} \} = 0 \; \; , \; \; \{ q_{i}, p_{j} \} = c 
 \delta_{ij} \; \; \; , \; \label{LaplBra1}
\end{eqnarray}
where the notation $\{a, b \}$ means the Laplace bracket defined in the main text. In other words, the 
matrix elements of the $\hat{M}^{\prime} \hat{J} \hat{M}$ matrix always coincide with the corresponding 
Laplace brackets, while numerical values of these brackets are determined from the matrix equation 
$\hat{M}^{\prime} \hat{J} \hat{M} = c \hat{J}$. Now, by taking adjoint of Eq.(\ref{LaplBra1}) one finds 
the following matrix equation: $\hat{M} \hat{J} \hat{M}^{\prime} = c \hat{J}$, which leads to the three 
group of equalities for the adjoint Laplace brackets:   
\begin{eqnarray}
  \{ q_{i}, q_{j} \}^{\star} = 0 \; \; , \; \;  \{ p_{i}, p_{j} \}^{\star} = 0 \; \; , \; \; \{ q_{i}, 
  p_{j} \}^{\star} = c \delta_{ij} \; \; \; , \; \label{adjnt3}
\end{eqnarray}  
where the sign `$\star$' means that inside of Lagrange brackets we have to apply our `swap of variables' 
described above. By using the explicit formulas for the Lagrange brackets and our recipes to construct 
the adjoint matrix we can write  
\begin{eqnarray}
  \{ q_{i}, q_{j} \}^{\star} = \Bigl[ \sum^{n}_{k=1} \Bigl( \frac{\partial \tilde{q}_{k}}{\partial q_{i}} 
  \frac{\partial \tilde{p}_{k}}{\partial q_{j}} - \frac{\partial \tilde{q}_{k}}{\partial q_{j}} 
  \frac{\partial \tilde{p}_{k}}{\partial q_{i}} \Bigl)\Bigr]^{\star} = 
  \sum^{n}_{k=1} \Bigl( \frac{\partial \tilde{q}_{i}}{\partial q_{k}} 
  \frac{\partial \tilde{q}_{j}}{\partial p_{k}} - \frac{\partial \tilde{q}_{j}}{\partial q_{k}} 
  \frac{\partial \tilde{q}_{i}}{\partial p_{k}} \Bigl) = [ \tilde{q}_{i}, \tilde{q}_{j}]  
\end{eqnarray}  
where the notation $[ a, b ]$ stands for the `regular' Poisson bracket. Analogous expressions can be 
derived for other fundamental Lagrange brackets $\{ q_i, p_j \}^{\star} = [ \tilde{q}_i, \tilde{p}_j 
]$ and $\{ p_i, p_j \}^{\star} = [ \tilde{p}_i, \tilde{p}_j ]$, where all Poisson brackets are 
calculated in the old dynamical variables. These formulas indicate clear that the adjoint of the 
Laplace bracket equals (remarkably and unambiguously) to the corresponding Poisson bracket. Briefly, 
this means that we have reduced calculations of the adjoints of the Laplace brackets to computations 
of the corresponding Poisson brackets. Note also that the right-hand sides of Eqs.(\ref{adjnt3}) do 
not change during the $\star$ procedure and we obtain the following numerical values for the 
fundamental Poisson brackets 
\begin{eqnarray}
 [ \tilde{q}_{i}, \tilde{q}_{j} ] = 0 \; \; , \; \;  [ \tilde{p}_{i}, \tilde{p}_{j} ] = 0 \; \; , 
 \; \; [ \tilde{q}_{i}, \tilde{p}_{j} ] = c \delta_{ij} \; \; \; , \; \label{adjnt32}
\end{eqnarray}   
which coincide with the expected numerical values. Thus, we have shown that the Poisson brackets coincide 
with the adjoints of the corresponding Laplace brackets (and vice versa). In other words, these two 
systems of brackets are closely related to each other and each of these brackets can equally be used to 
check and prove the canonicity of some new set of Hamilton dynamical variables. Based on these facts it is 
relatively easy (for simplicity, we choose $c = 1$ here) to prove the following theorem \cite{Gold}: if 
$u_1, u_2, \ldots, u_{2 n}$ are the $2 n$ independent functions of the variables $q_1, \ldots, q_n, p_1, 
\ldots , p_n$, then the two equations  
\begin{eqnarray}
 \sum^{2 n}_{k=1} \{ u_{k}, u_{i} \} [ u_{k}, u_{j} ] = \sum^{2 n}_{k=1} \{ u_{i}, u_{k} \} [ u_{j}, 
 u_{k} ] = \delta_{ij} \; \; , \; \; \label{adjnt33}
\end{eqnarray}   
are always obeyed for these functions. This equation(s) explicitly shows a very close relation between 
the Laplace and Poisson brackets and they essentially follow from the definitions of these two brackets.

The last remark, which we want to make here, describes the main difference which arises in definitions of 
the canonical transformations in the classical and quantum mechanics. The Jacobi matrix $\hat{M}$, which 
describes the canonical transformation of the Hamiltonian dynamical variables, can also be defined (with 
a few additional tricks) in quantum mechanics. However, the governing equation Eq.(\ref{Jacobi}) for the 
Jacobi $\hat{M}$ matrix in quantum mechanics includes the self-adjoint unit matrix $\imath \hat{J}$, which 
is not the unit simplectic $\hat{J}$ matrix as it was in classical mechanics. This means that in quantum 
mechanics the both newly-defined Poisson brackets and their numerical values will always include 
(explicitly) the imaginary unit $\imath$. In the fundamental Poisson bracket we can introduce the `new' 
momenta in the coordinate representation by including $\imath$ in its definition. On the other hand, the 
numerical value of the corresponding Poisson bracket must also include the imaginary unit $\imath$. Let us 
consider the following example. Suppose we have a point non-relativistic particle with the mass $m$ which 
is located at the point with the Cartesian coordinates $(x, y, z)$ and has the velocity $(v_x, v_y, v_z)$. 
One can introduce the momenta for this particle $p_i = m v_i$, where $i = (x, y, z)$. In classical mechanics 
we have three fundamental Poisson brackets: $[ x_{i}, p_{j}] = \delta_{ij}, [ x_i, x_j ] = 0$ and $[ p_i, 
p_j ] = 0$. The same momenta and numerical values of all non-zero Poisson brackets in quantum mechanics are 
defined as follows: $p_{k} = -\imath \hbar \frac{\partial}{\partial x_{k}}$ and $[ p_{i}, x_{j} ] = 
-\imath \hbar \delta_{ij}$, respectively \cite{DirQM}. Here $(i,j) = (x, y, z)$ and $\hbar = \frac{h}{2 
\pi} \approx 1.054571817 \cdot 10^{-34}$ $J \cdot sec$ is the reduced Planck constant which is also called 
the Dirac constant. An additional trick in this case is the explicit form of the coordinate operator 
$x_{k}$ in momentum representation: $x_{k} = -\imath \hbar \frac{\partial}{\partial p_{k}}$, or 
$\frac{\partial}{\partial p_{k}} = \frac{\imath}{\hbar} x_{k}$, where $k = x, y, z$. 

Note that the both quantum momenta and numerical values of the Poisson brackets contain the imaginary 
unit $\imath$. Actual presence of the self-adjoint matrix $\imath \hat{J}$ in the governing equation 
$\hat{M}^{\prime} (\imath \hat{J}) \hat{M} = \imath \hat{J}$ for the quantum Jacobi matrix $\hat{M}$ 
allows one to define the canonical transformations in quantum mechanics as the unitary transformations. 
In other words, canonical transformations of the Hamiltonian dynamical variables in quantum mechanics 
are represented by the unitary matrices $\hat{M}$ and $\hat{M}^{\prime} $ only. Thus, we have found 
the following relation between the classical and quantum Poisson brackets $[ f, g ]_{class} = 
\frac{\imath}{\hbar} [ \hat{f}, \hat{g} ]_{quant}$. This relation can be considered as a canonical 
transformation between the classical and quantum mechanics with the imaginary valence $c = 
\frac{\imath}{\hbar}$. The imaginary valence of this canonical transformation means that all dynamical 
variables in quantum mechanics must be represented by self-adjoint operators and canonical 
transformations between different sets of dynamical variables in quantum mechanics are always performed 
by the unitary matrices, which differ from the simplectic matrices used for the same purposes in 
classical mechanics. 

\section{Variational derivation of integral invariants for Hamiltonian systems}
\label{C}

As is well known, the Hamilton method has a number of significant advantages over other methods which are 
used to solve the same problems in classical mechanics, e.g., over the Lagrange method. One of these 
advantages is the method of integral invariants which is a `hidden' part of any true Hamilton approach. 
The method of integral invariants allows one to analyze and solve many problems in mechanics. Here we 
describe a variational derivation of integral invariants for arbitrary, in principle, Hamiltonian dynamical 
systems. First, let us consider variations of the two following actions (or action integrals) $W_L$ and 
$W_H$:  
\begin{eqnarray}  
  W_L = \int_{t_0}^{t_1} L\Bigl(t, q_{i}(t,\alpha), \dot{q}_{i}(t,\alpha)\Bigr) dt \; \; {\rm and} \; \; 
  W_H = \int_{t_0(\alpha)}^{t_1(\alpha)} L\Bigl(t, q_{i}(t,\alpha), \dot{q}_{i}(t,\alpha)\Bigr) dt \; , \; 
  \label{twoA} 
\end{eqnarray}  
where the notation $\alpha$ stands for the parameter. All coordinates $q_{i}(t,\alpha)$ and velocities 
$\dot{q}_{i}(t,\alpha)$ in these two actions depend on this parameter. In the second action ($W_H$) the lower 
and upper limits in the time-integral also depend upon this parameter $\alpha$. During variations of these 
two integrals we can always interchange the sings of variations $\delta$ and time derivative $\frac{d}{d t}$, 
since we can write 
\begin{eqnarray}  
  \delta \dot{q}_{i} = \delta \Bigl( \frac{d}{d t} q_{i}(t,\alpha) \Bigr) = \frac{\partial}{\partial \alpha} 
  \Bigl[ \frac{d}{d t} q_{i}(t,\alpha) \Bigr] \delta \alpha =  \frac{d}{d t} \Bigl[ \frac{\partial}{\partial 
  \alpha} q_{i}(t,\alpha) \delta \alpha \Bigr] = \frac{d}{d t} \delta q_{i}(t,\alpha) \; \label{interchange} 
\end{eqnarray}  
By using this equation we can derive the following formulas for variations of these two actions $W_L$ and 
$W_H$: 
\begin{eqnarray}  
 \delta W_L = \int_{t_0}^{t_1} \sum^{n}_{i=1} \Bigl[ \frac{\partial L}{\partial q_{i}} - \frac{d}{d t} 
 \Bigl(\frac{\partial L}{\partial \dot{q}_{i}} \Bigr)\Bigr] \delta q_i dt \; \; 
\end{eqnarray} 
and 
\begin{eqnarray}  
 \delta W_H = \Bigl\{ \sum^{n}_{i=1} \Bigl(\frac{\partial L}{\partial \dot{q}_{i}}\Bigr) \delta q_i - 
 \Bigl[ \sum^{n}_{i=1} \Bigl(\frac{\partial L}{\partial \dot{q}_{i}}\Bigr) \dot{q}_{i} - L \Bigr] 
 \delta t \Bigr\}\Bigg|^{t_1(\alpha)}_{t_{0}(\alpha)} + \int_{t_0(\alpha)}^{t_1(\alpha)} \sum^{n}_{i=1} 
 \Bigl[\frac{\partial L}{\partial q_{i}} - \frac{d}{d t} \Bigl(\frac{\partial L}{\partial \dot{q}_{i}} 
 \Bigr)\Bigr] \delta q_i dt \; . \; \label{deltH}
\end{eqnarray} 
Now, by introducing the following notations: $\frac{\partial L}{\partial \dot{q}_{i}} = p_{i}$ and 
$\sum^{n}_{i=1} \Bigl(\frac{\partial L}{\partial \dot{q}_{i}}\Bigr) \dot{q}_{i} - L = H$ we can re-write the 
last expression in the form: 
\begin{eqnarray}  
 \delta W_H = \Bigl[\sum^{n}_{i=1} p_{i} \delta q_i - H \delta t \Bigr]\Bigg|^{t_1(\alpha)}_{t_{0(\alpha)}} 
 + \int_{t_0(\alpha)}^{t_1(\alpha)} \sum^{n}_{i=1} \Bigl[ \frac{\partial L}{\partial q_{i}} - \frac{d}{d t} 
 \Bigl(\frac{\partial L}{\partial \dot{q}_{i}} \Bigr)\Bigr] \delta q_i dt \; \; , \; \; \label{deltH1}
\end{eqnarray} 
where the function $H$ is Hamiltonian of the system, while $p_{i}$ $(i = 1, \ldots, n)$ are the momenta. 

Let us define the optimal (or shortest) path between two spatial points $M_0$ and $M_1$. The optimal path 
satisfies the Lagrange equation $\frac{\partial L}{\partial q_{i}} - \frac{d}{d t} \Bigl(\frac{\partial 
L}{\partial \dot{q}_{i}} \Bigr) = 0$ and variation of the Lagrange action $W_L$ is always equal zero for 
the shortest path. It is clear that in this case the optimal path does not depend upon the parameter 
$\alpha$, i.e., for this path we can write the equation $q_i = q_i(t)$, where $i = 1, \ldots, n$ and all 
$q_i(t)$ functions do not depend on the parameter $\alpha$. If we consider the variation of the Hamilton 
action $W_H$, then the corresponding optimal path is also determined by the Lagrange equation. However, 
since in this case $t_0 = t_0(\alpha)$ and $t_1 = t_1(\alpha)$, then we obtain the whole $\alpha-$dependent 
family of optimal paths $q_i = q_i(t, \alpha)$, where $i = 1, \ldots, n$. On each of these optimal 
$\alpha-$dependent paths, the variation of $W_H$ is written in the form 
\begin{eqnarray}  
 \delta W_H(\alpha) = \delta \int_{t_0(\alpha)}^{t_1(\alpha)} L\Bigl(t, q_{i}(t,\alpha), 
 \dot{q}_{i}(t,\alpha)\Bigr) dt = \Bigl(\frac{\partial W_H(\alpha)}{\partial \alpha}\Bigr) \delta \alpha 
 = \Bigl[\sum^{n}_{i=1} p_{i} \delta q_i - H \delta t \Bigr]\Bigg|^{t_1(\alpha)}_{t_{0(\alpha)}} \; . 
 \; \label{deltH2}
\end{eqnarray}  
A different derivation of this formula can be found, e.g., \cite{GF}. For actual (or optimal) motion of the 
Hamiltonian system the variation of $W_H$ must be equal zero, i.e., $\delta W_H(\alpha) = 0$. 

Instead of the extended $(n + 1)$-dimensional coordinate space $\{ q_{i}(\alpha), t(\alpha) \}$, let us 
introduce the extended $(2 n + 1)-$dimensional phase space of the generalized coordinates $q_i$, momenta 
$p_i$ and time $t$ which we shall designate as the $\{ q_{i}(\alpha), p_{i}(\alpha), t(\alpha) \}$ space. 
In this phase space, we choose an arbitrary closed curve $C_0$, which is described by the equations: $q_i(0) 
= q_{i}(t_0, \alpha), p_i(0) = p_{i}(t_0, \alpha), t = t_0(\alpha)$, where $i = 1, \ldots, n$ and $0 \le 
\alpha \le \ell$. Note that for $\alpha = 0$ and for $\alpha = \ell$ we have the same point on the curve. 
This means that $q_{i}(0) = q_{i}(\ell)$ and $p_{i}(0) = p_{i}(\ell)$. At the next step, from each point of 
this closed curve $C_0$, as from the initial point, we draw the corresponding optimal path. Such an optimal 
path is uniquely determined (after we set the initial point on the curve $C_0$) from the system of 
Hamiltonian canonical equations. By choosing different initial points on the closed curve $C_0$ we obtain a 
tube of optimal (or shortest) paths each of which is also $\alpha-$dependent. The explicit equations for 
this tube of optimal paths are: $q_i = q_i(t, \alpha), p_i = p_i(t, \alpha)$, where $i = 1, \ldots, n, 0 
\le \alpha \le \ell$ and also $q_i(t, 0) = q_i(t, \ell)$ and $p_i(t, 0) = p_i(t, \ell)$ for $i = 1, \ldots, 
n$. Formally, each optimal path on the surface of this tube can be considered as some generatrix of the 
tube. 

On this tube, we select the second closed curve $C_1$, which also covers the tube and has only one common 
point with each generatrix. The equations of this closed curve $C_1$ are written in the form: $q_i(1) 
= q_{i}(t_1, \alpha), p_i(1) = p_{i}(t_1, \alpha), t = t_1(\alpha)$, where $i = 1, \ldots, n$ and $0 \le 
\alpha \le \ell$. Now, by integrating the equation $\delta W_H(\alpha) = 0$ over the parameter $\alpha$ 
in the range from $\alpha = 0$ to $\alpha = \ell$, one finds 
\begin{eqnarray}  
 0 &=& W_H(\ell) - W_H(0) = \int_{0}^{\ell} \Bigl[ \sum^{n}_{i=1} p_i \delta q_i - H \delta t 
 \Bigr]\Bigg|^{t_1}_{t_{0}} = \int_{0}^{\ell} \Bigl[ \sum^{n}_{i=1} p^{1}_i \delta q^{1}_i - H_1 \delta t_{1} 
 \Bigr] \nonumber \\
 &-& \int_{0}^{\ell} \Bigl[ \sum^{n}_{i=1} p^{0}_i \delta q^{0}_i - H_0 \delta t_{0} \Bigr] =\oint_{C_1} \Bigl[ 
 \sum^{n}_{i=1} p_i \delta q_i - H \delta t \Bigr] - \oint_{C_0} \Bigl[ \sum^{n}_{i=1} p_i \delta q_i - H 
 \delta t \Bigr] \; \; . \; \; \label{deltH3}
\end{eqnarray}  
or 
\begin{eqnarray}  
 \oint_{C_1} \Bigl[ \sum^{n}_{i=1} p_i \delta q_i - H \delta t \Bigr] = \oint_{C_0} \Bigl[ \sum^{n}_{i=1} p_i 
 \delta q_i - H \delta t \Bigr] \; \; . \; \; \label{deltH4}
\end{eqnarray}  
This means that the closed loop integral $I = \oint \Bigl[ \sum^{n}_{i=1} p_i \delta q_i - H \delta t \Bigr]$  
taken along an arbitrary closed contour does not change its value during an arbitrary displacement along the 
tube of straight paths. In other words, this integral $I$ is invariant, which is called the integral invariant 
of Poincare-Cartan (see, Eq.(\ref{7eq2}), in the main text) for the Hamiltonian dynamical system with the 
Hamiltonian $H$.

In pure mathematical language we have created and investigated the rotor tube for the differential form $\omega^{1} 
= {\bf p} \; d{\bf q} - H dt$, where ${\bf p} = (p_1, \ldots, p_n)$ and ${\bf q} = (q_1, \ldots, q_n)$. The rotor 
lines, each of which is the generatrix of this tube, are uniformly determined by the canonical Hamiltonian equations. 
In other words, the true Hamiltonian trajectories (or integral curves of the canonical Hamilton equations) are the 
generatrixes of the rotor tube for the differential form $\omega^{1} = {\bf p} \; d{\bf q} - H dt$. Note that about 
integral invariants for Hamiltonian systems it is possible to talk endlessly, but for now it is better to stop here. 
The last thing which we want to present in this Appendix is the explicit formula for the differential of the 
$\omega^{1}$ form, which is the 2-form $d \omega^{1}$ \cite{Fland}:
\begin{eqnarray}  
 d \omega^{1} = \sum^{n}_{i=1} \Bigl( d p_i \wedge d q_i - \frac{\partial H}{\partial p_i} d p_i \wedge d t 
  - \frac{\partial H}{\partial q_i} d q_i \wedge d t\Bigr) = 0 \; \; , \; \; \label{deltH5}
\end{eqnarray} 
where $\wedge$ is the standard notation of exterior product \cite{HCartan}, \cite{Fland}. The matrix $\hat{A}$ of
this 2-form $d \omega^{1}$ in the $(2 n + 1)-$dimensional phase space is: 
\begin{eqnarray}
  \hat{A} = \left( \begin{array}{ccc}
  0 & - E & \frac{\partial H}{\partial {\bf p}} \\
  E &  0  & \frac{\partial H}{\partial {\bf q}} \\ 
 - \frac{\partial H}{\partial {\bf p}} & - \frac{\partial H}{\partial {\bf q}} & 0 \\
\end{array} \right) \; \; , \; \, {\rm where} \; \; \; \left( \begin{array}{cc}
  0 & - E \\
  E &  0  \\ 
\end{array} \right) = \hat{J} \label{C1}
\end{eqnarray}
and $\hat{J}$ is the unit simplectic $2 n \times 2 n$ matrix which has been defined and discussed in 
the Appendix B. The rank of this matrix $\hat{A}$ equals $2 n$ (the rank of $\hat{J}$ matrix), i.e., 
it has one zero-eigenvalue. The corresponding eigenvector is $(-\frac{\partial H}{\partial {\bf q}}, 
\frac{\partial H}{\partial {\bf p}}, 1)$ and it determines the direction of rotor lines for the 
differential form $\omega^{1} = {\bf p} \; d{\bf q} - H dt$. More about the uniqueness of integral 
invariants, their classification and integral invariants of higher orders can be found, e.g., in 
\cite{Lee1947}.

\end{document}